\documentclass[preprint,11pt]{elsarticle}

\usepackage{lipsum}
\usepackage{mathtools}
\usepackage{graphicx}

\usepackage{amsmath}    
\usepackage{amssymb}
\usepackage{relsize}
\usepackage{enumitem}
\usepackage{bm}
\usepackage{hyperref}
\usepackage{latexsym}
\usepackage{verbatim}
\usepackage{float}
\usepackage{placeins}
\usepackage{color}
\usepackage[caption=false]{subfig}

\usepackage{lipsum}
\usepackage{mathtools}
\usepackage{graphicx}
\usepackage{amsmath}    
\usepackage{amssymb}
\usepackage{relsize}
\usepackage{enumitem}
\usepackage{bm}
\usepackage{hyperref}
\usepackage{latexsym}
\usepackage{verbatim}
\usepackage{float}
\usepackage{placeins}
\usepackage{color}
\usepackage[caption=false]{subfig}

\usepackage[a4paper, left=0.60in, right=0.60in, top=0.70in, bottom=0.75in]{geometry}

\def\beq{\begin{equation}}
\def\eeq{\end{equation}}

\def\beq{\begin{equation}}                          
\def\eeq{\end{equation}}                          
\def\bea{\begin{eqnarray}}                          
\def\eea{\end{eqnarray}}

\DeclareRobustCommand{\uvec}[1]{{%
  \ifcsname uvec#1\endcsname
     \csname uvec#1\endcsname
   \else
    \bm{\hat{\mathbf{#1}}}%
   \fi
}}

\begin{document}

\begin{frontmatter}



\title{Dynamical Swirl Structures Powered by Microswimmers in Active Nematics}


\author[inst1]{Partha Sarathi Mondal}
\author[inst1]{Pawan Kumar Mishra}
\author[inst2]{Tam\'{a}s Vicsek}
\author[inst1]{Shradha Mishra}

\affiliation[inst1]{organization={Department of Physics, Indian Institute of Technology (BHU) Varanasi},
            addressline={}, 
            city={Varanasi},
            postcode={00000}, 
            state={221005},
            country={India}}

\affiliation[inst2]{organization={Department of Biological Physics, Eötvös University},
            addressline={Pázmány Péter sétány 1A}, 
            city={Budapest},
            postcode={H-1117},
            country={Hungary}}

\begin{abstract}
Active nematics, in their pure form, have demonstrated a plethora of dynamic and steady-state behaviors, including large-scale dynamic structures, collective flows, and intricate multi-spatial temporal dynamics. This complexity further increases in the presence of external polar agents. We investigate active nematics interspersed with polar microswimmers, akin to active apolar cells infused with active impurities (microswimmers). Our comprehensive numerical study reveals that varying the microswimmers' motility induces a novel spatiotemporal state in the active nematics backdrop. This state is marked by macroscopic swirl-like structures and a reduction in the overall order of the active nematics. Interestingly, this state emerges at intermediate motility levels, where microswimmers form local clusters and exhibit coherent motion. However, at higher motility levels, the swirls become less coherent, and microswimmer clustering intensifies. We show that the effect of the polar microswimmers on active nematics can be interpreted as a spatiotemporally correlated colored noise on active nematics, which promotes bend instability in active nematics, leading to the observed swirling dynamics. Our findings indicate that the spatiotemporal states are highly sensitive to the microswimmers' motility, offering potential avenues for pathogen identification based on known motility characteristics.\\
\end{abstract}

\begin{highlights}
\item A class of active systems has aligning motile impurities which we interpret as microswimmers
\item Previous studies addressing inhomogeneities in active systems focused on quenched or diffusive impurities.
\item The presence of microswimmers leads to dynamic swirl structures and topological defects in the system.
\item Microswimmers influence active nematic in a way which is analogous to the effect of a spatiotemporally correlated coloured noise.
\item Our findings offer potential avenues for pathogen identification in active systems.
\end{highlights}

\begin{keyword}
Active Matter \sep Active Nematic \sep Polar Nematic Mixture \sep Spatiotemporally Correlated Colored Noise, Topological Defects, Swirl Structures, Bending in Active Nematics
\end{keyword}

\end{frontmatter}


\section{Introduction} \label{sec:secI}
Active matter systems are comprised of motile agents known as Self Propelled Particles (SPPs) \cite{vicsek1995, bechinger2016active,bar2020self,ganguly2013aspects,vicsek2012collective}, which absorb energy from their environment and convert it into mechanical motion. This distinctive interplay between energy dissipation and self-propulsion sets active matter systems apart from their equilibrium counterparts \cite{vicsek2012collective,toner2005,marchetti2013,ramaswamy2017active,das2020introduction,needleman2017active,fodor2016far,o2022time}. These systems have garnered significant interest due to their prevalence across various scales, ranging from microscopic to macroscopic levels. Examples include cytoskeletal filaments \cite{alberts2002self},  molecular motors \cite{kolomeisky2007molecular}, bacterial colonies \cite{niu2012bacterial,shapiro1995significances,ohgiwari1992morphological,wu2009periodic}, schools of fish \cite{jolles2017consistent}, herds of animals \cite{moreno2020search,gueron1996dynamics}, and flocks of birds \cite{wang2019collective}. The classification of self-propelled particles into active polar or active nematics categories, based on their symmetry, highlights substantial differences in their dynamics. This classification aids in selecting the appropriate category to address specific problems within these systems.\\
Numerous studies \cite{mishra2010fluctuations,pattanayak2018collection,singh2020phase,ginelli2010large,mishra2006active,mishra2014giant,thampi2014instabilities,das2017order,henkes2011active,mishra2022active,zhang2018interplay} have explored the collective dynamics of SPPs in both pure polar and pure nematic systems. However, the introduction of a foreign species into these systems adds a layer of complexity to their behavior. When species with differing alignment tendencies are mixed, the resulting system may exhibit features absent in pure systems. Interactions both within and between species in the mixture dictate how each species responds to the presence of the others, unveiling a host of novel characteristics.\\
Additionally, if the foreign agents are nonmotile, then the system falls into a distinct subclass, where the immobile species acts as a  disorder \cite{mccandlish2012spontaneous,chepizhko2013optimal,sandor2017dynamic,singh2021bond,kumar2022active,kummel2015formation,chepizhko2015active,vahabli2023emergence,das2018polar,das2020nonquenched,pattanayak2019enhanced}. The role of disorder of various origins in active systems has recently gathered significant interest as it mirrors the complexities encountered by SPPs in natural, heterogeneous environments, resulting in noticeable alterations in their dynamics compared to homogeneous environments. One key observation from these studies indicates that the presence of disorder can induce rotational tendencies among the SPPs within the systems. While most of these investigations have focused on quenched disorder, it was demonstrated in reference \cite{chepizhko2015active,yllanes2017many} that the influence of diffusing obstacles on the system could vary depending on the dynamics of the obstacles.\\
In this context, a new subclass known as Living Liquid Crystal (LLC) \cite{gruler1995migrating,zhou2014living,vats2023symbiotic,turiv2020polar,vats2024surface,kumar2013motility} has emerged, where bacteria function as polar particles immersed in a Liquid Crystal (LC). LLC systems demonstrate intriguing characteristics, such as the formation of unique structures \cite{zhou2014living,zhou2017dynamic,mushenheim2014dynamic,sokolov2015individual,chi2020surface}, accumulation of bacteria at defect cores \cite{genkin2017topological}, and enhanced dynamics of +1/2 defects in LC \cite{genkin2017topological}, among others. In reference \cite{sampat2021}, the authors explored the LLC system within the realm of active matter. The authors analyzed the behavior of the background active nematics in the presence of polar particles representing bacteria. By varying the bacterial density, they observed that the global orientational order of the nematic exhibits a nonmonotonic behavior, accompanied by complex patterns consisting of dynamic circular patterns in the orientation field of the nematic for intermediate densities of polar particles.\\
In this study, we investigate a mixture of active polar and nematic species akin to that of reference  \cite{sampat2021}. We address an important question: How does the activity of the polar particles affect the apolar particles? To this end, we focus on the small densities of polar species in the mixture and adjust the activity of the polar particles while keeping their density constant.\\
The primary observations of our study are as follows: (i) we observe a nonmonotonic behavior in the global orientational ordering of the active nematics upon tuning the activity of polar particles; (ii) the coupling between apolar and polar particles induces collective bending of the nematic, resulting in the formation of swirled and strongly bend structures in the nematic. Furthermore, we propose that (iii) the influence of polar particles on the active nematics can be interpreted as spatiotemporally correlated colored noise acting on the active nematic.\\
The remainder of the paper is organized as follows: In Section \ref{sec:secII}, introduces our model and details of the parameters utilized for numerical simulation. Section \ref{sec:secIII} is dedicated to discussing the results of our study. Finally, in Section \ref{sec:secIV}, we summarize our findings and investigate potential applications and future directions for further research.

\section{Model}\label{sec:secII}
\begin{figure}[htp]
    \centering
    \includegraphics[width=0.80\linewidth]{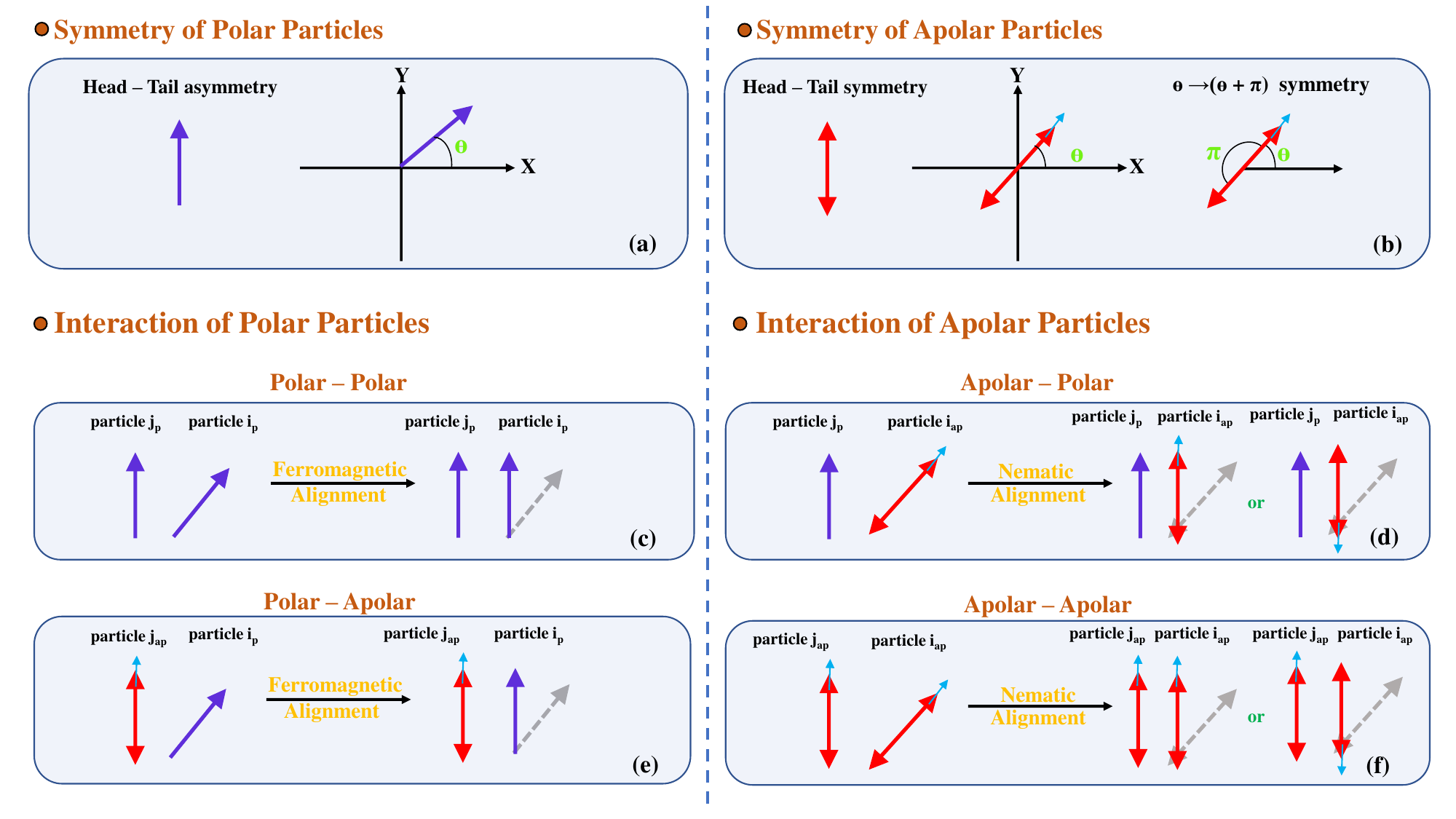}
    \caption{(color online) The cartoon illustration serves as a visual depiction of our model,  distinguishing polar particles through single-headed blue arrow and apolar particles through double-headed red arrow. The cyan-colored arrows on each apolar particle depict their instantaneous velocity direction. Subplots (a) and (b) convey insights into the symmetries and characteristics of these two particle types within our system.  In subplot (a), the arrow head signifies the self-propulsion direction of a polar particle, and its orientation is measured by the angle that the arrow head makes with the positive $X$-direction. In contrast, an apolar particle exhibits head-tail symmetry, facilitating its movement in both directions along its axis with equal probability. The orientation angle of an apolar particle is defined by the angle its instantaneous velocity direction forms with the positive $X$-direction as illustrated in subplot (b). This representation also highlights the intrinsic $\theta \to \theta + \pi$ symmetry exhibited by apolar particles. Subplots (c)-(f) elaborate on the interactions between different particle types in our system. Instances before and after interactions are marked by a black arrow symbolizing the arrow of time. Polar particles engage in ferromagnetic interactions with neighbors, seeking to align parallel to the instantaneous velocity direction of their neighbors. Subplots (c) and (e) depict the interaction of a single polar particle ($i_p$) with a polar ($j_{p}$) and an apolar ($j_{ap}$) neighbour, respectively. In both subplots, the dashed single-headed grey arrow on the right side of the black arrow represents the orientation of particle $i_p$ before interaction. Conversely, apolar particles participate in nematic interactions with neighbors, allowing alignment either parallel or antiparallel to the instantaneous velocity direction of each neighbor with equal probability. Subplots (d) and (f) illustrate the alignment mechanism of an apolar particle ($i_{ap}$) with a polar particle ($j_p$) and an apolar particle ($j_{ap}$), respectively. In both subplots, the dashed double-headed grey arrow on the right side of the black arrow represents the orientation of particle $i_{ap}$ before interaction.}
    \label{fig:1}
\end{figure}

Our study examines a two-dimensional system featuring a mix of polar and apolar active particles. Each particle (polar or apolar) in the system is characterized by a position vector $\boldsymbol{r}_{i,p/ap} = (x_{i,p/ap},y_{i,p/ap})$ relative to an arbitrary origin and an orientation vector $\theta_{i,p/ap}$, where, subscripts $p$, and $ap$ refer to polar and apolar particles, respectively. The quantification of orientation for both types of particles is elaborated in FIG.\ref{fig:1}. In the absence of perturbations, the polar particles exhibit unidirectional motion along the direction of their head and are referred to as polar swimmers, while the apolar particles display bidirectional motion along their long axis, adhering to the symmetry of $\theta \to \theta + \pi$. Self Propulsion velocities of polar and apolar particles are denoted by $v_p$ and $v_{ap}$, respectively. \\
The particles interact through the alignment interaction, with the neighbours within the interaction radius as introduced in reference \cite{vicsek1995}. The nature of alignment solely depends on the intrinsic symmetry of the particles under consideration. For polar particles, alignment is ferromagnetic, compelling them to move along the mean direction of motion of their neighbors. Conversely, apolar particles exhibit nematic alignment, randomly choosing between parallel and antiparallel alignment with the mean direction of orientation of their neighbors. Imperfections in alignment arise from errors made by the particles in following  their neighbors, introducing inherent fluctuations into the system, such that the orientation of each particle deviates by a small angle from the mean of its neighbours. Such noise is named as scalar noise in previous studies \cite{chate2008collective,nagy2007new}.
The specifics of particle symmetry and interactions are elucidated in FIG.\ref{fig:1}. Further details outlining the dynamical equations of both particle types are provided in \ref{app:A}.\\
The system is simulated in a square box of size $L \times L$ subjected to periodic boundary conditions to study bulk properties of the mixture and minimize boundary effects. The density of apolar particles ($\rho_{ap} = \frac{N_{ap}}{L^2}$) is kept constant at 1.0, while the density of polar particles ($\rho_p = \frac{N_{p}}{L^2}$) is maintained at a low value, $\rho_p \in [0.02, 0.03, 0.04, 0.05]$.  In this setup, the self-propulsion velocity of apolar particles ($v_{ap}$) is kept fixed at $0.25$, while we vary the velocity of polar particles ($v_p$) from $0$ to $0.25$, encompassing the activity ratio $\mu = \frac{v_p}{v_{ap}}$ within the interval $[0, 1]$. At low velocity, the  polar particles have very slow dynamics. To overcome this, we add a small diffusive part with magnitude $V_R = 10^{-3}$ in the position update of the polar particles.\\ 
Eq.(\ref{equn:a1}-\ref{equn:a4}) are solved numerically with step size $\Delta t = 1$. One simulation step is counted when the position and orientation of all the particles (polar+apolar) are updated once. The results presented are based on simulations using a $100 \times 100$ system size unless otherwise specified. For this system size, the system is simulated for a total of $1.7 \times 10^5$ time steps, of which in the initial $1.3 \times 10^5$ steps, the system is allowed to reach  the steady state and the remaining $0.4 \times 10^5$ steps used for calculation of the observable. Additionally, some results are showcased for a system size  $130 \times 130$, wherein the total number of simulation steps is increased to {\bf $2.5 \times 10^5$} for each parameter set $[v_p, \rho_p]$. The average over at least {\bf $15$} independent realizations are performed to ensure better statistics. Other fixed parameters in our model include $R_p = R_{ap} = 1.0$ (interaction radii) and $\eta = 0.2$ (strength of noise).

\begin{figure*}
\centering
\includegraphics[width=0.47\textwidth]{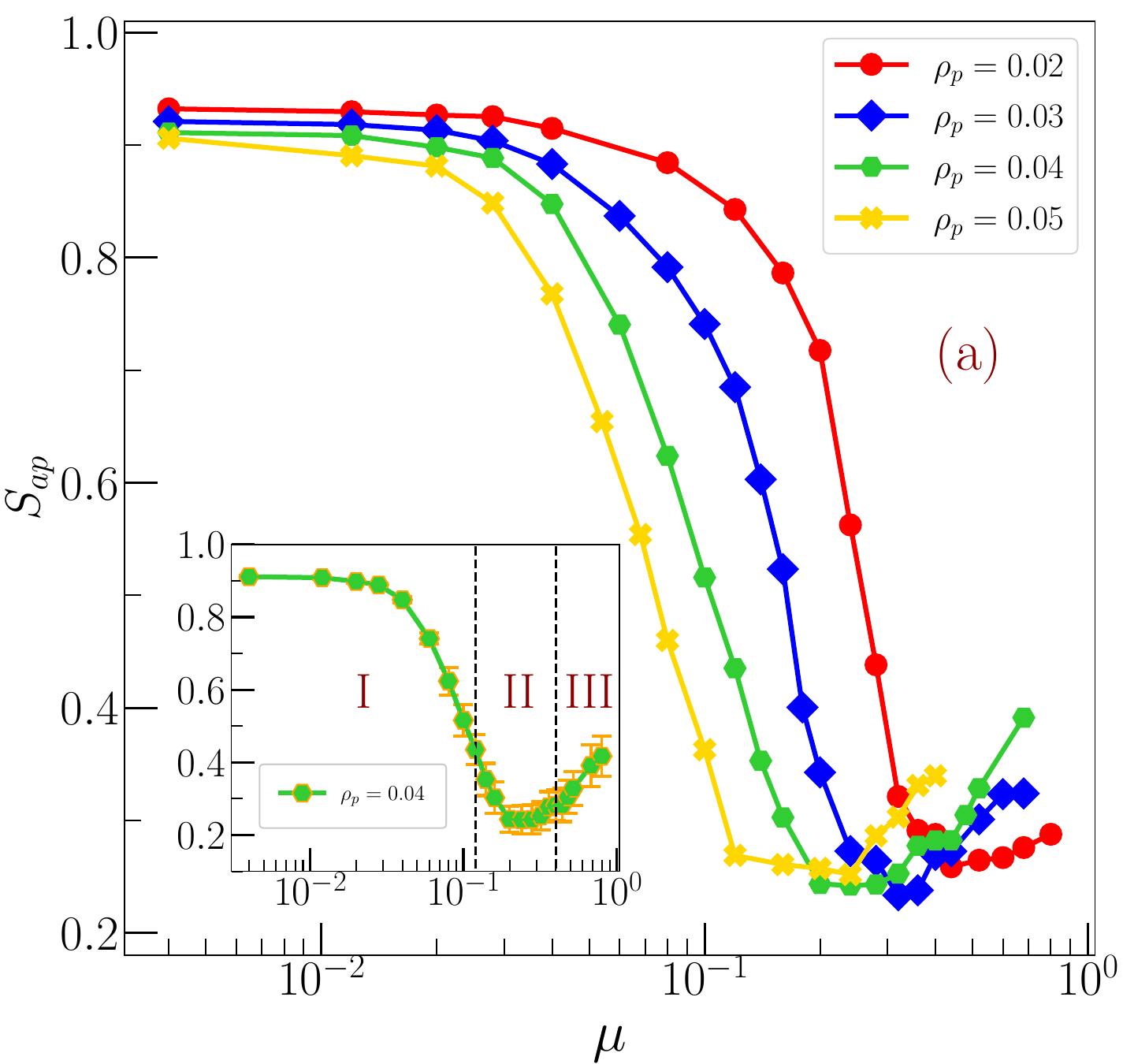}

\caption{(color online) The figure illustrates the variation of $S_{ap}(\mu)$ with $\mu$. This non-monotonic behavior is consistent across different values of $\rho_p = 0.02, 0.03, 0.04, 0.05$. (Inset) Three distinct states are labeled as I, II, and III. Vertical dashed lines indicate the boundaries between the different states for $\rho_p = 0.04$. The errorbars show the uncertainty in the value of $S_{ap}$. System size, $L=100$;}
\label{fig:2}
\end{figure*}

\begin{figure*}
\centering

\subfloat[]{
\begin{minipage}{0.27\textwidth}
  \includegraphics[width=\textwidth]{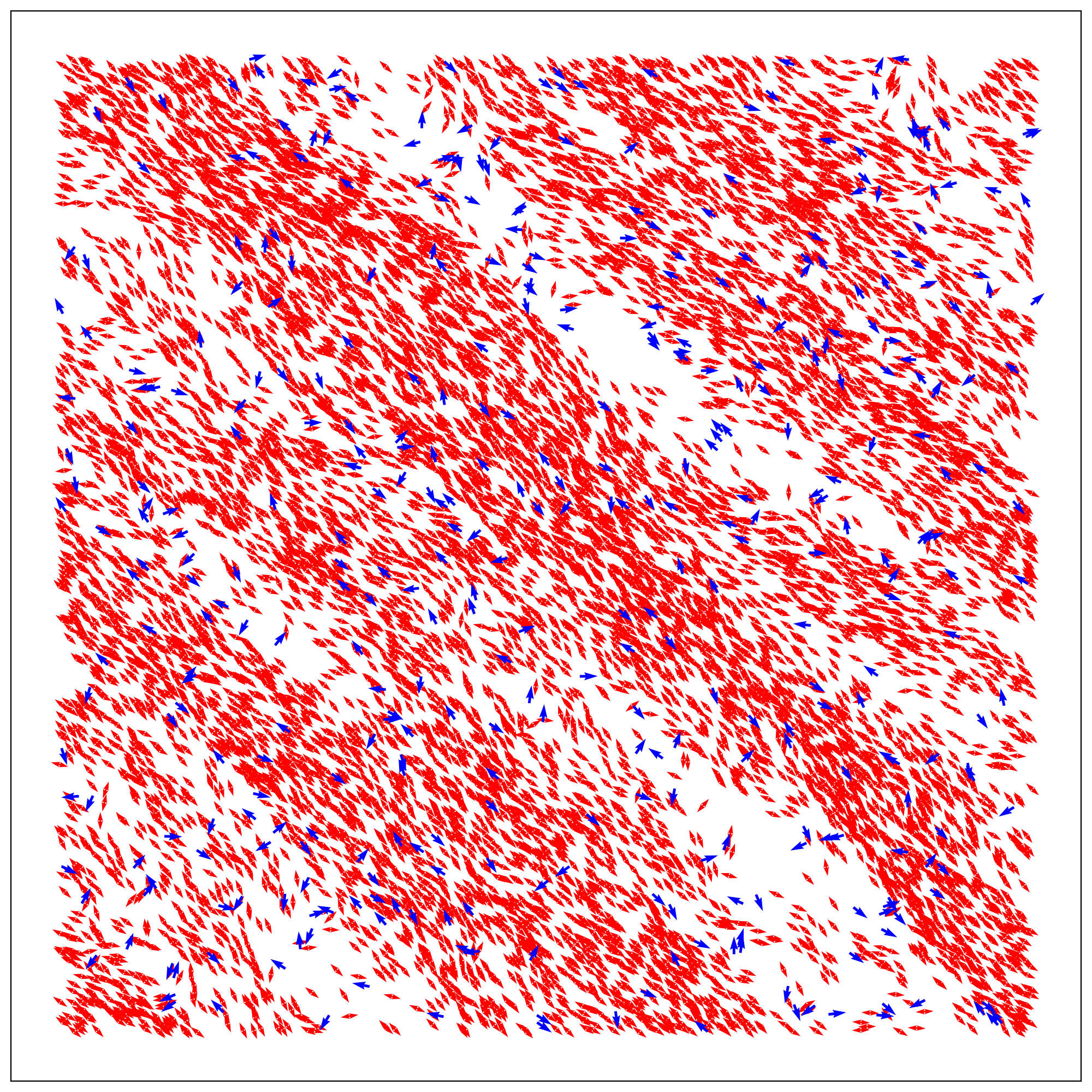} \\[5pt]
  \includegraphics[width=\textwidth]{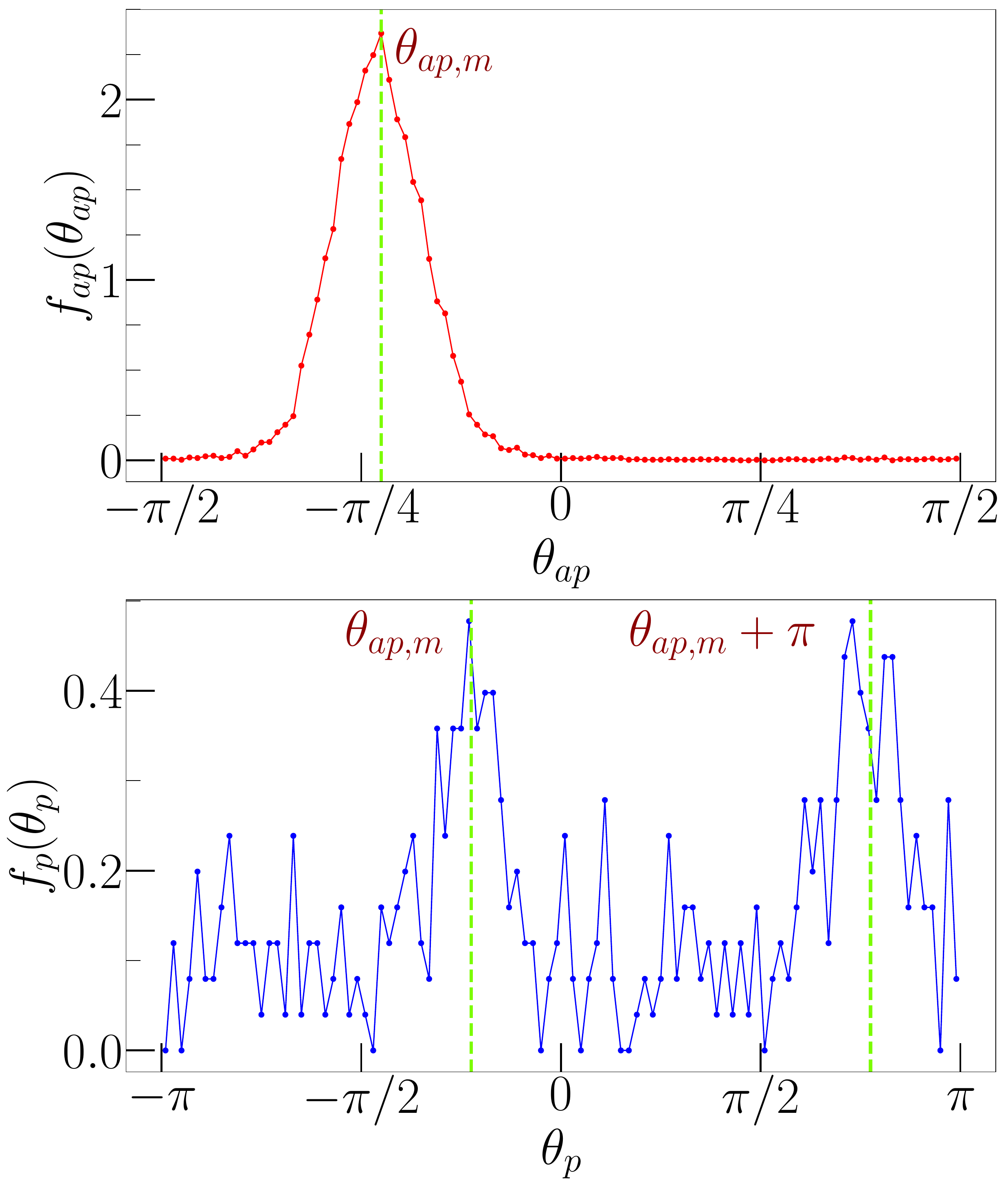}
\end{minipage}%
}
~
\subfloat[]{
\begin{minipage}{0.27\textwidth}
  \includegraphics[width=\textwidth]{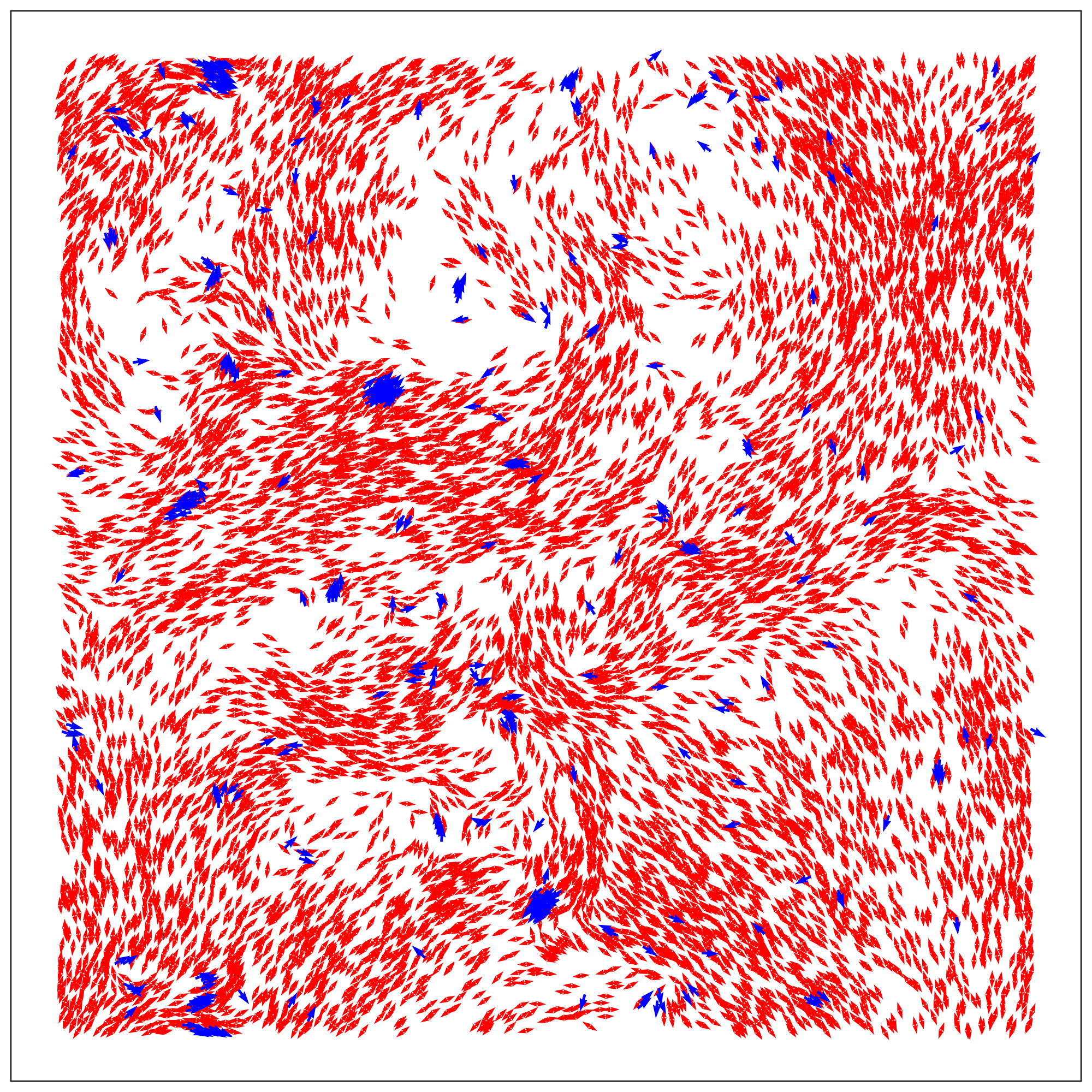} \\[5pt]
  \includegraphics[width=\textwidth]{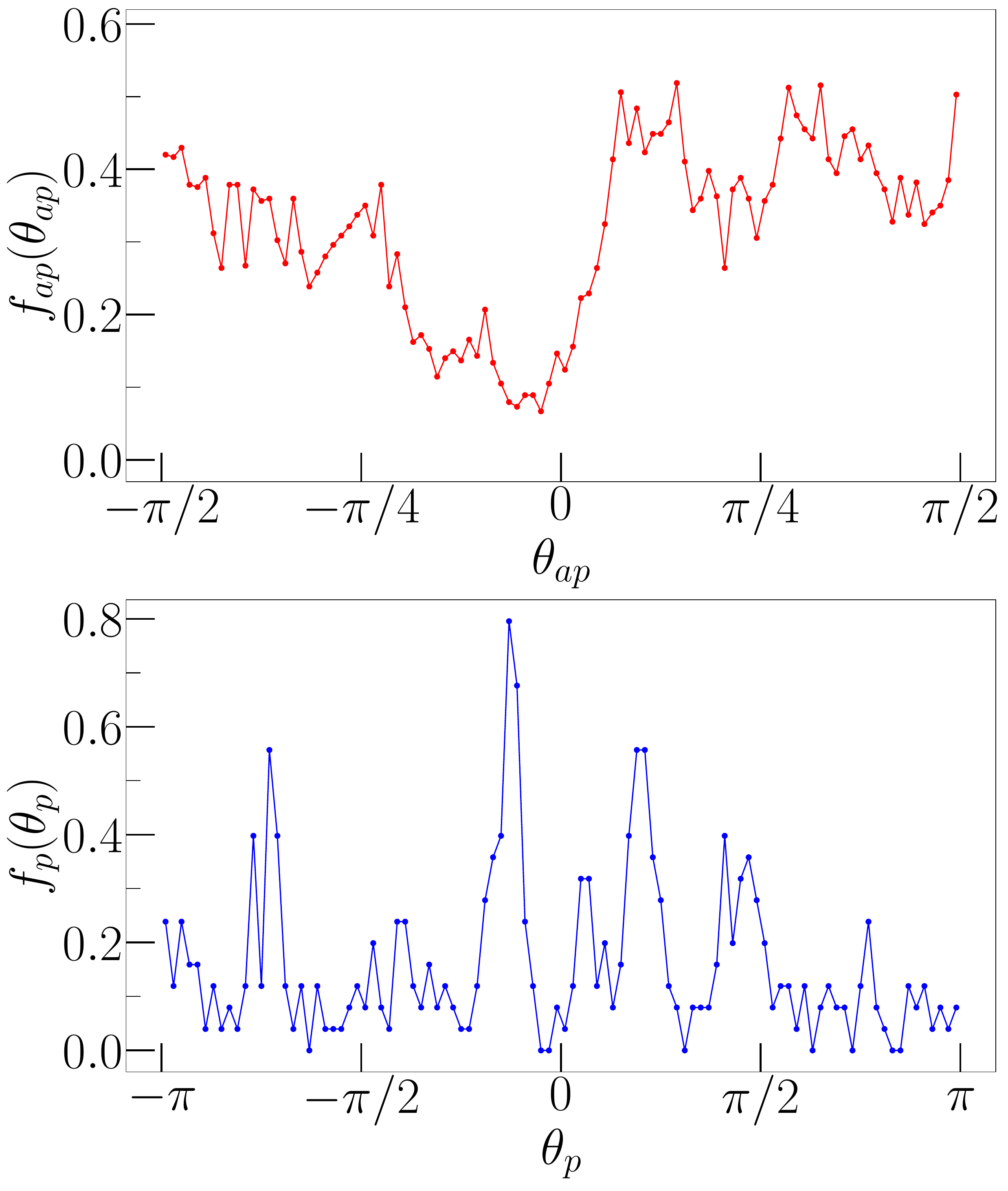}
\end{minipage}%
}
~
\subfloat[]{
\begin{minipage}{0.27\textwidth}
  \includegraphics[width=\textwidth]{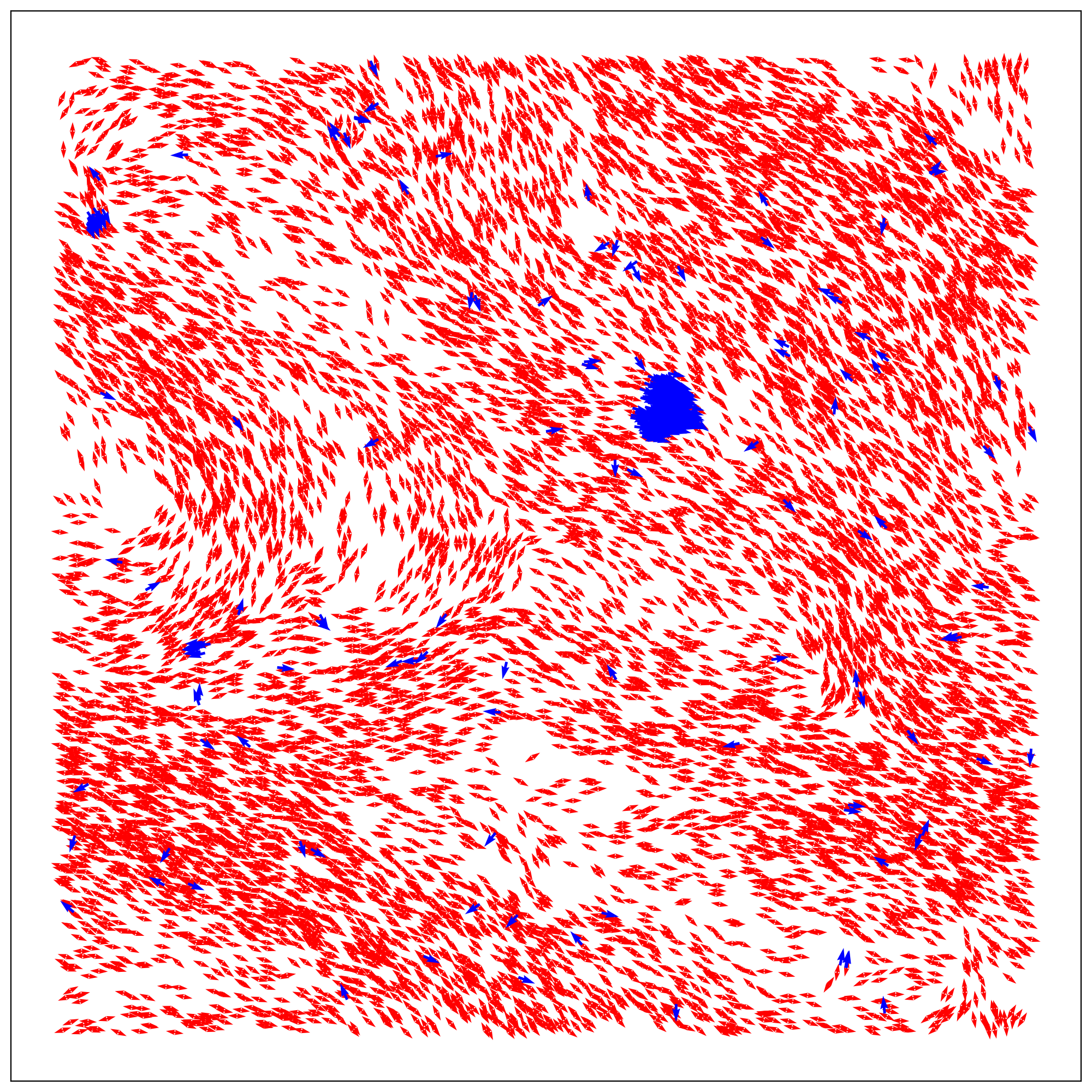} \\[5pt]
  \includegraphics[width=\textwidth]{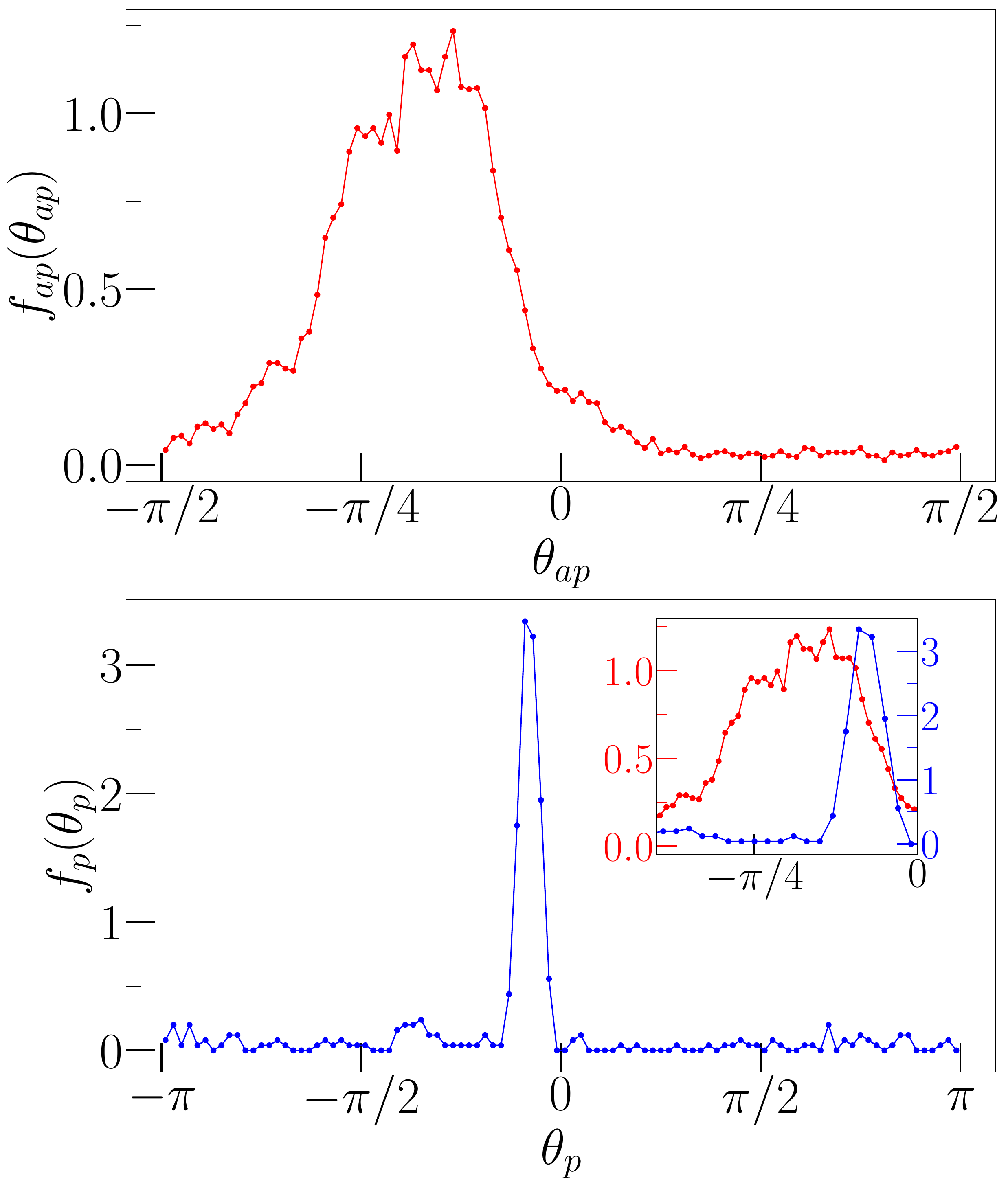}
\end{minipage}%
}
\caption{(color online) The top panel displays snapshots of the system in three different States: (a) State-I: $\mu = 0.004$, (b) State-II: $\mu = 0.20$, and (c) State-III: $\mu = 0.68$. The OPDF of apolar, $f_{ap}(\theta_{ap})$ vs. $\theta_{ap}$, and polar, $f_p(\theta_{p})$ vs. $\theta_{p}$ (corresponding to the snapshots in the top panel) is shown below the snapshot (middle panel-apolar particles, $\&$ bottom panel-polar particles) for each State. In subplot (a), the green dashed line in the plot $f_{ap}(\theta_{ap})$ vs. $\theta_{ap}$ marks the position of the maxima, referred to as $\theta_{ap,m}$. In the $f_p(\theta_{p})$ vs. $\theta_{p}$ plot, two dashed green lines show that the two peaks are positioned approximately at $\theta_{ap,m}$ and $\theta_{ap,m}+\pi$. In subplot (c), the inset shows the overlapping peaks of in the OPDF of polar and apolar particles, indicating that the enhanced ordering in State-III originates from the coherent motion of the polar clusters; $\rho_p = 0.04$ and $L = 100$. }
\label{fig:3}
\end{figure*}

\begin{figure*}[ht]
\begin{center}
\includegraphics[width=0.95\linewidth]{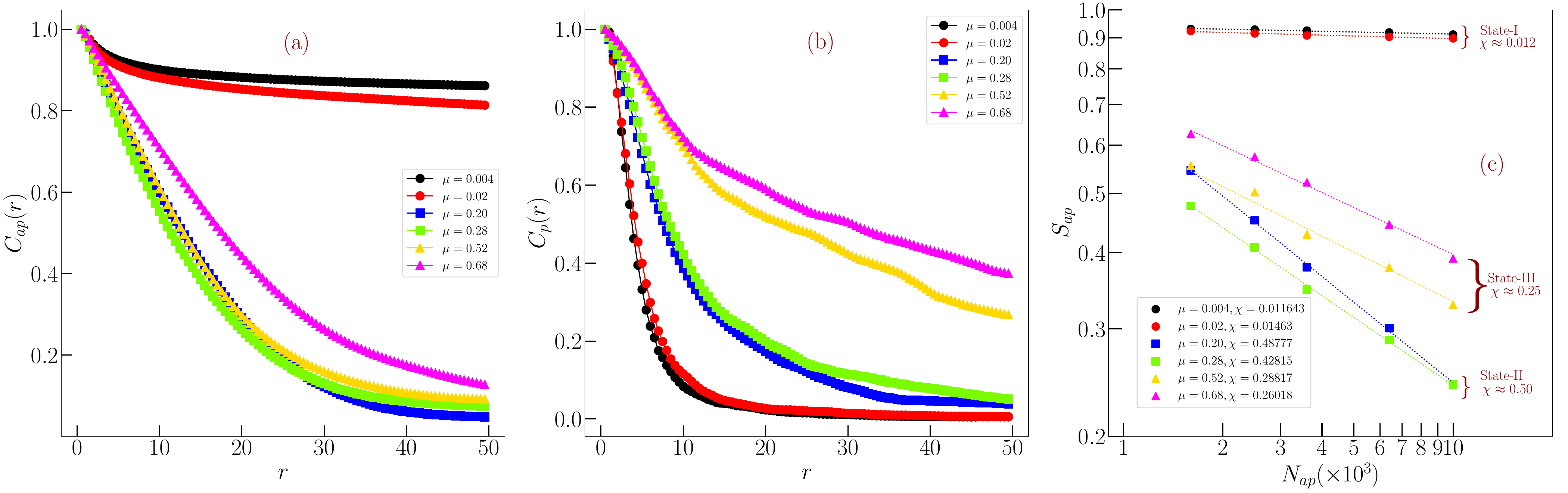}
\caption{(color online) The figure illustrates : the behavior of the two-point correlation functions for (a) Apolar Particles ($C_{ap}(r)$ versus $r$) and (b) Polar Particles ($C_p(r)$ versus $r$) in different States: State-I ($\mu =$ $0.004$ $\&$ $0.02$), State-II ($\mu =$ $0.20$ $\&$ $0.28$), and State-III ($\mu =$ $0.52$ $\&$ $0.68$); (c) System size dependence of Nematic Order Parameter, $S_{ap}$, plotted in log-log scale. The data shown for different values of $\mu$ in  State-I ($\mu =$ $0.004$ $\&$ $0.02$), State-II ($\mu =$ $0.20$ $\&$ $0.28$), and State-III ($\mu =$ $0.52$ $\&$ $0.68$) shows power-law variation. The dotted lines show power-law fit: $S_{ap} \approx N_{ap}^{\chi}$. Rest of the parameters are same as FIG.\ref{fig:3}.}
\label{fig:4}
\end{center}
\end{figure*}

\section{Results}\label{sec:secIII}
\subsection{Nematic Order Parameter ($S_{ap}$)} \label{sec:secIIIA}

To quantify the degree of order in the apolar particles, we use the nematic scalar order parameter defined as:

\begin{equation*}
    S_{ap} =\Bigg<2\sqrt{\bigg(\overline{\cos^2\theta}  -\frac{1}{2}\bigg)^2 + \bigg(\overline{\sin \theta \cos\theta}\bigg)^2} \Bigg>
\end{equation*}

where the `bar' denotes an average over all the particles, and $<...>$ denotes an average over time and 15 different realizations. The uncertainty in the order parameter value is represented as the errorbar which is estimated by the standard deviation of $S_{ap}$ over different realizations.  A value of $S_{ap} \approx 1$ indicates the presence of global orientational order among the apolar particles, while $S_{ap} \approx 0$ corresponds to a state with no global orientational order. FIG.\ref{fig:2} presents a plot of $S_{ap}$ as a function of $\mu$ for four different densities of polar particles ($\rho_p = 0.02, 0.03, 0.04$, and  $0.05$). In FIG.\ref{fig:2}, three distinct regimes can be easily identified based on the values of $S_{ap}$, referred to as different States. We have labeled these three states in the inset of FIG.\ref{fig:2}.\\
First, we analyze the static properties of the three states using the following observables:\\
1. The probability density function of orientation (OPDF) of apolar ($f_{ap}(\theta_{ap})$ vs. $\theta_{ap}$) and polar particles ($f_p(\theta_{p})$ vs. $\theta_{p}$). OPDF $f(\theta) \Delta\theta$ represents the probability that a randomly selected particle will have orientation in the range $\bigg[\theta-\frac{\Delta\theta}{2},\theta+\frac{\Delta\theta}{2}\bigg)$. The details of the calculation of OPDF of polar and apolar particles are provided in \ref{app:B} \\
2. Spatial two-point correlation functions of the orientation of polar ($C_p(r)$) and apolar ($C_{ap}(r)$) particles. These correlations are defined as:
	
	\begin{equation}
	C_p(r)=\bigg<\cos(\theta_{p}(r_0+r)-\theta_{p}(r_0)) \bigg>
		\label{equn:5}
	\end{equation}
	
	\begin{equation}
	C_{ap}(r)=\bigg<\cos(2\theta_{ap}(r_0+r)-2\theta_{ap}(r_0)) \bigg>
		\label{equn:6}
	\end{equation}
where $<....>$ implies an average over all the particles as well as time in steady state and 15 independent realizations.
We extract the correlation lengths, $L_p$ and $L_{ap}$ for polar and apolar particles, respectively, as the points (value of r) where the correlation function $C_p(r)$ and $C_{ap}(r)$ crosses the $0.5$ value. These correlation lengths provide a measure of the characteristic length scales over which the orientation of the polar and apolar particles in the mixture are correlated.\\
The snapshots of the system configuration in each state and the corresponding OPDF of apolar and polar particles are shown FIG.\ref{fig:3}(a-c). The plots of the correlation function $C_{ap}(r)$ and $C_{p}(r)$, for apolar and polar particles, respectively, in different states are depicted in FIG.\ref{fig:4}(a-b). In State-I, for low $\mu$ values, the system demonstrates a strong nematic ordering, characterized by a high $S_{ap}$ value and a single sharp peak in the OPDF of apolar particles depicted in FIG.\ref{fig:3}(a). The persistent alignment of apolar particles is further evident from the slow decay of $C_{ap}(r)$, displaying a power-law behavior with a very small exponent for small $r$ and saturating at large $r$ as can be seen in FIG.\ref{fig:4}(a). However, in the case of polar particles, the effect of activity is overshadowed by the diffusion term, causing random motion of isolated polar particles, resulting in the presence of multiple small distinct peaks in OPDF of polar particles, as shown in FIG.\ref{fig:3}(a). The swift decline of the correlation function, $C_p(r)$, further underlines the minimal interaction between polar particles. However, the OPDF of polar particles displays two relatively stronger peaks at $\theta_{ap,m}$ and $(\theta_{ap,m} + \pi)$, suggesting that due to alignment interactions, individual polar particles orient themselves parallel to the local nematic orientation direction. Hence, at lower values of $\mu$, the orientation of polar particles is slaved by the global nematic orientation. Due to small $\mu$, polar particles are almost immobile and act like obstacles for apolar particles. Hence the dynamic nature of polar particles is suppressed for very small $\mu$. This leads to the formation of stripe structure in apolar particles,  as shown in FIG.\ref{fig:3}(a), instead of macroscopic bands observed in studies of pure active nematics \cite{mishra2006active,sampat2021,chate2006,dey2012spatial}. Similar stripe patterns are found previously for the collection of polar self-propelled particles at high velocity \cite{nagy2007new}.\\ 
As $\mu$ increases, the influence of polar particle activity becomes more pronounced, causing interactions and hence alignment among the polar particles. This leads to the emergence of small clusters comprising only a few polar particles. This developing spatial correlation among the polar particles is evident from $C_p(r)$, while the multiple randomly positioned peaks in the OPDF of polar particles indicate the random movement of these clusters, as shown in FIG.\ref{fig:3}(b). In State-II, the random motion of these polar particle clusters significantly impacts the nematic ordering of apolar particles. Consequently, there is a considerable decrease in $S_{ap}$, highlighting the breakdown of orientational order among the apolar particles, resulting in a broadened OPDF of apolar particles as shown in FIG.\ref{fig:3}(b). The disordered nature of the apolar particles is further reflected in the sharp decay of $C_{ap}(r)$, illustrating the lack of spatial correlations among them. The system snapshot in FIG.\ref{fig:3}(b) visually represents this disordered nature of apolar particles. In this State, the system configuration, as shown in FIG.\ref{fig:3}(b), underscores the dominant influence of polar-apolar interactions over apolar-apolar interactions, significantly impacting the overall behavior of the system.\\
Continuing to increase $\mu$, the polar particle clusters exhibit growth in size and the emergence of coherent motion, evidenced by the sharp peak in the OPDF of polar particles, (FIG.\ref{fig:3}(c)) and the slow decay of the two-point correlation function, $C_p(r)$ (FIG.\ref{fig:4}(a)), eventually terminating at a non-zero value. In this State, most of the polar particles are contained in one or two coherently moving clusters, while a small fraction of polar particles form tiny clusters moving randomly. 
This steady motion of polar clusters contributes to increased nematic ordering in State-III, as evident from the rise in $S_{ap}$ and the reappearing peak in OPDF of apolar particles shown in  FIG.\ref{fig:3}(c)). This increase in ordering ($S_{ap}$) is solely because of the interaction of apolar particles with the coherently moving polar clusters. The coinciding peaks in the OPDF of apolar and polar particles (inset of FIG. \ref{fig:3}(c)) further validate this observation.\\
Further, to unravel the internal structure of the nematic ordering in different states, we conducted a comprehensive study of its system size dependence. Interestingly, our investigation revealed  a power-law variation of $S_{ap}$ with $N_{ap}$, expressed as $S_{ap} \approx N_{ap}^{-\chi}$ {\cite{chate2006}}, depicted in FIG.\ref{fig:4}(c). In State-I, the system exhibited a Quasi Long Range Order, evident from $\chi < \dfrac{1}{16}$, which align perfectly the power-law decay observed in $C_{ap}(r)$. Upon entering State-II, $\chi$ increases to approximately $0.50$. This pronounced change in the value of $\chi$ signifies a transition from an ordered to a disordered state for the orientation of the apolar particles.  In State-III, we found that $\chi$ approached $0.25$, indicating a slightly enhanced ordering compared to State-II. Despite this, the arrangement of apolar particles bears a closer resemblance to State-II characteristics than to State-I.\\
Overall, the analysis in this section provides insight into the stationary properties of the three states based on the behavior of the background nematic,  shedding light on the non-monotonic behavior of nematic ordering as the system crosses from State-I to State-II and then to State-III. Animations of the system in three different states are provided in supplementary movies SM1-SM3 (\ref{sec:sm1}).\\
A comprehensive exploration of the dynamics of polar particles in different states is presented in \ref{app:C}.

\subsection{Dynamic Swirl State} \label{sec:secIIIB}

\begin{figure}
\centering

\subfloat[]{\includegraphics[width=0.40\textwidth]{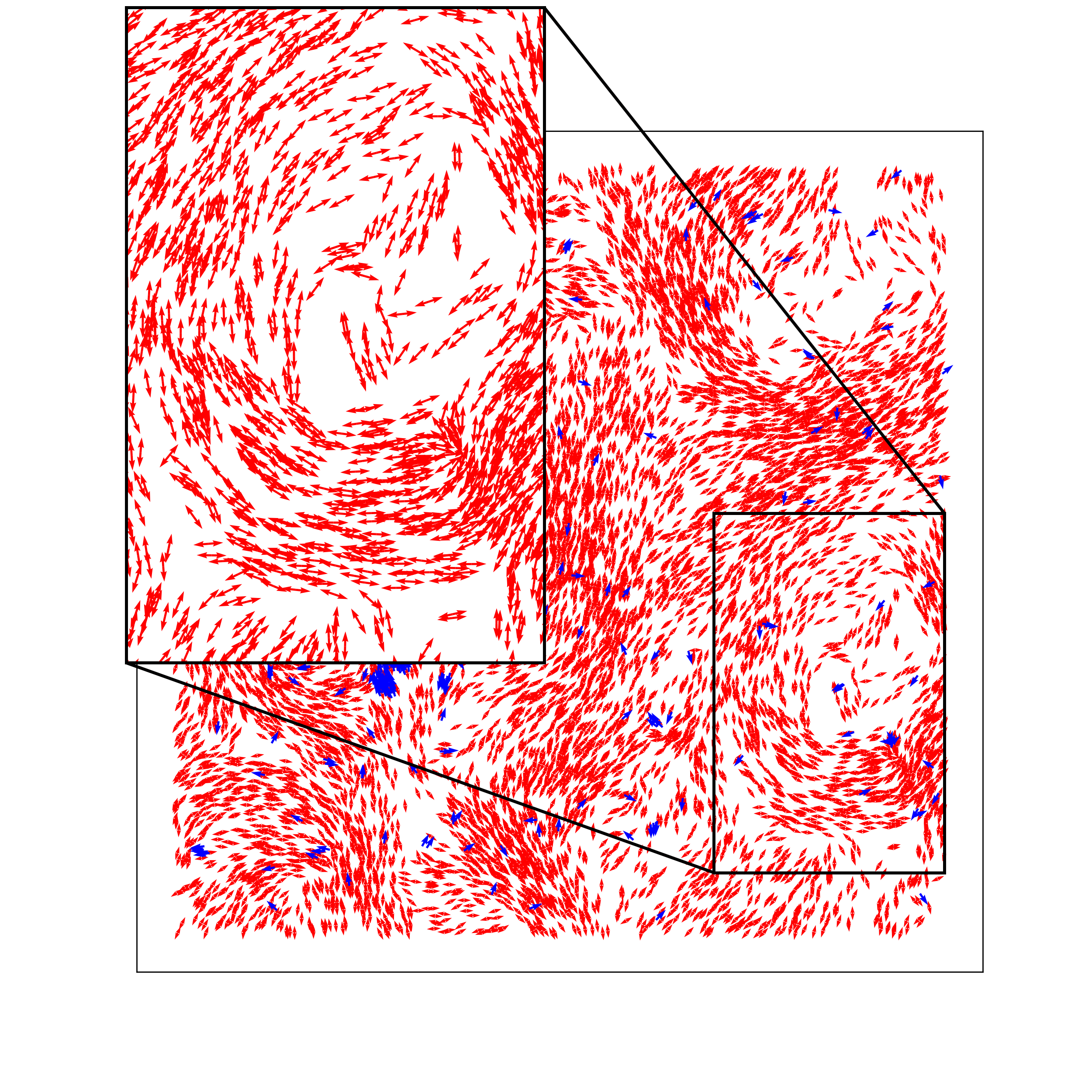}}
~
\subfloat[]{\includegraphics[width=0.40\textwidth]{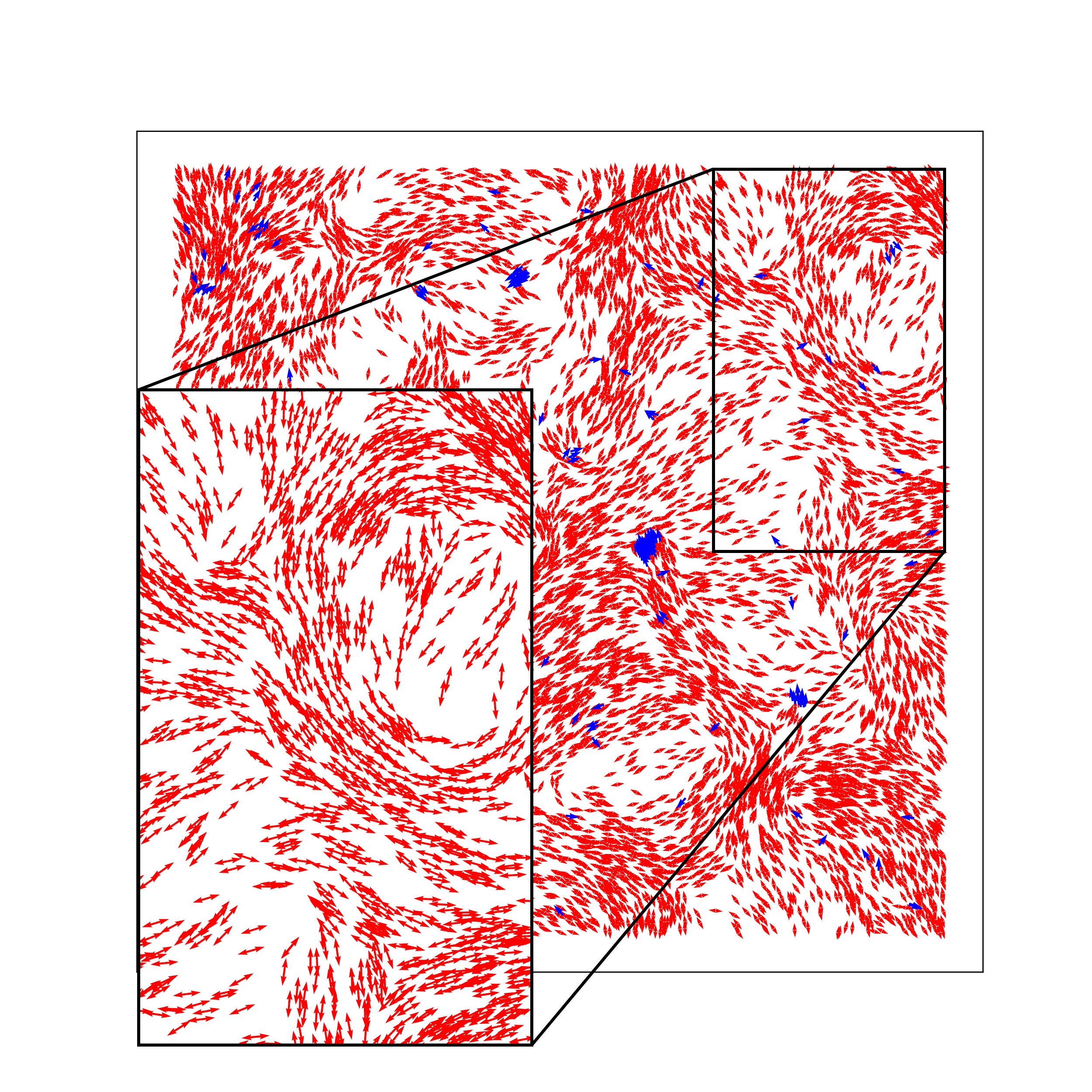}}

\caption{(color online) Visuals of Circular Defects in State-II and State-III of the Mixture. Snapshots (a) and (b) showcases the system configuration for  $\mu =$ $0.20$ (State-II) $\&$ $0.68$(State-III) respectively.  The inset provides a zoomed view of the region inside the rectangular box. All other parameters are same as in FIG.\ref{fig:3}. }
\label{fig:5}
\end{figure}

\begin{figure}[htp]
\begin{center}
\includegraphics[width=0.80\linewidth]{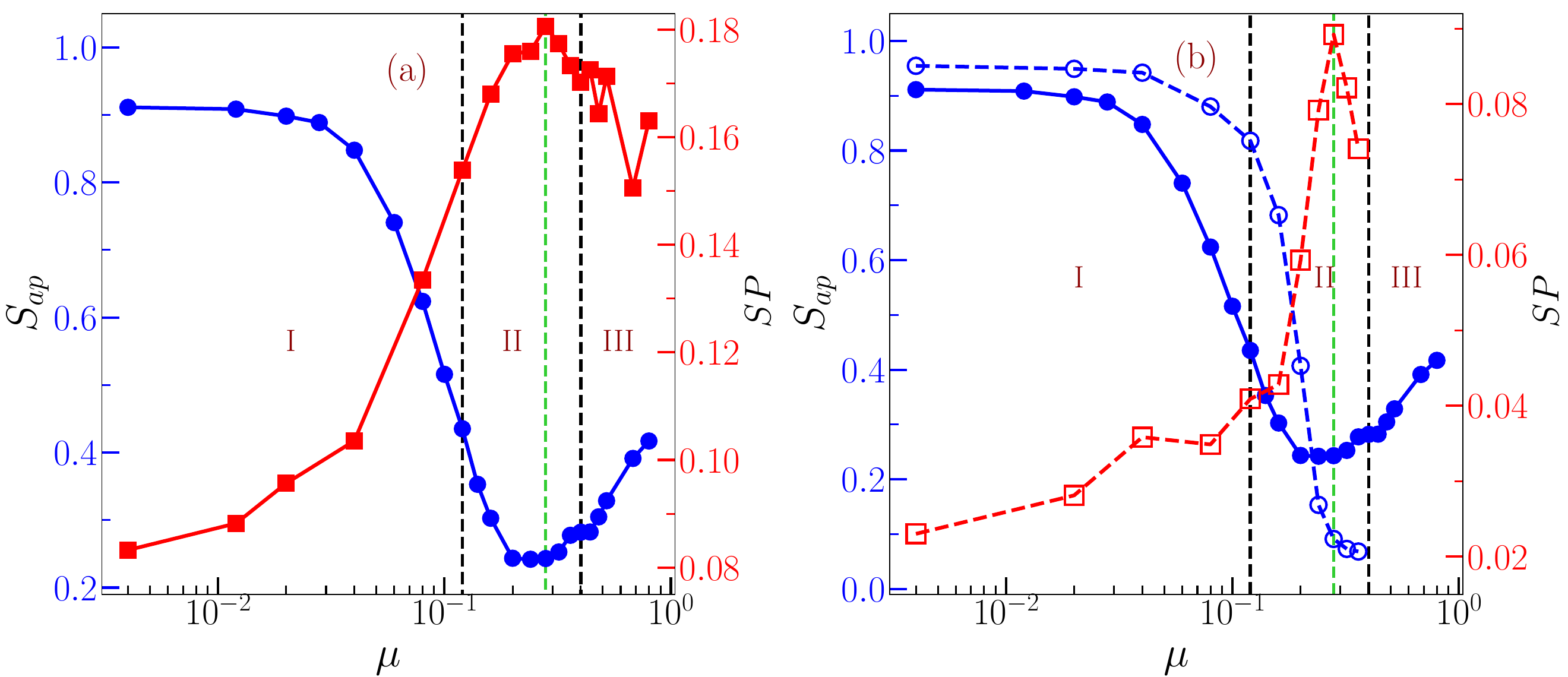}
\caption{(color online) The diagram depicts the Swirling Parameter ($SP$) variations across different states for various $\mu$ values, alongside the behavior of $S_{ap}$ for the original model in (a), and the model with colored noise in (b). In panel (b), the filled symbol (with solid line) and empty symbols (with dashed line) show the results for the original model and active nematics with colored noise, respectively.
The vertical black dashed lines denote the boundary between different states, and the vertical green dashed line highlights the correspondence between the peak of $SP$ and the minima of $S_{ap}$, indicating the increased strength of swirled structures in State-II. The rest of the parameters are the same as in FIG.\ref{fig:3}. }
\label{fig:6}
\end{center}
\end{figure}

As previously discussed, the increase in $\mu$ disrupts the strong nematic ordering among apolar particles observed in State-I due to the influence of polar particles. Consequently, both State-II and State-III, exhibit system configurations resembling a disordered state concerning the orientation of apolar particles. However, it is essential to note that these seemingly disordered States, State-II and State-III, are significantly different from the typical disorder states at large noise strengths characterised by the random orientation of particles. These states are highly dynamic, exhibiting robustness and complexity that are not commonly found in disordered states by high noise levels.\\
In State-II and State-III, we observe the formation of $\pm \dfrac{1}{2}$ topological defects corresponding to apolar particles, a characteristic feature of nematically interacting systems. The characteristics of these defects are very different in active nematics in contrast to the corresponding equilibrium system \cite{giomi2014defect,doostmohammadi2018active,zhang2018interplay}. Further details on the properties of these defects in active nematics are provided in \ref{sec:sm2}. Additionally, intriguing higher-order swirl-like structures emerge within the nematic system. These swirl structures consist of a high concentration of apolar particles along the perimeter or edge, with apolar particle expulsion occurring at the core, as depicted in FIG.\ref{fig:5}. The swirls undergo periodic cycles of formation, dissociation, and reformation, contributing to the highly dynamic and complex nature of the apolar system in State-II and State-III. Similar structures were reported in \cite{sampat2021}, where the authors studied a system by varying the density of polar particles at fixed activity in a setup similar to ours. Further, the configuration of the apolar particles in the State-II, shows that the system consists of regions of strong order separated by regions of strong bend distortions and swirls. The system confiuration is similar to the `Mixed Phase' observed in ref.
 \cite{pakpour2024delay}, which examines the behavior of the polar flock subjected to a time delay in the particle-particle interaction.\\
We refer to the state exhibiting the ongoing periodic formation and dissociation of swirl structures as the ``Dynamic Swirl State". It is similar to the active turbulence phenomena observed in active nematics. In active nematics, the active turbulence phase emerges as a function of the activity of nematic particles, characterized by the spontaneous formation and annihilation of $\pm \dfrac{1}{2}$ defects \cite{saw2017topological,thampi2014vorticity,alert2022active,giomi2015geometry}. Similarly, by increasing the activity of polar particles in the active nematics system, we observe the continuous formation and dissociation of swirl-like structures. In this section, we focus on examining the stability of these swirl structures, intending to gain a deeper understanding of their behavior, dynamics, and their overall impact on the system.\\
Swirl structures are regions exhibiting strong bend distortions within the active nematics system. The presence of these structures signifies the local rotational tendencies in the orientation field of apolar particles, resulting in a non-zero curl of the local nematic order parameter of apolar particles denoted by the vector $\boldsymbol{J} = (J_x,J_y)$, where, $J_x$ and $J_y$ are the two components of the local order parameter. We define an observable termed Swirling Parameter ($SP$), which is calculated as the width of the PDF of $\boldsymbol{\nabla} \times \boldsymbol{J}$. The $SP$ provides a quantitative measure of the strength of these swirl structures. The details on the calculation of the $SP$ is provided in the \ref{app:D}.\\
When plotted against $\mu$, $SP$ shows a non-monotonic behavior exhibiting a peak in State-II, as shown in FIG.\ref{fig:6}. The low value of $SP$ in State-I is due to the presence of strong nematic ordering and the minimal rotational tendencies of the apolar particles only at the strip edges. In contrast, in State-II and State-III, the presence of swirl structures results in a much larger value of $SP$. Further, the interaction between the swirl structures and the polar particle clusters plays a crucial role in determining the fate of the system in State-II and III. In State-II, the polar clusters consist of only a few particles, making them highly susceptible to changes on interacting with the swirled structures. FIG.S2 in \ref{sec:sm3} provides a visual description of this interaction: upon interaction with a swirl, a tiny polar cluster dissociates, while a slightly larger cluster merely undergoes a change in its orientation. This influence of the swirl structures on the dynamics of the polar clusters indirectly sets a positive feedback loop that enhances the stability of the swirl structures in this state. However, in State-III, the polar clusters grow much larger in size, bestowing them with the capability to distort and ultimately disrupt these swirl structures. Consequently, the stability of the swirl structures in State-III diminishes. One such interaction between a large polar cluster and a swirl structure in State-III is displayed through a series of snapshots in FIG.S3 in \ref{sec:sm4}. A detailed discussion on the stability of the swirl structures in different states is provided in \ref{sec:sm5}. \\

\FloatBarrier
\FloatBarrier

Furthermore, we observe the entrapping of isolated polar particles within void regions at the core of the swirl structures, visually represented through a series of snapshots in FIG.S2 in \ref{sec:sm3} and FIG.S3 in \ref{sec:sm4}. These particles seem to lack the necessary drive to escape and continue to undergo frequent reorientations. This behavior highlights the profound impact of the swirl structures on the individual particles.

\subsection{Mechanism of the emergence of the Swirl structures} \label{sec:secIIIC}
In this section, our primary focus is to discern the underlying mechanisms leading to the formation of swirls in active nematics under the influence of polar microswimmers. Our investigation employs a two-step approach. Firstly, we aim to unveil how the presence of polar particles stimulates the apolar system to exhibit intriguing behaviors across different States. Subsequently, we delve into understanding how this leads to the observed formation of swirled structures.

\subsubsection{Polar particles act like  spatially-temporally correlated colored noise for apolar system}\label{sec:secIIIC1}
Our observations thus far emphasize the crucial role played by the interactions between polar and apolar particles in determining the system's dynamics across different states. Here, we delve deeper into this interaction by proposing that polar particles serve as a form of spatiotemporally correlated colored noise for the apolar particles. The rationale supporting this interpretation is detailed in the \ref{app:E1}. Before proceeding with our discussion to evaluate the validity of this interpretation, it is essential to explore the concept of colored noise within the context of the different states.\\
In State-I, where polar particles exhibit limited motion and primarily reorient within specific regions of space, their influence can be analogized to a quenched random field distributed throughout the space i.e. a colored noise with minimal spatiotemporal correlation. In State-II, polar particles form multiple small clusters moving randomly, hence showing persistent motion over a larger length and time scale compared to State-I. Thus, in this State, the correlation length and time of the colored noise are significantly larger. In State-III, most of the polar particles are contained inside one or two large polar clusters moving coherently through the system, and actively contribute to enhancing the nematic ordering of apolar particles through alignment interactions. Hence, in this State, the effect of polar particles can be considered as a moving field with spatial and temporal correlation. However, this proves to be tough, if not unattainable, to accurately simulate with colored noise. Consequently, our comparative analysis between the two setups remains constrained to State-I and State-II exclusively. \\
In our model, since we are trying to mimic the effect of polar particles with colored noise, we feed the data of correlation length ($L_p$) and correlation time ($\tau_p$) of polar particles for different $\mu$ into the colored noise (correlation length $L_c = L_p$, and correlation time $\tau_c = \tau_p$ as shown in Fig.\ref{fig:c2}). As discussed in the \ref{app:E1}, the presence of polar particles in the system influences the interaction among the apolar particles in the system. Consequently, the colored noise in the active nematics is introduced to perturb the alignment of apolar particles with their neighbors within the interaction radius based on their distance. The details of the equation of motion of apolar particles with colored noise are provided in \ref{app:E2}.\\
Our observations with this setup are as follows: For $L_c$ and $\tau_c$ values corresponding to State-I, we observe strong nematic ordering of the apolar particles, and we observe the formation of stripes like structures of the apolar particles similar to those observed in the State-I in our original model.  As $L_c$ and $\tau_c$ increase, the nematic ordering of the system diminishes, as shown in FIG.\ref{fig:6}(b). Finally, the system transitions to a disordered state for $L_c$ and $\tau_c$ values corresponding to State-II. In this case, we observe the formation of swirled structures akin to State-II in our original setup. However, these structures are smaller in size and transient in nature i.e. they persist in the system for very short period of time. The snapshots of the apolar particle system with colored noise are shown in \ref{app:E3}\\
Further, to characterize these structures, we calculate the $SP$ (for this spatiotemporal colored noise model), which exhibits a peak at values corresponding to State-II, as shown in FIG.\ref{fig:6}(b). However, the magnitude of $SP$ in this model is small, indicating the weak nature of these structures. This suggests that, although using a colored noise with fixed $L_c$ and $\tau_c$ is appropriate to reproduce some aspects of the system's behavioral transition from State-I to State-II with increasing $\mu$, it is not entirely accurate. The motion of polar particles (and clusters) depends significantly on the status of nematic particles in their neighborhood.  Hence, rather than adopting a mean value of $L_c$ and $\tau_c$ throughout the system depending on $\mu$, $L_c$ and $\tau_c$ should be functions of the local nematic ordering i.e. $L_c = L_c(\mu, Q)$ and $\tau_c = \tau_c (\mu, Q)$, where,  $Q$ is the nematic tensor order parameter defined in the neighborhood of each particle. However, determining the functional dependence of $L_c$ and $\tau_c$ on the local nematic ordering, $Q$, remains an intricate task, adding a layer of complexity to our understanding.\\
Next, we explain how the polar particles change the mechanics of well formed nematics as we increase their velocity.

\begin{figure}[htp]
\begin{center}
\includegraphics[width=0.45\linewidth]{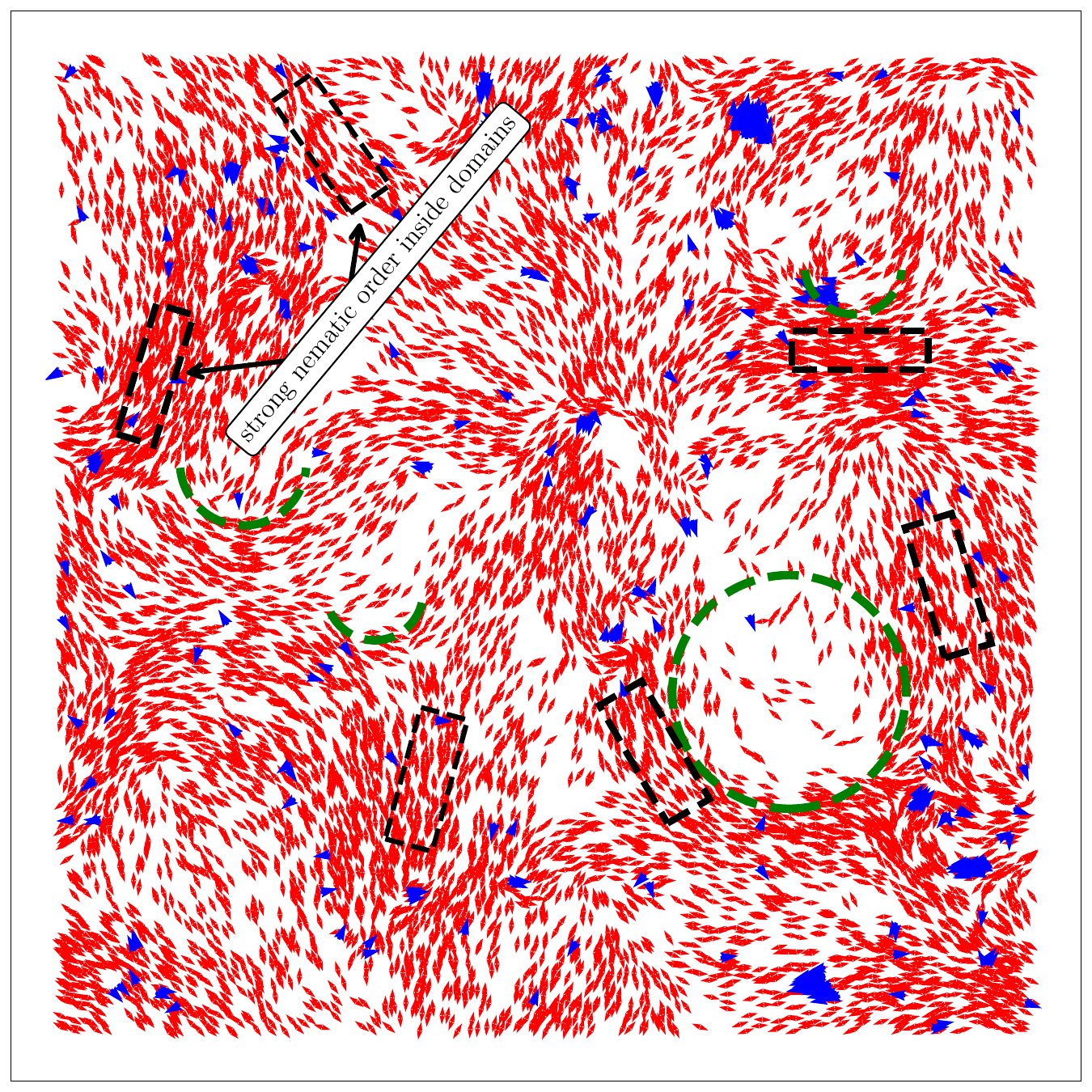}
\caption{(color online) The figure depicts the configuration of the system during State-II. Green dashed arcs delineate regions with pronounced bend distortions, while a green dashed circle indicates the presence of a swirl structure within the apolar particles. Black dashed squares emphasize domains characterized by robust nematic ordering. These domains are entirely isolated, and the ordering direction within each domain is completely independent. The figure highlights the significant correlation between the positions and orientations of the domains and the locations of swirls and pronounced bend structures. The parameters are the same as in FIG.\ref{fig:3}. }
\label{fig:7}
\end{center}
\end{figure}

\subsubsection{Polar Particles promote collective bending of apolar particles}\label{sec:secIIIC2}
\begin{figure}[htp]
\begin{center}
\includegraphics[width=0.80\linewidth]{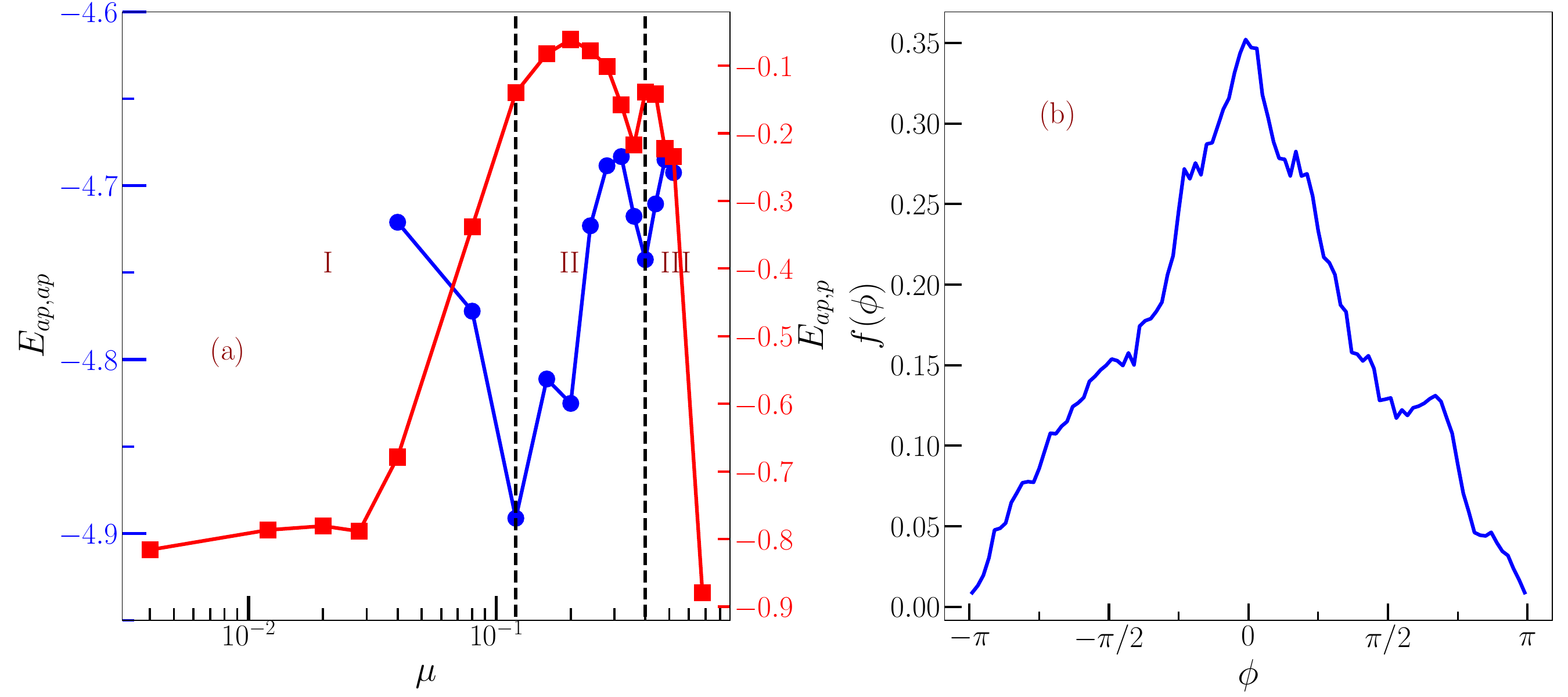}
\caption{(color online) The figure depicts (a) the variation of $E_{ap,ap}$ and $E_{ap,p}$  vs. $\mu$ across different States; (b) the PDF of $\phi = (\theta_{i,ap}-\theta_{j,p})$, $f(\phi)$, in State-II, giving a quantitative measure of relative orientation of the apolar particles with the polar particles inside the interaction radius. The parameters are the same as in FIG.\ref{fig:3}.}
\label{fig:8}
\end{center}
\end{figure}

As previously discussed, the influence of polar particles on apolar particles disturbs the alignment between the apolar particles and can be mimicked by introducing spatiotemporally correlated colored noise in the apolar-apolar interaction. Experimental findings \cite{saw2017topological} on the system comprising microtubules and kinesin molecular motors, exhibiting behaviors akin to active nematics, reported the facilitation of collective in-plane bending of cells by weakening cell-cell interactions.
In a recent study \cite{zhang2018interplay}, it is reported that the elasticity of a liquid crystal system consisting of a dense suspension of semiflexible biopolymers can be manipulated over a relatively wide range of elastic moduli.\\
Corroborating with these observations, we propose an explanation for the mechanism underlying the formation of swirled structures in our system: as $\mu$ increases, polar particles starts to form clusters. As these clusters traverse the ordered nematic, they induce bend-type distortions in the nematic ordering. Simultaneously, inherent $\pm \dfrac{1}{2}$ defects in active nematics consist of bend-type distortions. When another cluster of polar particles moves through regions with bend distortion or near a defect structure, the distortion is amplified, squeezing apolar particles together and creating void spaces in the neighborhood of the bend. Ultimately, the collective action of multiple polar clusters leads to the formation of strong bend distortions which look like macroscopic swirl like structures in the system. \\
To substantiate this argument, we conducted simulations starting with strong nematically ordered initial condition, identical to State-I,  and initial condition of polar identical to State-II and III. The observations show that the polar clusters drive the formation of swirls consisting of strong bend distortions in the nematic, aligning with our argument. A zoomed in animation of the system, showcasing the formation of a swirl induced by the polar clusters, is provided in the supplementary movie SM4 (\ref{sec:sm1} ).\\
Furthermore, we observed a distinct pattern among the apolar particles: the presence of strongly ordered domains. These domains are entirely disconnected from one another, with each domain's orientation being independent of the others, resulting in a lack of global ordering in the system. In FIG.\ref{fig:7}, we have highlighted some of these domains, demonstrating their correlation with regions exhibiting strong bend distortions and swirl structures. \\
To characterize these domains, we calculate the following interaction energies:
\begin{center}
 $E_{ap,ap} = -\mathlarger{\mathlarger{\sum}}_{i,ap} \mathlarger{\mathlarger{\sum}}_{|\boldsymbol{r}_{i,ap}-\boldsymbol{r}_{j,ap}| < R_{0,ap}} cos(2\theta_{i,ap}-2\theta_{j,ap})$   
\end{center}
 and 
\begin{center}
 $E_{ap,p} = -\mathlarger{\mathlarger{\sum}}_{i,ap} \mathlarger{\mathlarger{\sum}}_{|\boldsymbol{r}_{i,ap}-\boldsymbol{r}_{j,p}| < R_{0,ap}} cos(2\theta_{i,ap}-2\theta_{j,p})$   
\end{center}
where, $E_{ap,ap}$ and $E_{ap,p}$ accounts for the apolar-apolar and apolar-polar interaction, respectively. FIG.\ref{fig:8}(a) shows the variation of $E_{ap,ap}$ vs. $\mu$ in State-II and State-III, exhibiting a minimum in deep State-II, signifying the presence of strongly ordered domains of apolar particles in the system. The variation of $E_{ap,p}$ vs. $\mu$ is illustrated in FIG.\ref{fig:8}(a). It exhibits a 
 maximum in State-II (approximately 0), indicating a misalignment of apolar particles with polar particles in their vicinity. Additionally, the Probability Density Function (PDF) of
$\phi = (\theta_{i,ap}-\theta_{j,p})$, $f(\phi)$, displays a higher order peaks at a finite $\phi = \pm \pi/4$ and $\pm \pi/2$, depicted in FIG.\ref{fig:8}(b). This accounts for the observed peak in the plot of $f(\phi)$ $vs.$ $\phi$ and explains the origin of bend distortions induced in the well-ordered nematic by polar clusters. The polar particle clusters in State-II approach the strongly ordered nematic at an angle to the ordering direction. Consequently, a passing polar cluster induces distortion in the strongly ordered nematic, resulting in bend distortion.

\FloatBarrier
\FloatBarrier

\subsection{Number Fluctuation} \label{sec:secIIID}
\begin{figure}[htp]
\begin{center}
\includegraphics[width=0.45\linewidth]{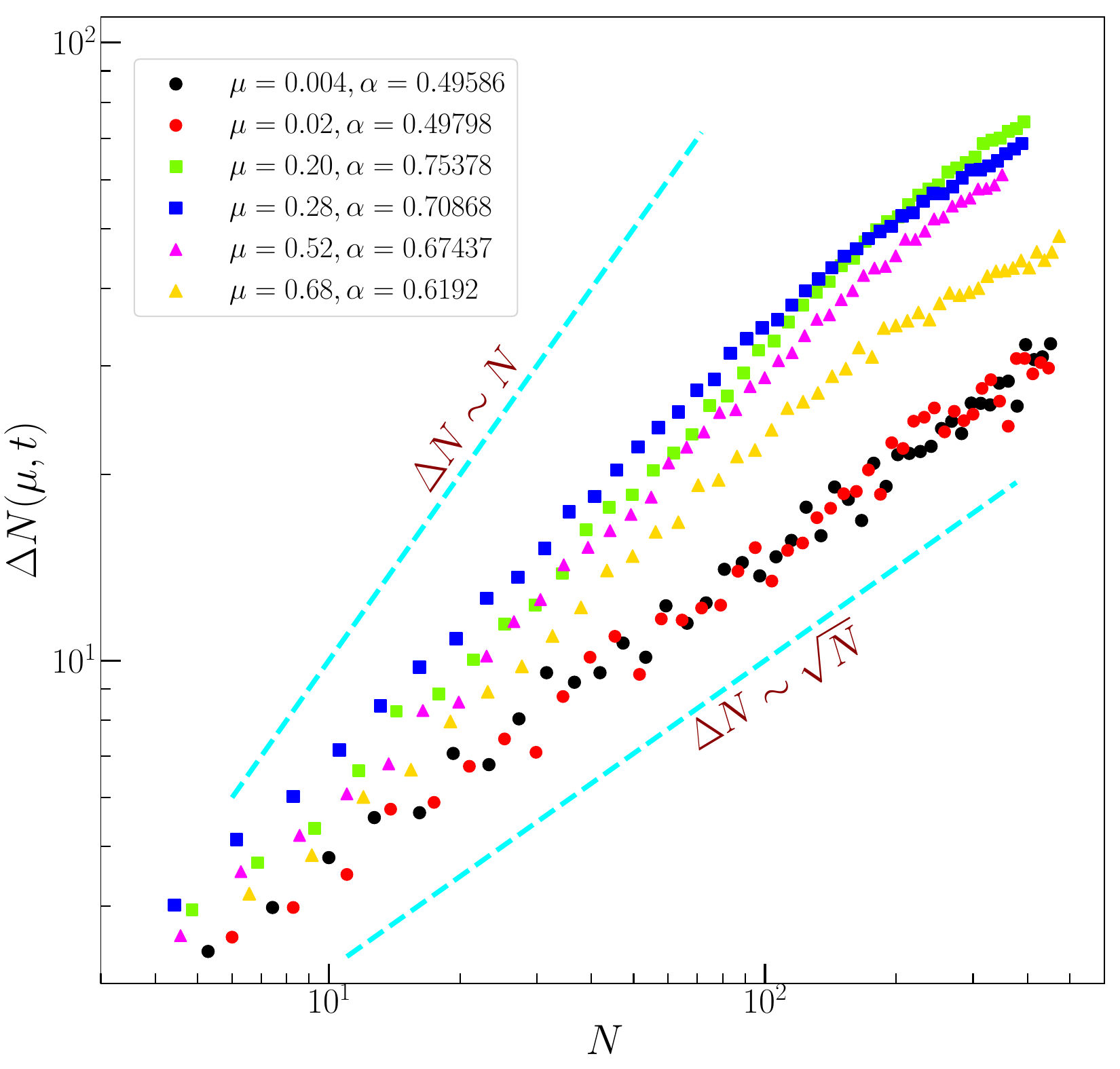}
\caption{(color online) The figure depicts the behavior of number fluctuation, $\Delta N$ in different States: State-I ($\mu =$ $0.004$ $\&$ $0.02$), State-I ($\mu =$ $0.20$ $\&$ $0.28$) and  State-I ($\mu =$ $0.52$ $\&$ $0.68$). The plot is shown on a log-log scale. The bottom and top dashed lines in `cyan' color indicate lines of slope $\frac{1}{2}$ and $1$, which correspond to the equilibrium system and the pure active nematics system, respectively. System size, $L=130$. The rest of the parameters are the same as in FIG.\ref{fig:3}. }
\label{fig:9}
\end{center}
\end{figure}

From an observational point of view, strong Number Fluctuation is one of the novel features distinguishing active systems from their equilibrium counterparts. For 2D active nematics, it is predicted numerically to vary as: $\Delta N \sim N^{\alpha}$ with $\alpha \sim 1$, both theoretically \cite{chate2006,ramaswamy2003active} as well as experimentally\cite{narayan2007long} compared to $\Delta N \sim \sqrt{N}$ in equilibrium systems, hence called Giant Number Fluctuation (GNF). Later, the study in ref.\cite{mishra2014aspects} showed that the presence of an external field makes the orientational fluctuations in active nematics finite resulting in the suppression number fluctuations in active nematics. In our study, we calculated the number fluctuation of apolar particles by measuring the number of particles in a subsystem of size $l \times l$ such that $l < \frac{L}{2}$ to take care of PBC, and then $\Delta N$ is calculated as Root Mean Square (RMS) fluctuation of N. We observed a variation of $\Delta N$ $vs.$ $N$  as : $\Delta N \sim N^{\alpha}$ as shown in FIG.\ref{fig:9}. \\
In State-I of our mixture, $\alpha$ takes value close to $\approx 0.50$, suggesting a uniform arrangement of apolar particles throughout the system. This suppression of GNF becomes apparent from the snapshots of the system shown in FIG.\ref{fig:3}(a), which shows strip like high density structures in apolar system in contradiction to well known phenomena of band formation in active nematic. This phenomenon can be attributed to apolar-polar interactions. Although on a global scale the polar particles are disordered, the OPDF of polar and apolar particles shows that there is a strong correspondence between them. On a micro scale, the apolar-polar interaction suggests that the presence of polar particles acts like a local orienting field to the apolar particles and this field changes direction randomly because of frequent reorientation of polar particles. This leads to suppression of GNF.\\
However, as we go from State-I to State-II, the formation of randomly moving polar clusters and the emerging circular vortices in apolar particles cause enhancement of number fluctuation in the system compared to State-I and the value of $\alpha$ is close to $\approx 0.75$. Upon entering State-III, the coherent motion of the polar clusters leads to suppression the number fluctuation in this state compared to State-II and $\alpha$ takes the value $\approx 0.70$.\\
The Dynamic Swirl State we observe here is not only a dynamic state in nematic ordering, but also has large density fluctuation, which is very non-trivial as in previous studies GNF is observed in ordered nematic. 

\FloatBarrier
\FloatBarrier

\section{Discussion} \label{sec:secIV}

To summarize, in this study, we explore the characteristics of a mixture of apolar rods and polar swimmers on a two dimensional substrate by keeping the density of both species fixed ($\rho_{ap}$ $\&$ $\rho_p$) and varying the activity ratio, $\mu$, within the low noise regime. On varying $\mu$, the system is found to exist in three distinct states based on the strength of the nematic ordering of apolar particles. The dynamics of the polar particles are influenced by two competing factors: self-propulsion and diffusion. At low value of activity ratio $\mu$, in State-I, diffusion dominates, causing diffusive motion of isolated polar particles, whereas the apolar particles display strong nematic ordering. As we increase $\mu$,  the activity of the polar particles starts playing a dominant role, causing alignment interaction between the particles, leading to the formation of small clusters. These clusters undergo random motion, disrupting the nematic ordering of apolar particles and causing the polar particles' mean-squared displacement (MSD) to become super-diffusive in State-II. As $\mu$ further increases, these clusters grow larger in State-III. In this state, we observe the formation of one or two large polar clusters, which move coherently through the system enhancing the nematic ordering of apolar particles. However, there are a few small polar clusters that move randomly and prevent complete alignment.\\
In State-II and III, which are disordered from the perspective of global nematic ordering of apolar particles, we observe the formation of swirled structures, accompanied by half-integer defects. The fate of polar particles in this state is strongly influenced by these swirl structures and vice versa.  To quantify the swirled structures, we introduce a parameter which we call the Swirling Parameter ($SP$). The $SP$ proves valuable in characterizing the swirling pattern of apolar particles and provides insights into the strength and stability of the swirls in different parameter regimes of the system. The non-monotonic behavior of $SP$ and the observed peak in State-II highlight the intricate interplay between the polar clusters and the swirled structures, offering valuable insights into the underlying mechanisms driving the system dynamics. The dynamic swirl state appears when polar particles assemble in many small clusters.  The structures become weak when individual polar particles are moving randomly, as well as when they form macroscopic clusters. Additionally, we observe the trapping of isolated polar particles at the core of swirl structures.\\ 
These swirled structures undergo a continuous cycle of formation, dissociation, and subsequent reformation over time, creating a dynamical state, which we term the ``Dynamic Swirl state". We observe that the ``Dynamic Swirl state" bears a resemblance to the active turbulence observed in active nematics \cite{saw2017topological,thampi2014vorticity,alert2022active,giomi2015geometry}. Active turbulence, marked by the spontaneous creation and annihilation of half-integer defects, emerges as a function of the activity of apolar particles in pure active nematics systems. In our study, the transition to the Dynamic Swirl state arises by increasing the activity of polar particles, leading to the persistent formation and dissolution of swirl structures in apolar particles. The presence of polar particles breaks the nematic symmetry of the system and leads to the dynamic state with continuous formation and dissociation of swirl structures.
However, it is essential to emphasize that these two phenomena originate from entirely different underlying mechanisms, and the comparison drawn here serves as a simplified analogy rather than an in-depth equivalence. Further, active turbulence is observed in dense active nematic, whereas we observe these dynamic swirl structures in low density active nematics.\\
Furthermore, our attempt to unravel the underlying mechanism of the formation of the swirled structures uncovers a subtle intricacy: the influence of polar particles on apolar particles can be mimicked through colored noise, wherein the correlation length and time dynamically depend on the local nematic ordering and activity of the apolar particles. This opens up a new way of understanding the complex pattern in the mixture. Such colored noise highlights the subtleties inherent in capturing the dynamic interplay between apolar and the polar particles.\\
A distinctive feature of our model is the nonreciprocal interaction between the polar and apolar particles in the system: polar particles align ferromagnetically with apolar particles, while apolar particles align nematically with polar particles. In a recent study \cite{fruchart2021non}, it is reported that non-reciprocity in the interaction between the particles can have a substantial effect on the dynamics of the interacting many body systems. Therefore, it would be interesting to investigate how non-reciprocity impacts the behavior of the two types of particles in our model and how this effect varies with the activity levels of polar and apolar particles.\\
 Following the results of this study, several other questions remain open. An important characteristics of liquid crystal systems is the presence of topological defects of charge $\pm \frac{1}{2}$. However, our microscopic approach in this study does not allow us to explore the effect of the polar particles on the dynamics of the defects (see \ref{sec:sm2} ). It requires an in depth analysis of the kinetic of the system by constructing a coarse grained version of the model which is a work in progress. Further, our model can also be extended to include chirality in polar particles, as most of the microswimmers in nature are chiral. Our preliminary observations (see \ref{sec:sm6}), reveals a trade-off between the chirality ($\omega$) and activity of polar particles ($v_p$), and within a specific range of parameter values this stabilizes the swirl structures. However, a detailed exploration of the parameter space ($v_p,\omega$) is necessary to develop an understanding of the effect of chirality which is beyond the scope of this study. \\
At the experimental level, our study assumes significant importance, as it opens up an unexplored direction with promising possibilities. Through simulations, we introduce an active background nematic, a departure from the traditional LLC system. Yet, the essence of LLC remains evident as the activity of polar particles gives rise to novel orientational patterns, which would otherwise elude observation. The dependence of stability and scale of these intricate patterns on the magnitude of polar particle activity provides us with a powerful tool for detecting bacteria with precision. Additionally, we unveil the emergence of dynamic swirl structures, offering a potential means to trap bacteria at their core, a practical application worth exploring, and considerable potential for bacterial manipulation.

\section{Acknowledgement}
The authors thank Venceslas Ngounou for useful discussions. P.S.M, P.K.M and S.M.  thank Jacques Prost for useful discussions and  Sriram Ramaswamy for critical reading of the manuscript and providing useful comments. S.M. thank Pranay Bimal Sampat for the useful discussions at the start of the project.  P.S.M., P.K.M. and S.M., thanks PARAM Shivay for computational facility under the National Supercomputing Mission, Government of India at the Indian Institute of Technology, Varanasi and the computational facility at I.I.T. (BHU) Varanasi. P.S.M. and P.K.M.  thank UGC for research fellowship. S.M. thanks DST, SERB (INDIA), Project No.: CRG/2021/006945, MTR/2021/000438  for financial support. T.V. acknowledges funding from the European Union's Horizon 2020 research and innovation programme under the Marie Sklodowska-Curie grant agreement No 955576.

\appendix
\section{Dynamical Equations of  apolar and polar particles}\label{app:A}
{\em Update rules for apolar particle}:- Equations for position and orientation update of apolar particles are given by, 
\begin{equation}
	\boldsymbol{r}_{i,ap}(t+1) = \boldsymbol{r}_{i,ap}(t)+  \boldsymbol{v}_{i,ap}(t) \Delta t 
	\label{equn:a1},
\end{equation}
\begin{equation}
	\theta_{i,ap}(t+1) = \dfrac{1}{2}\bigg< \theta(t) \bigg>_{R_{0,ap}} + \zeta_{i,ap}(t)
	\label{equn:a2}
\end{equation}

In Equation (\ref{equn:a1}), $v_i(t)$ is the instantaneous velocity of the $i^{th}$ apolar particle at time $t$. In Equation (\ref{equn:a2}), first term represents the alignment term and second term is a delta correlated white noise with zero mean and uniform distribution in the range $\bigg[-\frac{\eta}{2},+\frac{\eta}{2}\bigg]$. The alignment term, which is the mean orientation of particles inside the interaction radius of apolar particle, is evaluated as
\begin{center}
	$\bigg< \theta(t) \bigg>_{R_{0,ap}} = \arg \bigg(  \mathlarger{\mathlarger{\sum}}_{|\boldsymbol{r}_{i,ap}-\boldsymbol{r}_{j,ap}| < R_{0,ap}} e^{2 i \theta_{j,ap}(t)}  + \mathlarger{\mathlarger{\sum}}_{|\boldsymbol{r}_{i,ap}-\boldsymbol{r}_{j,p}| < R_{0,ap}} e^{2 i \theta_{j,p}(t)} \bigg) $
\end{center}
In the above expression, first and second term represents the interaction of $i^{th}$ apolar particle with apolar and polar neighbors respectively. The instantaneous velocity vector of the $i^{th}$ apolar particle at time $(t+1)$ is given by,
\begin{center}
	$\boldsymbol{v}_{i,ap}(t+1) = R_{i,ap}(t+1).v_{ap}\bigg(\cos(\theta_{i,ap}(t+1),\sin(\theta_{i,ap}(t+1)\bigg)$
\end{center}

The term $R_{i,ap}$ is chosen randomly between $\pm 1$ with equal probability to incorporate it's bidirectional motion along the long axis of apolar particles.

{\em Update rules for polar particle}:- The position and orientation update equation of polar particles are given by,
\begin{equation}
    \boldsymbol{r}_{i,p}(t+1) = \boldsymbol{r}_{i,p}(t)+  \boldsymbol{v}_{i,p}(t) \Delta t+ \sqrt{2 V_R \Delta t} \boldsymbol{\xi_i} 
	\label{equn:a3}
\end{equation}
\begin{equation}
	\theta_{i,p}(t+1) = \bigg< \theta(t) \bigg>_{R_{0,p}} + \zeta_{i,p}(t)
	\label{equn:a4}
\end{equation}
In equation (\ref{equn:a3}) third term is an additional term added to standard self-propulsion dynamics of polar particle's position update. It accounts for diffusion, $\boldsymbol{\xi_i} \equiv \bigg(\cos(\phi_i(t)), 
 \sin(\phi_i(t)) \bigg)$  is a random unit vector with $\phi_i(t)$ is randomly chosen from the range $[0,2\pi]$ and $V_R$ is the diffusion constant.\\
In Equation (\ref{equn:a4}), the second term is a delta-correlated white noise with zero mean, chosen uniformly from the range $\bigg[-\frac{\eta}{2},+\frac{\eta}{2}\bigg]$. The first term represents alignment and captures the average of  instantaneous motion direction of all other particles within the interaction radius of the $i^{th}$ polar particle. It is evaluated as,
\begin{center}
	$\bigg< \theta(t) \bigg>_{R_{0,p}} = \arg \bigg(  \mathlarger{\mathlarger{\sum}}_{|\boldsymbol{r}_{i,p}-\boldsymbol{r}_{j,ap}| < R_{0,p}} \boldsymbol{v}_{j,ap}(t)  + \mathlarger{\mathlarger{\sum}}_{|\boldsymbol{r}_{i,p}-\boldsymbol{r}_{j,p}| < R_{0,p}} \boldsymbol{v}_{j,p}(t) \bigg) $
\end{center}
where the first and second terms correspond to the interactions of the $i^{th}$ polar particle with apolar and polar particles in its neighborhood, respectively. The instantaneous velocity vector of the $i^{th}$ polar particle at time $(t+1)$ is,
\begin{center}
	$\boldsymbol{v}_{i,p}(t+1) = v_p\bigg(\cos(\theta_{i,p}(t+1)),\sin(\theta_{i,p}(t+1))\bigg)$
\end{center}

\section{Probability Density Function of Orientation(OPDF) of apolar and polar particles}\label{app:B}
Consider a system consisting of N particles and their orientation at a given instant of time $t$ is denoted by $\{\theta_i, i=1, N\}$ where $-\theta_1 \leq \theta_i \leq +\theta_1$. In our model, $(\theta_1,N)$ = $(\pi,N_{p})$ and $(\pi/2,N_{ap})$ for the polar and apolar particles, respectively. The steps to calculate the probability density function of orientation (OPDF) of the system of particles are described as follows:
\begin{itemize}
    \item[(i)]{Divide the total range of $\theta \in [-\theta_1,+\theta_1]$ in $N_{bin}$ number of bins of width $\Delta\theta$ and calculate the number of particles in each bin.}
    \item[(ii)]{Let, $n_i$ is the number of particles in the $i^{th}$ bin and $\theta^i$ is the center of the $i^{th}$ bin. The probability that a randomly chosen particle will have an orientation in the range $(\theta^i - \frac{\Delta\theta}{2}, \theta^i + \frac{\Delta\theta}{2})$ is $P(\theta^i) = \frac{n_i}{N}$.}
    \item[(iii)]{The OPDF, $f(\theta)$ is defined as the probability per unit length of the interval $(\theta^i - \frac{\Delta\theta}{2}, \theta^i + \frac{\Delta\theta}{2})$ i.e. $f(\theta^i) = \frac{P(\theta^i)}{\Delta\theta}$.}
    \item[(iv)]{The normalization condition is given by, $\mathlarger{\int}_{-\theta_1}^{\theta_1} f(\theta)d\theta = \mathlarger{\sum}_{i=1}^{N_{bin}} f(\theta^i)\Delta\theta = \Delta\theta \mathlarger{\sum}_{i=1}^{N_{bin}} f(\theta^i) = 1$} 
\end{itemize}
In Fig.3 of the main text, presents the OPDF for apolar and polar particles, depicted in the middle and bottom panels, respectively. For each case, the OPDF is derived from a single time snapshot, shown in the top panel. The justification for utilizing just one snapshot to calculate the OPDF is as follows: In State-I, the polar particles reorient randomly, and in State-II, small polar clusters move without a fixed direction. Consequently, over time, while the location of peaks in the OPDF shifts, the general features remain constant. Averaging over time would thus flatten the OPDF, resulting in loss of detail and obscuring insights into the system's behavior. Similarly, for apolar particles, the dynamic nature of swirl structures, through their formation, dissociation, and reformation, in State-II and III suggests that time-averaging would erase these distinctive features.\\
Therefore, we present the OPDF of both polar and apolar particles at a single time instant, corresponding to the snapshot displayed in the top panel.

\section{Characteristics of Polar Particles}\label{app:C}
In this section, we employ several tools to gain insights into the dynamics of polar particles and its impact on the nematic in different states of the mixture:
\begin{enumerate}
\item {The order parameter for Polar particles, 
\begin{equation}
<\phi_p(\mu)> = \bigg<\frac{1}{N_p v_p} \bigg\vert \mathlarger{\sum}_{i=1}^{N_{p}}\boldsymbol{v}_{i,p}(t) \bigg\vert \bigg> 
\label{equn:b1}
\end{equation}
where $N_p$ is the total number of polar particles and $<...>$ implies average over time in steady state as well as over independent realizations. $<\phi_p(\mu)>$ quantifies the degree of ordering of polar particles within the system. A value close to 1 indicates a highly ordered state of the polar particles, while a value close to zero suggests a disordered spatial arrangement of polar particles.}

\item {The Mean Square Displacement (MSD) of polar particles, represented as
\begin{equation}
\Delta_p (\mu,t) = \bigg< \dfrac{1}{N_p}\mathlarger{\sum}_{i=1}^{N_{p}} |\boldsymbol{r}_{i,p}(t_0 + t)-\boldsymbol{r}_{i,p}(t_0)|^2     \bigg> 		
\label{equn:b3}    
\end{equation}
where, $<....>$ carries the same meaning as in Eq.(\ref{equn:b2}).	
MSD, displayed in FIG.\ref{fig:c1}(a), provides insights into the nature of trajectories of polar particles. It  exhibits power law behavior as $\Delta_p (\mu,t) \sim t^{\beta}$ where the exponent $\beta$  is indicative of the type of motion exhibited by the polar particles: $\beta = 1$ and $\beta = 2$ correspond to diffusive and ballistic motion, respectively, while $1 < \beta < 2$ indicates super diffusion. }

\item {Velocity Auto Correlation Function (VACF) of polar particles, 
\begin{equation}
\gamma_p(\mu,t) = \bigg< \frac{1}{v_0^2 N_p} \mathlarger{\sum}_{i=1}^{N_{p}} \boldsymbol{v}_{i,p}(t_0 + t) \cdot  \boldsymbol{v}_{i,p}(t)  \bigg> 			
\label{equn:b2}
\end{equation}
where $<....>$ implies average over reference times, $t_0$, in the steady state as well as independent realizations. VACF, demonstrated in FIG.\ref{fig:c1}(b), examines the correlation between the velocities of polar particles at different time intervals during it's motion and it reveals the timescale over which a polar particle retains its orientation. We fit $\gamma_p(\mu,t)$  with exponential, $exp\bigg[{-\dfrac{t}{\tau_p(\mu)} }\bigg]$  and calculated the correlation time $\tau_p(\mu)$. }

\setcounter{figure}{0}
\renewcommand{\thefigure}{C\arabic{figure}} 
\begin{figure*}
  \centering
  \includegraphics[width=0.80\textwidth]{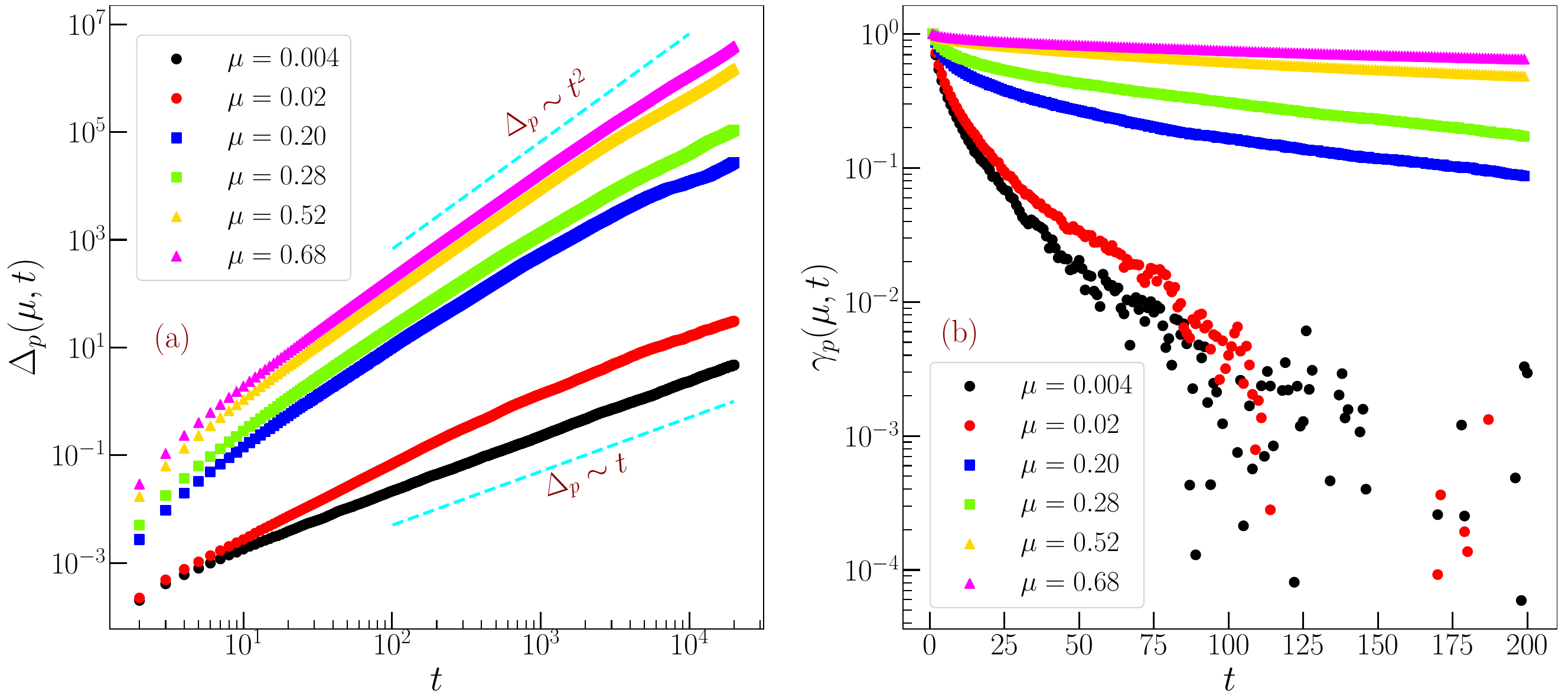} 
  \caption{(color online) The figure illustrates the behavior of the  (a) Mean Square Displacement (MSD), $\Delta_p(\mu,t)$, and (b) the Velocity Auto Correlation Function (VACF), $\gamma_p(\mu,t)$, of polar particles in different states of the system: State-I ($\mu = 0.004$ $\&$ $0.02$), State-II ($\mu = 0.20$ $\&$ $0.28$), and State-III ($\mu = 0.52$ $\&$ $0.68$). The MSD plot is shown on log-log scale and the VACF plot is shown on the log-y scale. In sublot (a), the two dashed lines in cyan represent the lines of slope $\beta = 1$ and $2$ corresponding to diffusive and ballistic motion, respectively. The rest of the parameters are the same as in FIG.\ref{fig:3}}
  \label{fig:c1}
\end{figure*}

\renewcommand{\thefigure}{C\arabic{figure}} 
\begin{figure*}
  \centering
  \includegraphics[width=0.45\textwidth]{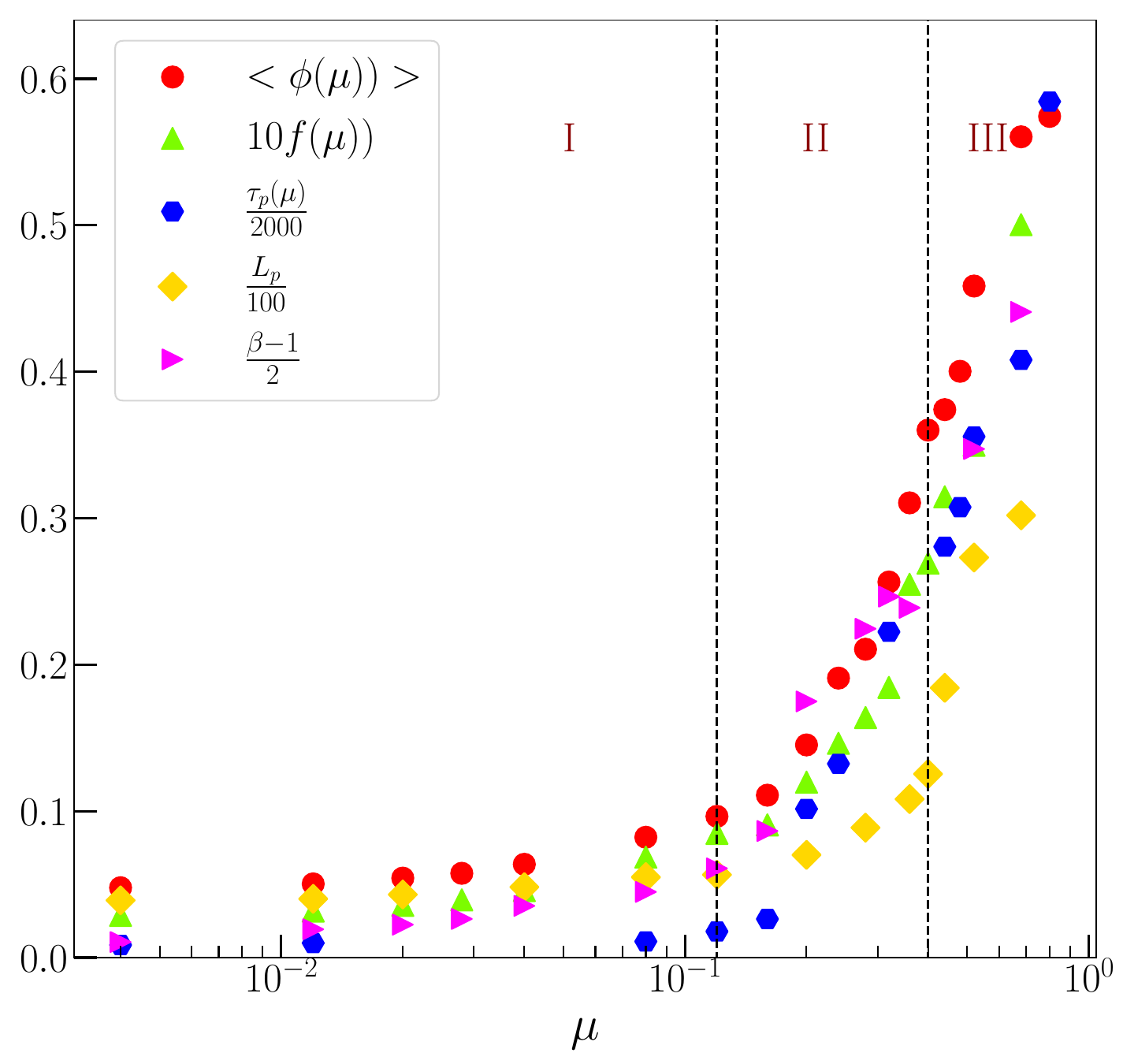} 
  \caption{(color online) This figure illustrates behavior of various observables for polar particles across different States. Two vertical dashed black lines serve as a visual guide, indicating the boundaries between these distinct States. To ensure clarity and meaningful comparison, different quantities, namely $<\phi(\mu)>$, $f(\mu)$, $\tau_p(\mu)$, $L_p$, and $\beta$, are appropriately scaled on the Y-axis, as indicated by the accompanying legends. The parameters used in this plot are the same as in FIG.\ref{fig:3}.}
  \label{fig:c2}
\end{figure*}

\item {The fraction of the total number of polar particles, on average, that reside within the interaction radius of each polar particle can be quantified using the following expression:
\begin{equation}
f(\mu) = \frac{<n>}{N_p},
\label{equn:b4}
\end{equation}
where $<n>$ denotes the average number of polar particles found within the interaction radius of the each polar particle. This quantity provides a measure of the average size of the polar particle clusters present in the system.  
}
\end{enumerate}

We show the variation of $<\phi_p(\mu)>$, $f(\mu)$, $\tau_p(\mu)$, $L_p(\mu)$ and  $\beta$   with activity ratio, $\mu$ in FIG.\ref{fig:c2}.

In State-I, the small magnitude of $\mu$ results in the diffusion term dominating over the activity term. As a consequence, the motion of polar particles closely resembles that of random walkers in two dimensions. This behavior is evident from the rapid decay of $\gamma_p(\mu,t)$, as shown in FIG.\ref{fig:c1}(b), with an extremely small correlation time $\tau_p(\mu)$, indicating the absence of temporal correlations in their motion. Furthermore, the low value of $f(\mu)$ suggests that, on average, only few polar particles resides within the interaction radius of another polar particle. The combination of a small correlation length $L_p$, a short correlation time $\tau_p(\mu)$ (FIG.\ref{fig:c2}), and the low value of $f(\mu)$ supports the notion that polar particles do not form clusters but rather engage in  random walk independent of others within their species. The diffusive nature of this behavior is further substantiated by the exponent $\beta \approx 1$ observed in the Mean Square Displacement (MSD), $\Delta_p (\mu,t)$, as shown in FIG.\ref{fig:c1}(a). Additionally, the presence of small random peaks in $P(\theta_{p})$ vs. $\theta_{p}$ and the visual snapshots of the system configuration, as shown in Figure \ref{fig:3}(a), reinforce the diffusive characteristics observed in this state (see the supplementary movie SM1 in \ref{sec:sm1}). \\
In State-II,  increase in the magnitude of $\mu$ leads to the emergence of spatial (FIG.\ref{fig:4}(b)) and temporal (FIG.\ref{fig:c1}(b)) correlations in the motion of polar particles. This is reflected in the growth of the correlation length $L_p$ and correlation time $\tau_p$, indicating the formation of clusters among the polar particles, as shown in FIG.\ref{fig:c2}. The fraction $f(\mu)$ increases, indicating a higher average number of polar particles within the interaction radius of each polar particle (\ref{fig:c2}). Consequently, the order parameter $<\phi_p(\mu)>$ increases, signifying a higher degree of ordering among the polar particles. The animation of the system (supplementary movie SM2 in \ref{sec:sm1}), in State-II clearly shows the dynamic nature of these clusters, characterized by their formation and dissolution. When a particle is part of a cluster, it exhibits persistent motion aligned with the overall direction of the cluster. However, upon leaving the cluster, the particle undergoes diffusion until it encounters and aligns with another cluster. As a result, the overall motion of polar particles displays superdiffusive behavior in the MSD, FIG.\ref{fig:c1}(a), with an exponent $\beta > 1$, FIG.\ref{fig:c2}. Snapshots of the system, as shown in FIG.\ref{fig:3}(b), configuration highlight the uncorrelated motion of the clusters and the disruption of strong orientational order among the apolar particles.\\
In State-III, the fraction $f(\mu)$ significantly increases, indicating the formation of larger polar particle clusters, as shown in FIG.\ref{fig:3}(c). In this State, we observe one or two large cluster of polar particles moving coherently and few small clusters moving randomly (see supplementary movie SM3 in \ref{sec:sm1}). This is supported by the presence of a large correlation length $L_p$, suggesting that the majority of particles are bound within these clusters. The motion of polar particles in this state becomes persistent, exhibiting a ballistic Mean Square Displacement (MSD) with an exponent $\beta \approx 2$. Consequently, polar particles move consistently in the direction of their head, influencing the  apolar particles coming in their path. Occasionaly a cluster of polar particles splits in two to three clusters, when it encounters a dense nematically ordered region. In this manner the dynamic nature of this state is maintained.   The polar order parameter $<\phi_p(\mu)>$ assumes a much higher value, approximately 0.50, indicating the existence of correlations within the motion of the polar clusters. In State-III, the dynamics of the system are predominantly governed by these polar clusters, which in turn  induce alignment among the apolar particles in their wake, leading to enhanced ordering among the apolar particles compared to State-II. However, the presence of small random-moving clusters prevents the overall global alignment of the nematic. \\
{\em Apolar particles disturb the polar ordering} :-
Further to investigate the influence of apolar particles on the dynamics of polar particles, we conducted simulations of a system of polar swimmers while keeping other parameters constant. The observations revealed higher values of the order parameter $<\phi_p(\mu)>$, with the difference increasing as $\mu$ increases. This finding suggests that the presence of apolar particles in the system suppresses the ordering among the polar particles in all three States, consequently affecting the nematic ordering of the apolar particles.\\

\section{Calculation of Swirling  Parameter (SP)}\label{app:D}
We first discretize the system into an underlying lattice. We consider a small $b \times b$ box with the lattice point at its center, such that $b<<L$(system size). For our calculation, we used $b = 4$. Inside each box we calculate the components of the local ordering  due to apolar particles, $\boldsymbol{J} = (J_x, J_y)$ as :
\begin{center}
	$J_x = \bigg< \dfrac{1}{n} \mathlarger{\sum}_{i,ap = 1}^{n} \cos(2 \theta_{i,ap})  \bigg>$,   $J_y = \bigg< \dfrac{1}{n} \mathlarger{\sum}_{i,ap = 1}^{n} \sin(2 \theta_{i,ap})  \bigg>$ 
\end{center}

where, $n$ is the number of apolar particles inside a box whose $x$ and $y$ coordinate lies in the range $x \in \bigg[x_c -\dfrac{b}{2},x_c +\dfrac{b}{2} \bigg]$ and $y \in \bigg[y_c -\dfrac{b}{2},y_c +\dfrac{b}{2} \bigg]$ with $(x_c,y_c)$ being the center of the box.
The swirling nature of the orientational field of the apolar particles can be estimated by calculating the curl of  $\boldsymbol{J}$ as,
$ \boldsymbol{R} = (\boldsymbol{\nabla} \times \boldsymbol{J})_z = \bigg( \dfrac{\partial J_x}{\partial y} - \dfrac{\partial J_y}{\partial x}  \bigg)$. The vector nature of $\boldsymbol{R}$ will give positive and negative values in two dimensions.\\
The partial derivatives $\dfrac{\partial J_y}{\partial x}$,$\dfrac{\partial J_x}{\partial y}$ are calculated using central difference method as described below,
\begin{center}
	$\dfrac{\partial J_y(i,j)}{\partial x} = \dfrac{J_y(i+1,j) - J_y(i-1,j)}{2b}$
\end{center}
\begin{center}
	$\dfrac{\partial J_x(i,j)}{\partial y} = \dfrac{J_x(i,j+1) - J_x(i,j-1)}{2b}$	
\end{center}

Next, we calculate the probability density function (PDF) of $\boldsymbol{R}$ and average it over 15 independent realizations. For all values of $\mu$, the PDF is symmetric about $\boldsymbol{R} = 0$.  That represents the equal probability of clockwise and anti-clockwise pattern of circular structure. \\
The PDF of $\boldsymbol{R}$, $f(\boldsymbol{R})$, in three different states is shown in FIG.\ref{fig:E1}(a). The detailed steps of calculation of $f(\boldsymbol{R})$ is same as that of OPDF provided \ref{app:B}. It can be seen that the width of the PDF changes non-monotonically as we go from State-I to State-II and then to State-III. The width of the PDF gives information about the quantitative measure of the swirl structures present in the system. To characterize it, first, we fit the PDF with the Lorentzian distribution, as shown in FIG.\ref{fig:E1}(b) for one set of parameters, given by
\begin{center}
	$F(x) = A \dfrac{\lambda}{(x-x_0)^2 + \lambda^2}$
\end{center}
where, $x_0$ is the median and $\lambda$ is the scale parameter. The scale parameter $\lambda$ carries the information about the width of $f(\lambda)$ i.e. larger $\lambda$ means larger width of the $f(\lambda)$.  
The value of $\lambda$ obtained from this fit is our order parameter, which we termed Swirling Parameter $SP$.

\setcounter{figure}{0}
\renewcommand{\thefigure}{E\arabic{figure}}
\begin{figure}[htp]
	\begin{center}
		\includegraphics[width=0.80\linewidth]{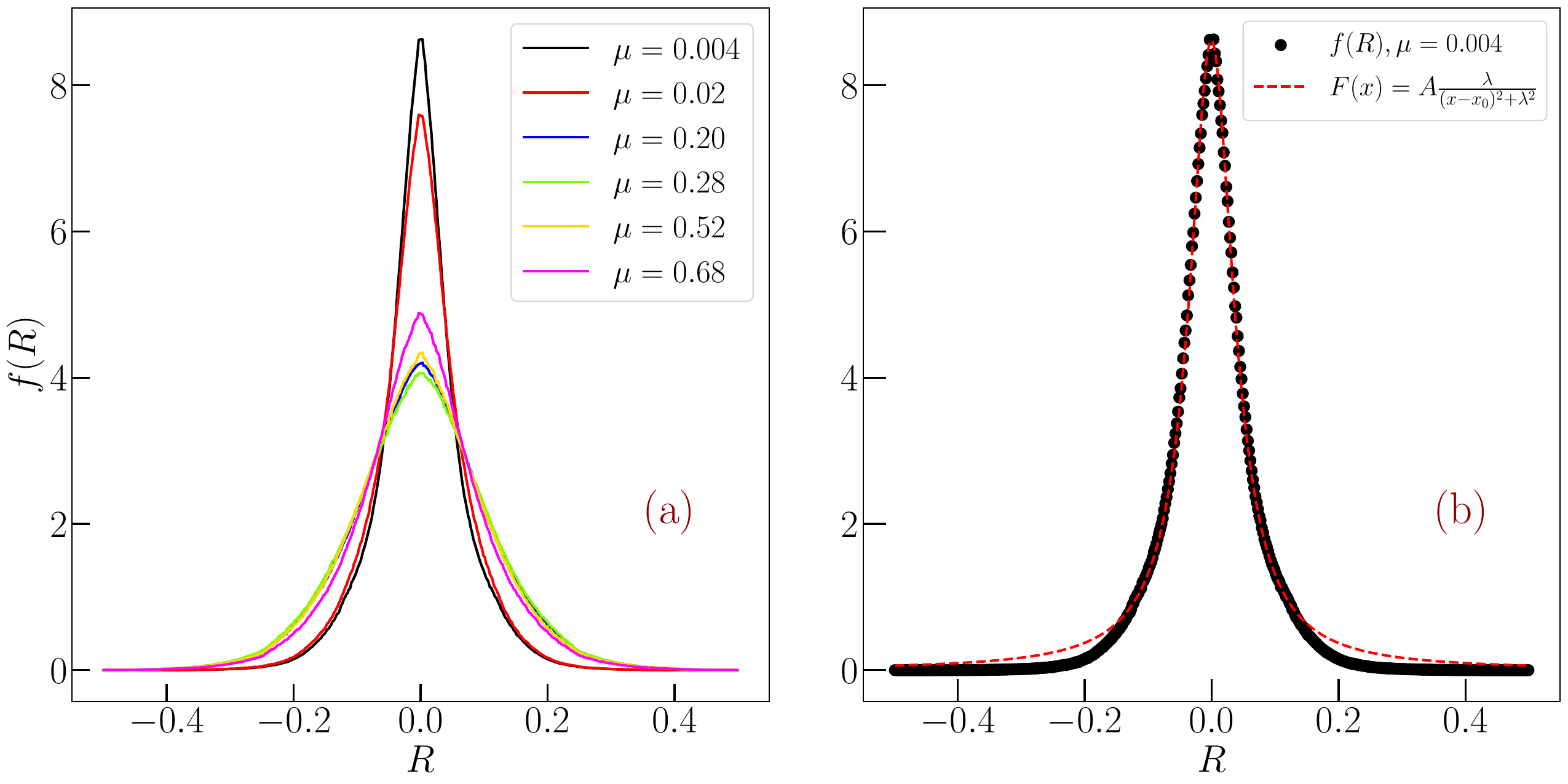}
		\caption{(Color Online) This figure illustrates (a) The PDF of $R$, $f(R)$, for different $\mu$ values in three different States: State-I ($\mu =$ $0.004$ $\&$ $0.02$), State-II ($\mu =$ $0.20$ $\&$ $0.28$), and State-III ($\mu =$ $0.52$ $\&$ $0.68$). $f(R)$ is averaged over 10 independent realizations; (b) Fitting of PDF $f(R)$ vs. $R$ with Lorentzian distribution. Parameters are the same as FIG.\ref{fig:3} in the main text. }   
		\label{fig:E1}
	\end{center}
\end{figure}

\section{Modelling apolar particles with spatiotemporally correlated colored noise}\label{app:E}
\subsection{Interpretation of the effect of polar particles on apolar particles as colored noise}\label{app:E1}
Here we delve into the interplay between polar and apolar particles within a system, particularly emphasizing the influence of polar particles on the apolar particles. Within our theoretical framework, each particle, based on its interaction radius, engages with surrounding particles, with the interaction's nature being contingent upon the particle's symmetry—polar or apolar. This context raises a pertinent question: Can an apolar particle, devoid of any polar particles within its interaction radius, still experience the effects of polar particles? The answer is, yes. This is due to cascading of information inherent to all interacting active matter systems. This phenomenon in the context of our model happens in such a way  that an  apolar particle can be affected by polar particles even in their absence within the interaction radius. This is achieved through a domino effect, where information is transferred from one particle to another via nearest-neighbor interactions. Hence the  alignment among apolar particles is invariably disturbed by the presence of polar particles in the system. This comes in addition to the  noise already present in the system. This disturbance is not static; it changes with the motion of polar particles clusters, leading to a dynamic and correlated perturbation  in both space and time.\\
There are several studies examining the effect of quenched obstacles (disorder) on active matter systems \cite{mccandlish2012spontaneous,chepizhko2013optimal,sandor2017dynamic,singh2021bond,kumar2022active,kummel2015formation,chepizhko2015active,vahabli2023emergence}. In the quenched case, the obstacles are fixed in space and time. Therefore, the spatiotemporal dependence of the effect of the obstacle on the particles is due to the motion of the particles. In contrast, in our case, the obstacles (polar particles) themselves are motile, introducing additional spatiotemporal dependence. Thus, the effect of the obstacles (polar particles) on the apolar particles strongly depends on how these polar particles or clusters move. This intrinsic distinction sets our model apart from other active matter systems with quenched obstacles.\\
Further, recent studies \cite{vahabli2023emergence,chepizhko2015active} have revealed that disorder in active matter systems can induce collective rotational motions. In our model, the introduction of polar particles into the system leads to the emergence of swirl-like patterns among the apolar particles, visible within a specific ranges of polar particle activity. The apparent similarity of the these observations prompts us to raise the question:  Can the influence of polar particles on apolar particles be effectively emulated by applying a spatiotemporally correlated colored field to the latter? Given that in our foundational model, the primary mode of interaction is alignment among particles, addressing disorder as a field might prove to be a more apt approach than considering physical disorder directly.

\subsection{Equations of motion of active nematic with colored noise}\label{app:E2}
Here, we propose a theoretical framework for studying a collection of apolar particles subjected to spatiotemporally correlated colored noise. The mathematical model governing the dynamics of these particles, focusing on their position and orientation evolution, is as follows.\\
The posiotion update of the $i^{th}$ particle at subsequent time steps is defined by:
\begin{equation}
	\boldsymbol{r}_{i}(t+\Delta t) = \boldsymbol{r}_{i}(t)+  \boldsymbol{v}_{i}(t) \Delta t
	\label{equn:c1}
\end{equation}
where, $\boldsymbol{v}_{i}(t)$ is the instantaneous velocity of the $i^{th}$ particle at time $t$.\\
The orientation update mechanism for each particle involves alignment with nearby apolar particles within the interaction radius, modulated by the colored noise. The orintation update rule of each particle is the following:
\begin{equation}
	\theta_{i,ap}(t+1) = \dfrac{1}{2}
\bigg< \theta(t) \bigg>_{R_0} + \zeta_{i,ap}(t)
	\label{equn:c2}
\end{equation}
Where, the $1^{st}$ term on the right hand side captures the aforementioned effects and the second term represents a delta correlated Gaussian white noise with zero mean and uniformly distributed in the range $[-\frac{\eta}{2},\frac{\eta}{2}]$. The $1^{st}$ term is evaluated as follows:
\begin{center}
	$\bigg< \theta_{i,ap}(t) \bigg>_{R_0,ap} = \arg \bigg(  \mathlarger{\mathlarger{\sum}}_{|\boldsymbol{r}_{i,ap}-\boldsymbol{r}_{j,ap}| < R_{0,ap}} e^{i[2\theta_{j,ap}(t)+\xi(|\boldsymbol{r}_{i,ap}-\boldsymbol{r}_{j,ap}|,t)]} \bigg) $
\end{center}
In the above expression $R_0$ is the interaction radius and $\xi(r,t)$ represents the colored noise, with its spatial and temporal correlation described as: $\bigg<\xi(r,t)\xi(r=0,t=0) \bigg> = A e^{-\frac{r}{L_c}}e^{-\frac{t}{\tau_c}}$, where $A$ is the strength of the noise and , $L_c$ and $\tau_c$ are respectively the correlation length and correlation time of the noise, respectively.

\pagebreak
\subsection{Comparision of the structure of different states in the original model and active nematics with colored noise}\label{app:E3}
\renewcommand{\thefigure}{E\arabic{figure}}
\FloatBarrier
\begin{figure}[H]
\centering
		\text{\textbf{\color{black}Colored Noise}}\hspace{0.30\linewidth}\text{\textbf{\color{black}Original Model}}\\
\subfloat[]{\includegraphics[width=0.40\textwidth]{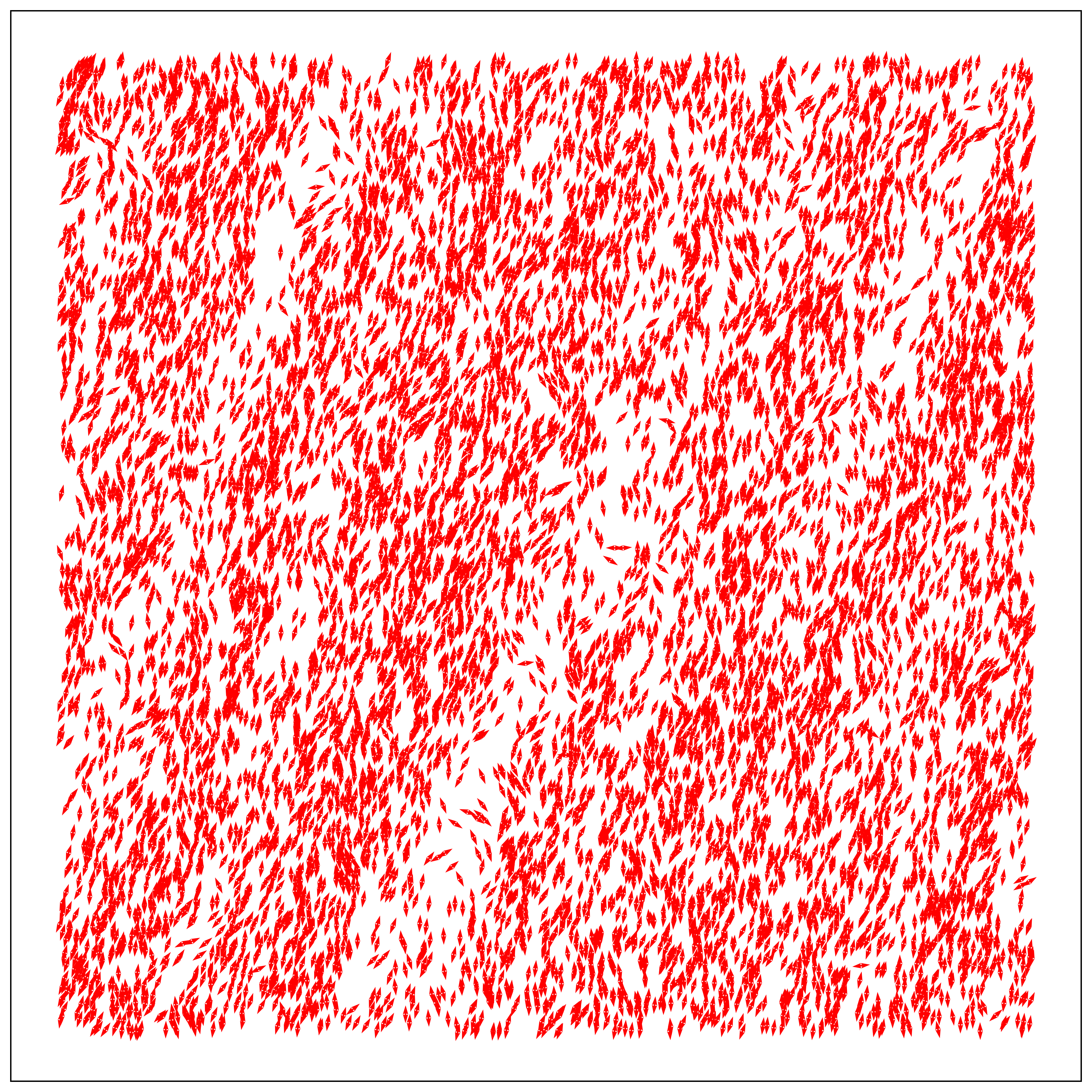}}
~
\subfloat[]{\includegraphics[width=0.40\textwidth]{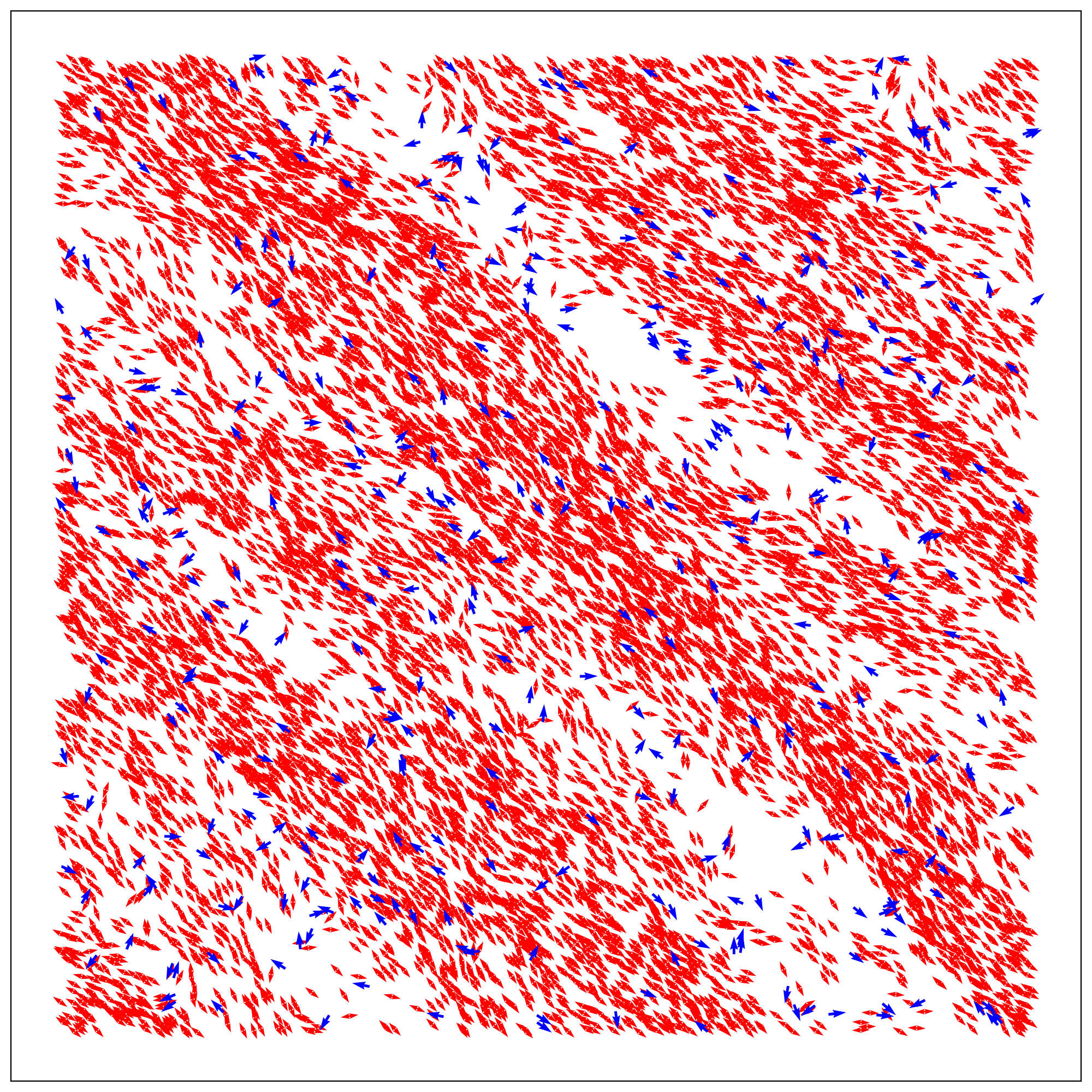}}

\subfloat[]{\includegraphics[width=0.40\textwidth]{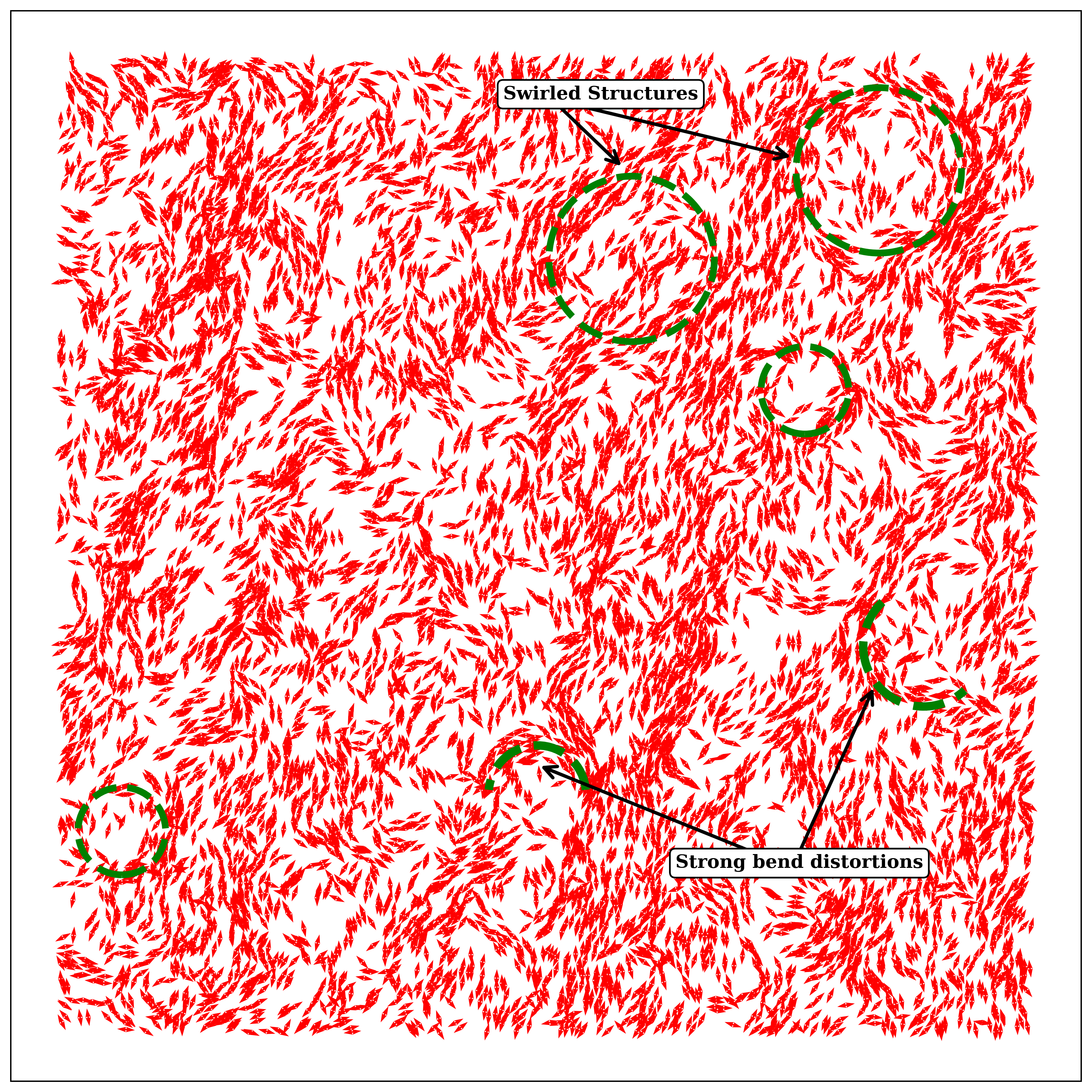}}
~
\subfloat[]{\includegraphics[width=0.40\textwidth]{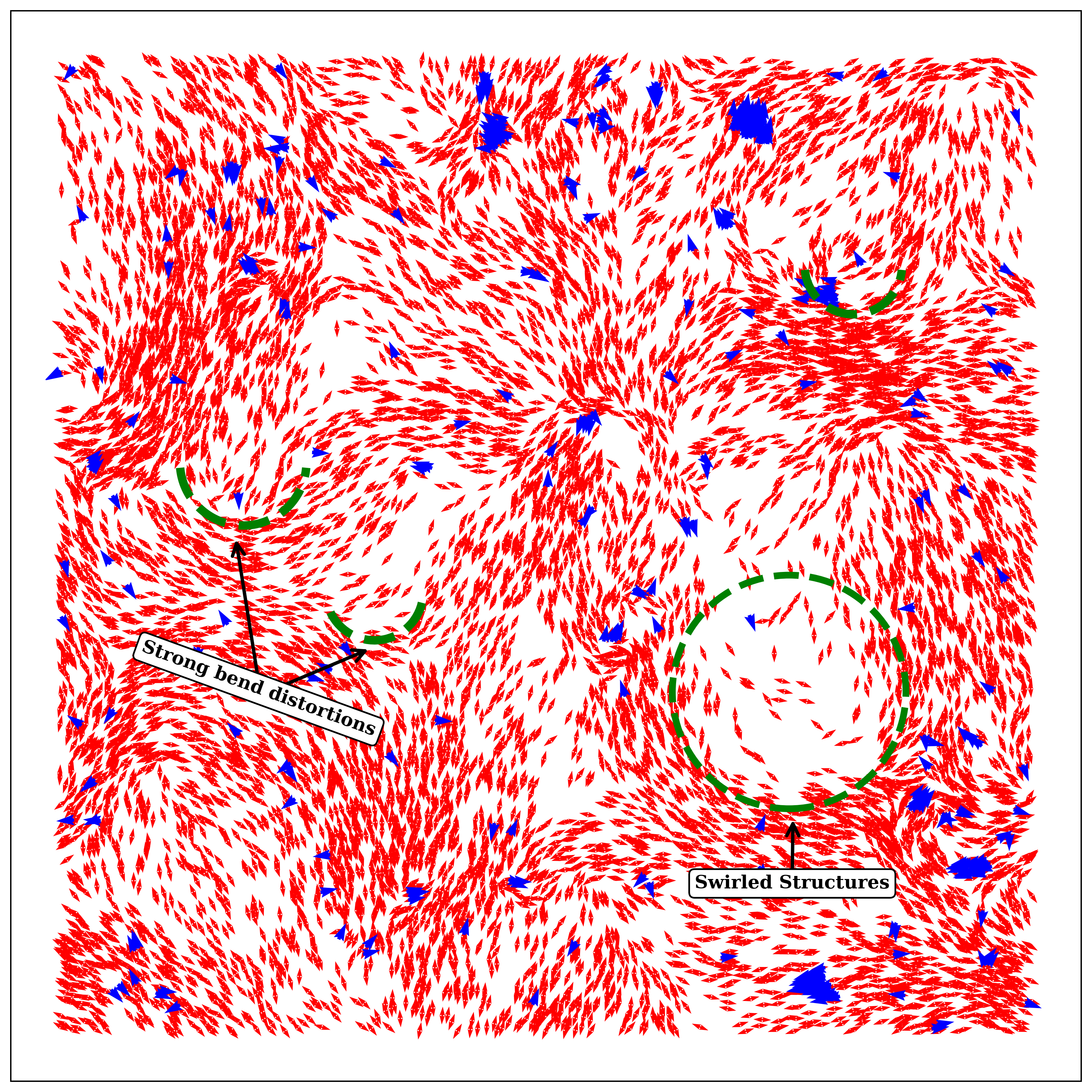}}

\caption{(color online) The figure depicts a visual comparison between the structures observed in the State-I and State-II in both the original model and the model of active nematics with a colored noise as discussed in the main text. In the first row, snapshots of State-I in the two models reveal a similar structure with the formation of stripe-like patterns in both the models. The second row features snapshots of State-II, we observe similarities such as the emergence of swirl and strong bend structures in the model with a colored field, akin to the original model. Some of these structures are highlighted in both the snapshots. Despite these shared features, distinctive visual differences are apparent between the two models. These differences offer valuable insights, as expounded upon in Sec.III(C) of the main text. The remaining parameters remain consistent with those detailed in FIG.\ref{fig:3} of the main text.}
\label{fig:e3}
\end{figure}
\FloatBarrier

\pagebreak
\section{Supplementary Material}\label{app:F}
\subsection{Description of the Supplementary Movies}\label{sec:sm1}
The supplementary movies (SM1-SM3) provide a visual representation of the dynamics of polar and apolar particles in different states, as discussed in Sec.III(A).  The movies are captured within the time steps $1.3 \times 10^5$ to $1.5 \times 10^5$. The specific  details of each movie are as follows:
\begin{itemize}
    \item {\textbf{SM1:}  This movie shows the dynamics of the particles in the system in State-I  for the activity ratio, $\mu = 0.004$. Rest of the parameters are same as in FIG.3 in the main text. (SM1.mp4)\\
    Link: \url{https://youtu.be/5u_nVrtrET0}
    }
    
    \item {\textbf{SM2:} This movie illustrates the dynamics  of the particles in State-II with the activity ratio, $\mu = 0.20$. Rest of the parameters are same as in FIG.3 in the main text. (SM2.mp4)\\
    Link: \url{https://youtu.be/iyld_KGpv-Y}
    }
    \item {\textbf{SM3:} This movie depicts the dynamics  of the particles in the system in State-III with the activity ratio, $\mu = 0.52$.  Rest of the parameters are same as in FIG.3 in the main text. (SM3.mp4)\\
    Link: \url{https://youtu.be/EsXY8qbFYaI}
    }
\end{itemize}
Additionally, the supplementary movie SM4 provides a visual description of the formation of a swirl structure in active nematics through their interaction with randomly moving polar clusters, as discussed in Sec.III(C2). The simulation begins with the initial condition of the active nematics identical to the steady state configuration in State-I (see FIG.3(a) in the main text), and the initial condition of polar particles similar to the steady state configuration in State-II (see FIG.3(b) in the main text), for a system size of $L = 100$. The animation is zoomed in to a small part of the whole system for convenience. (SM4.mp4)\\
Link: \url{https://youtu.be/HYGKscAr1Ag}

\subsection{Topological  defects in active nematics}\label{sec:sm2}
Topological defects with charges of $\pm \frac{1}{2}$ are distinguishing features in nematic systems \cite{de1993physics,schan,priestly2012introduction}. In equilibrium settings, such defects diffuse through the system, eventually annihilating in pairs of opposite charges as propelling the system towards an equilibrium state \cite{muzny1992direct,HaraP,Mondal_2024,harth2020topological}. The emergence of these defect pairs is associated with an energy cost, preventing their spontaneous formation in equilibrium systems. Therefore, as the system evolves, the quantity of defects decreases over time. The time scale of achieving a defect-free equilibrium state extends with the size of the system, leading to what is known as Quasi Long Range ordering. However in active nematic systems, owing to the local input of energy, spontaneous proliferation of defects are possible and the $\pm \frac{1}{2}$ defects persist even in the steady state. The asymmetrical shape, coupled with activity-induced stress, results in $+\frac{1}{2}$ defects exhibiting self-propelled particle-like motion within active nematic systems\cite{doostmohammadi2018active}.\\
In our study, we identify the presence of $\pm \frac{1}{2}$ defects in the steady state during State-II and State-III. These defects are visually identifiable, as demonstrated in FIG.\ref{fig:s1}. However, conducting a quantitative analysis of these defects presented challenges. Given that the particles in our model are point-like without physical shape, and we operate at a relatively low density ($\rho_{ap} \approx 1$), there are numerous void spaces within the system. This complexity impedes the effective application of algorithms for locating the defects and tracking it over time \cite{canova2016competing,jensen1992phenomenological,wenzel2021defects,delmarcelle1994topology}. Consequently, our exploration does not delve deeply into the quantitative dynamics of $\pm \frac{1}{2}$ defects in the system.

\FloatBarrier
\setcounter{figure}{0}
\renewcommand{\thefigure}{S\arabic{figure}}
\begin{figure}
\centering

\subfloat[]{\includegraphics[width=0.27\textwidth]{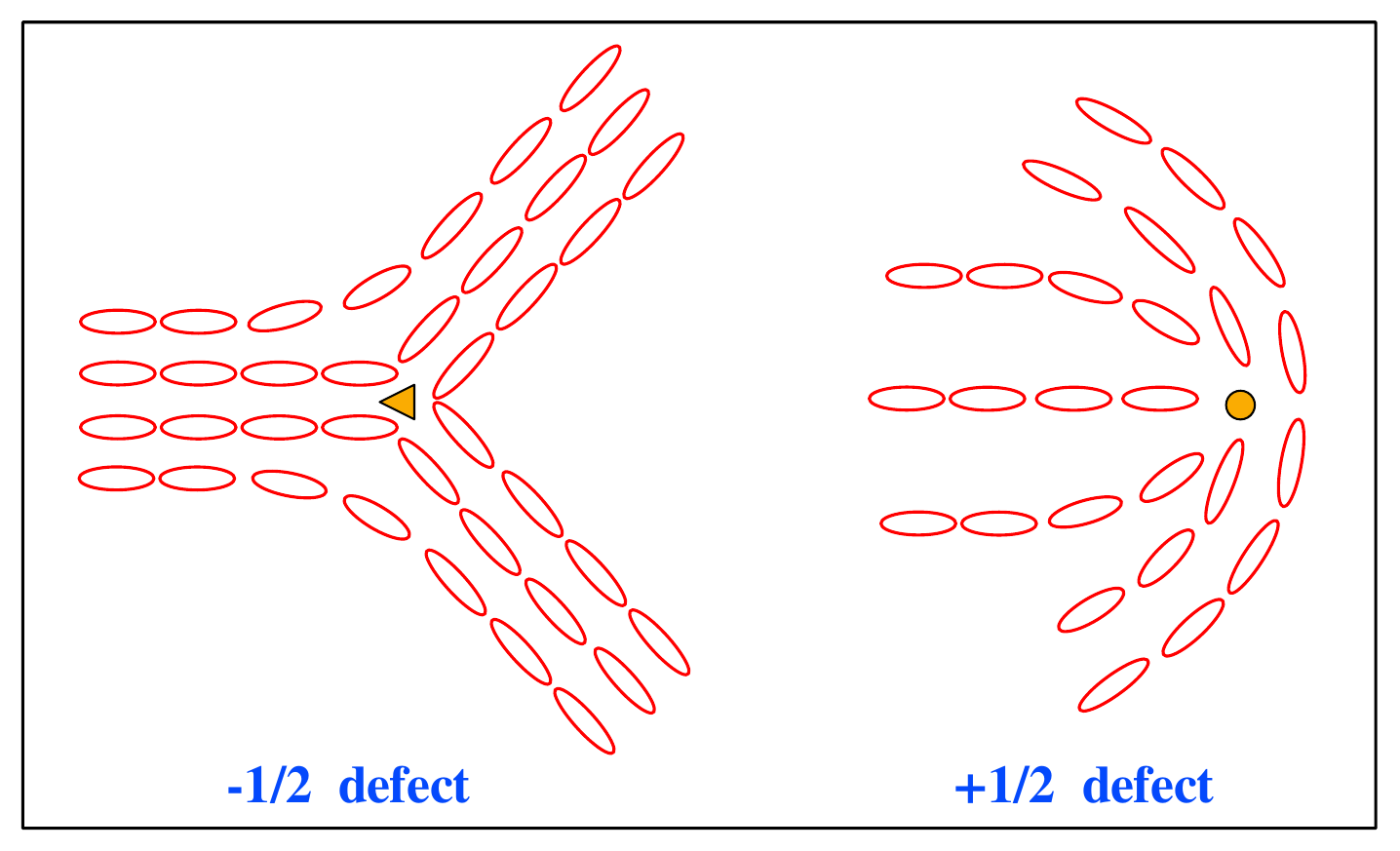}}
~
\subfloat[]{\includegraphics[width=0.27\textwidth]{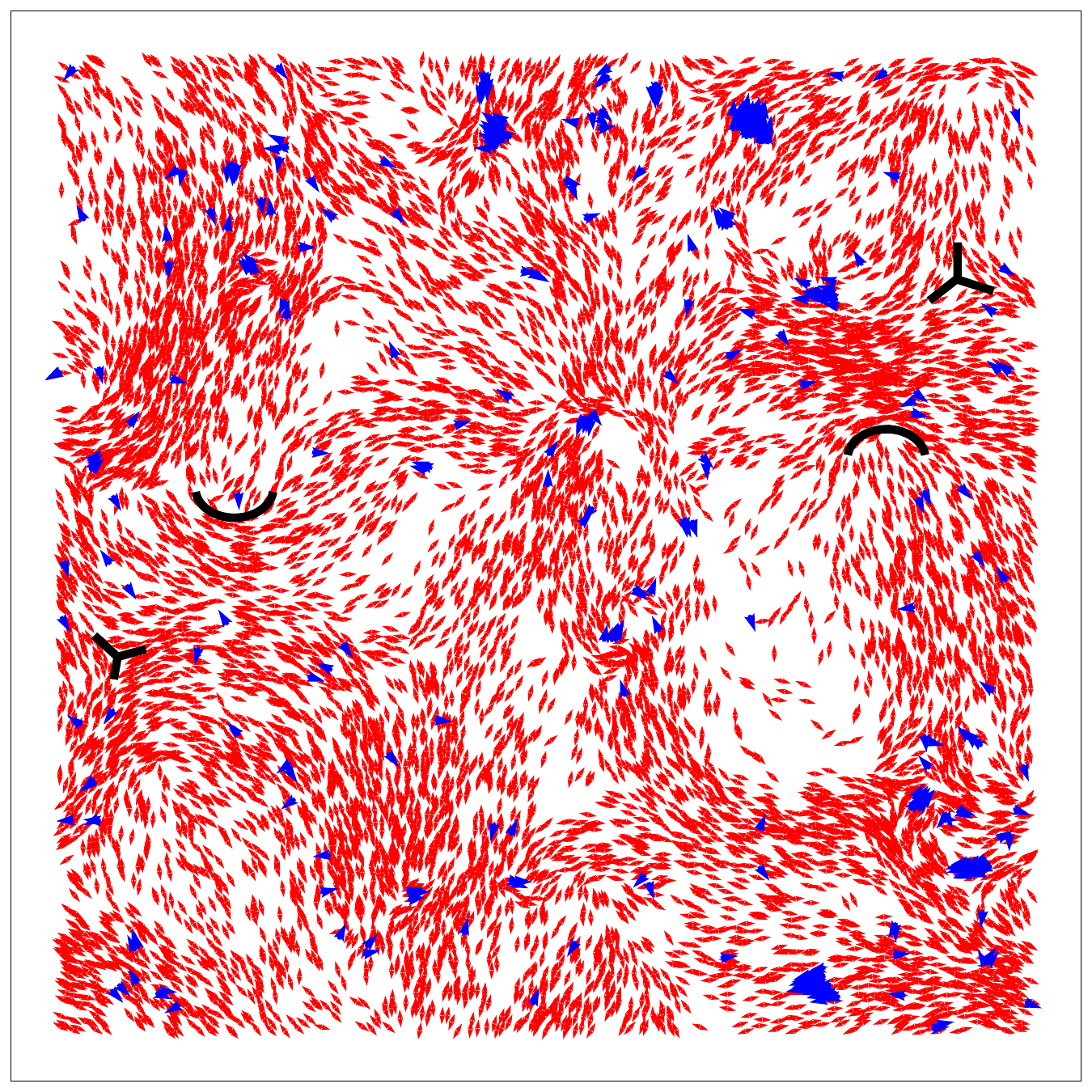}}
~
\subfloat[]{\includegraphics[width=0.27\textwidth]{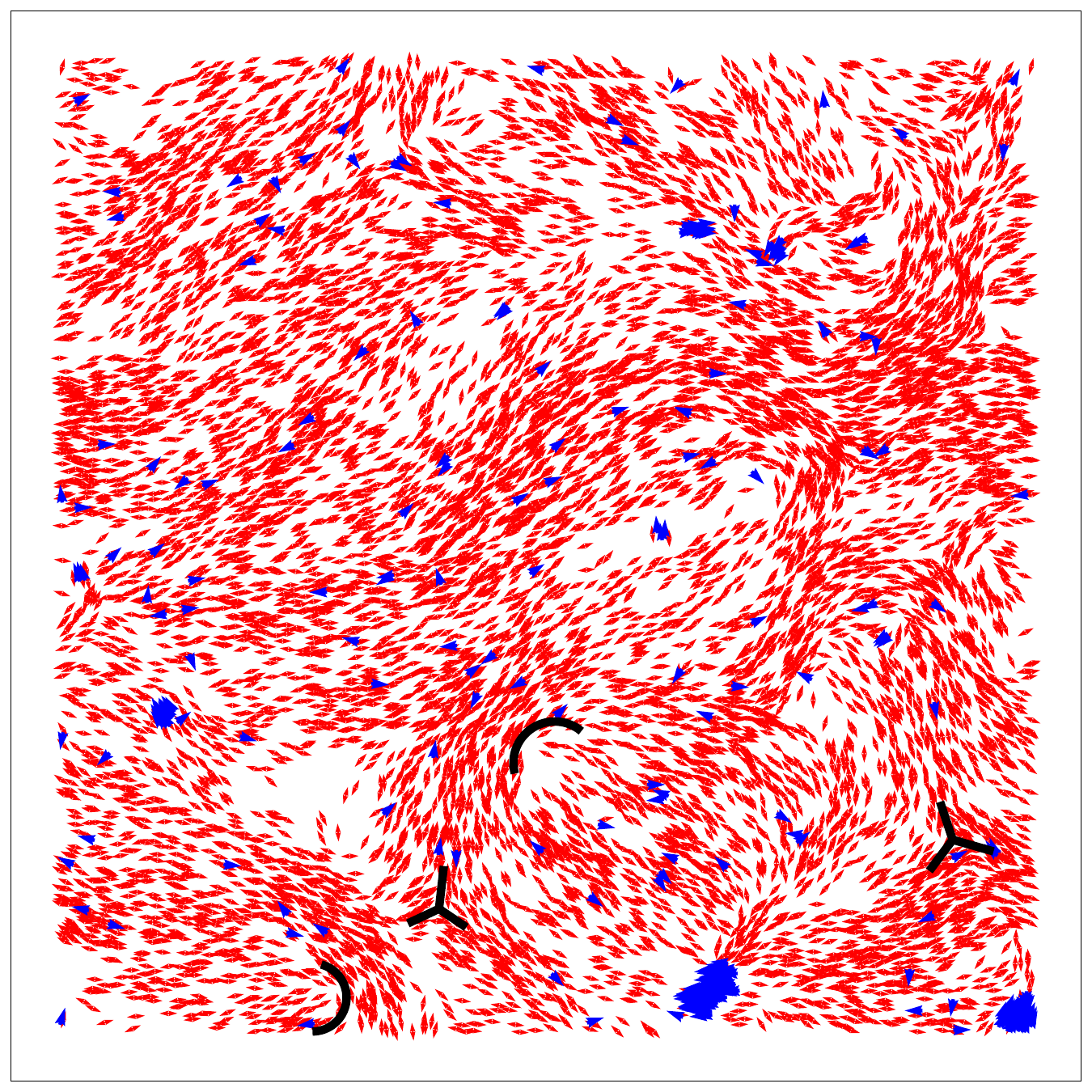}}

\caption{(color online) The figure depicts the $\pm \frac{1}{2}$ defects in active nematic systems. The structure of $+\frac{1}{2}$ and $-\frac{1}{2}$ defects are described in (a); In (b) and (c) some $+\frac{1}{2}$ and $-\frac{1}{2}$ defects are marked with black circulr arc and black 'Y' symbols. }
\label{fig:s1}
\end{figure}
\FloatBarrier

\subsection{Interaction of Polar clusters with Swirl Structures in State-II}\label{sec:sm3}

\FloatBarrier
\renewcommand{\thefigure}{S\arabic{figure}}
\begin{figure}[H]
\centering

\subfloat[]{\includegraphics[width=0.35\textwidth]{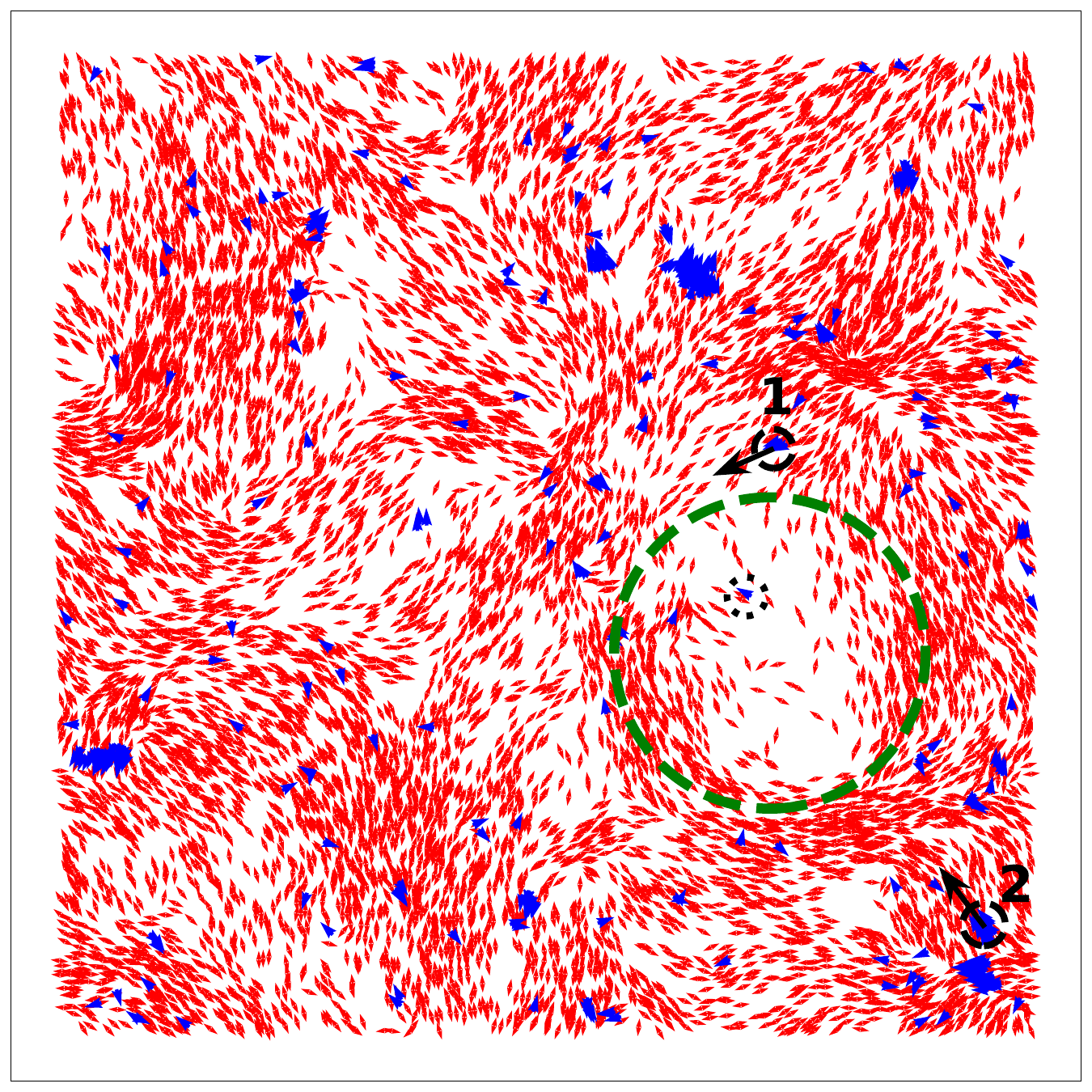}}
~
\subfloat[]{\includegraphics[width=0.35\textwidth]{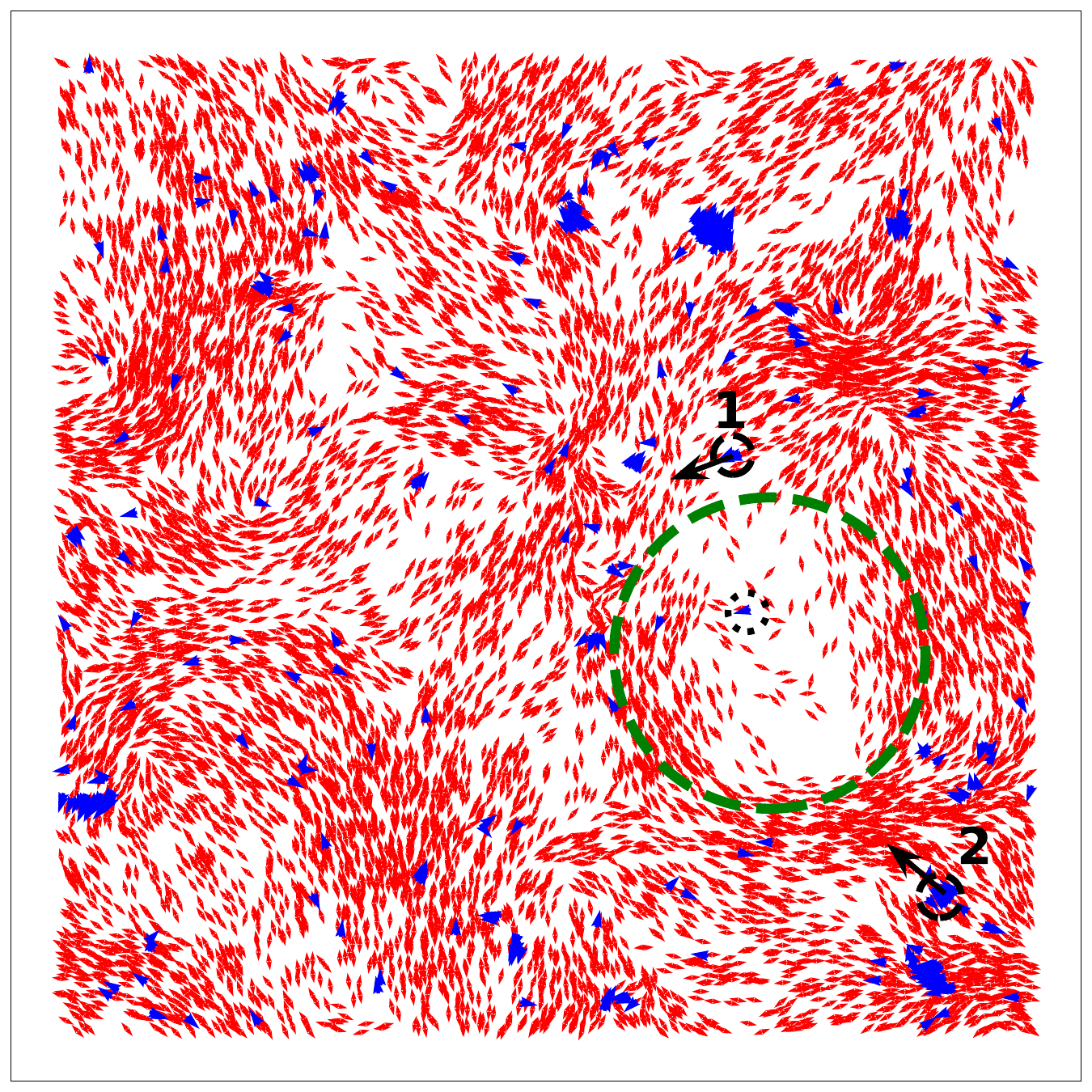}}

\subfloat[]{\includegraphics[width=0.35\textwidth]{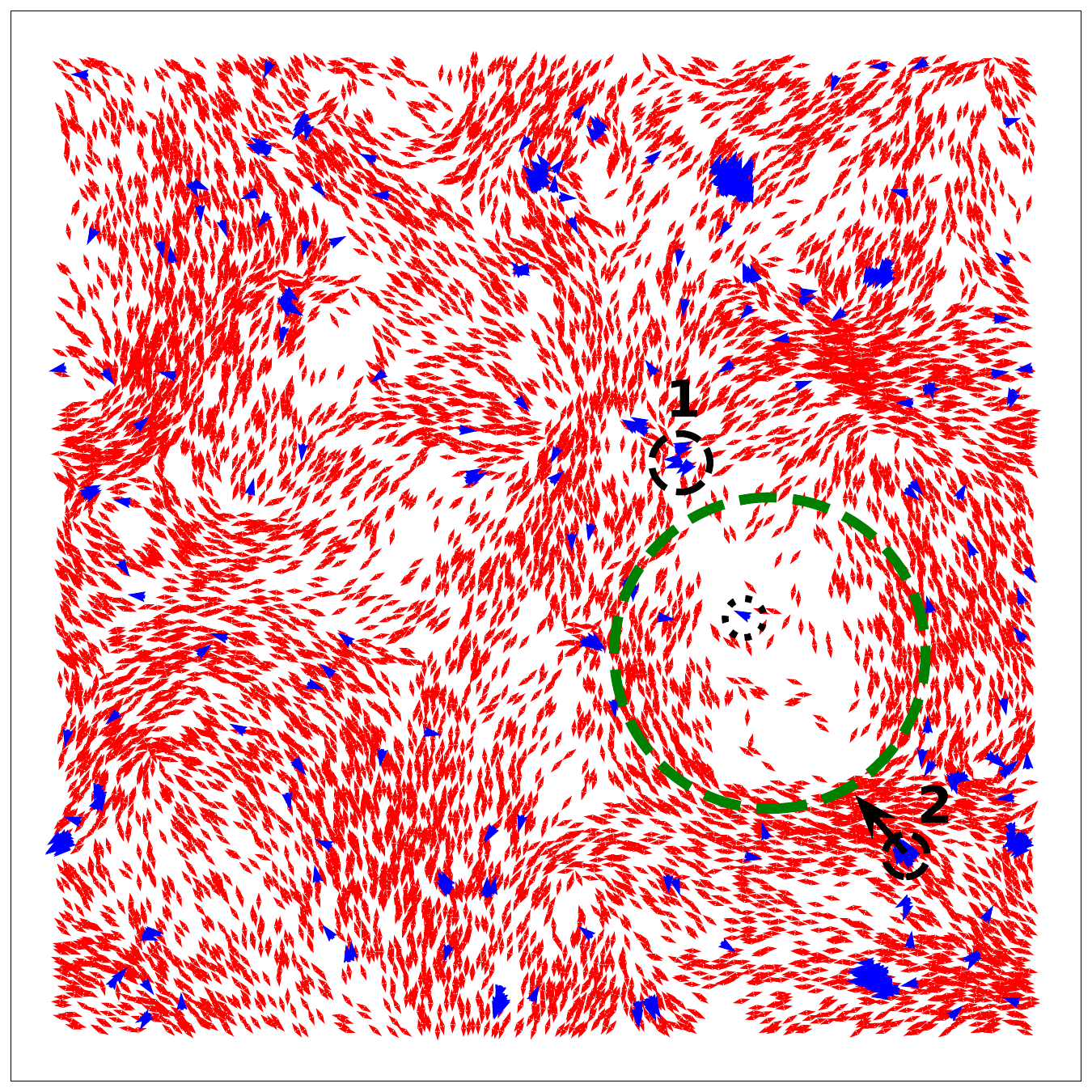}}
~
\subfloat[]{\includegraphics[width=0.35\textwidth]{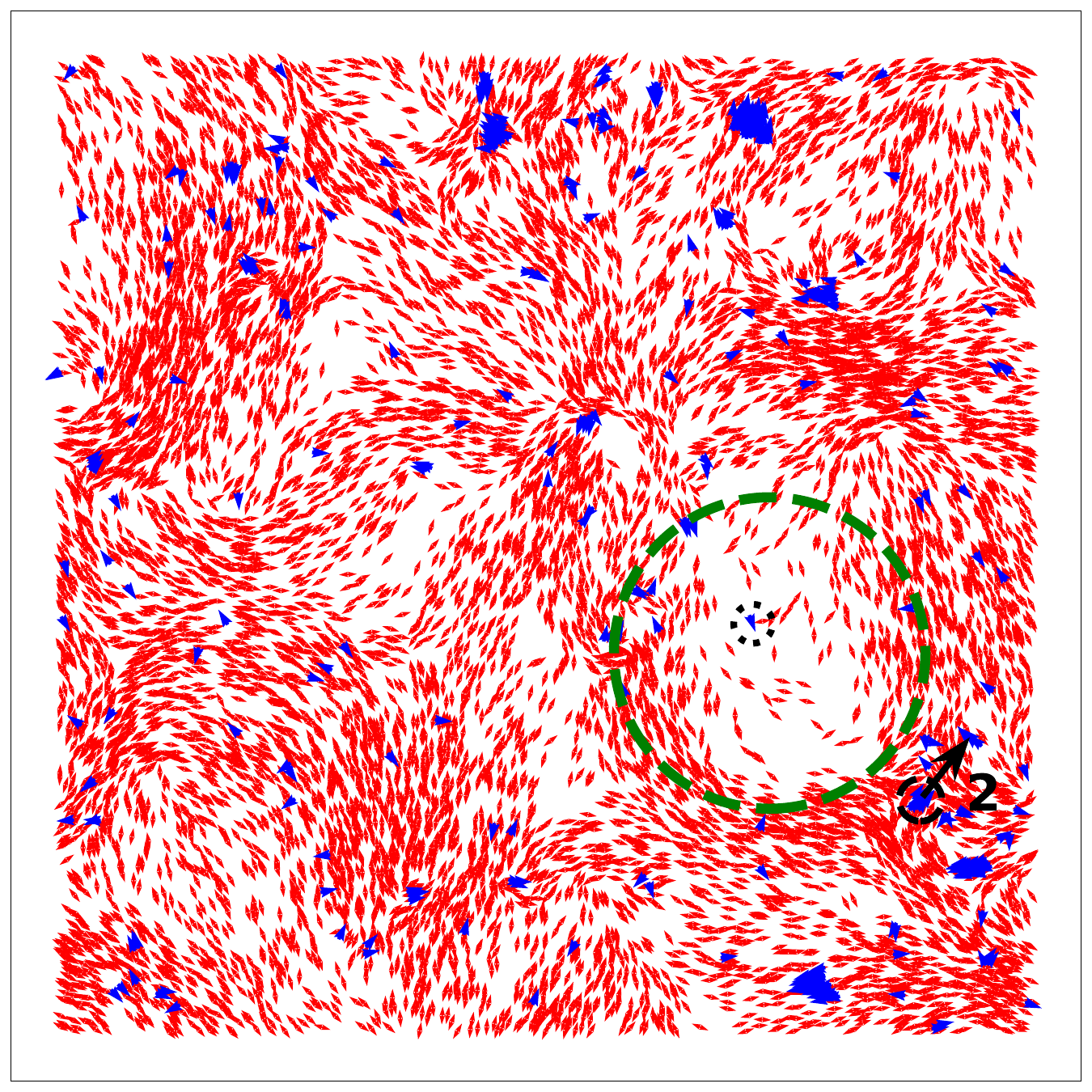}}

\caption{(color online) This figure captures an interaction between  Polar cluster and swirl structures in State-II of the mixture. Two distinct polar particle clusters are identified as 1 and 2, and their direction of motion is illustrated by black arrows. The boundary of the circular Vortex structure is indicated by a dashed green circle. The series of snapshots (a), (b), (c), and (d) depict the progressive evolution of the system over time. Upon interaction with the circular defect, the smaller cluster (1) undergoes dissociation, while the larger cluster (2) merely gets reoriented. Moreover, throughout all the snapshots, a dashed black circle highlights the presence of an isolated polar particle, which becomes entrapped at the center of the vortex and exhibits random motion. The parameters used in this simulation are the same as those of FIG.3 in the main text.}
\label{fig:F1}
\end{figure}
\FloatBarrier

\subsection{Interaction of Polar clusters with Swirl Structures in State-III}\label{sec:sm4}

\FloatBarrier
\renewcommand{\thefigure}{S\arabic{figure}}
\begin{figure}[H]
\centering

\subfloat[]{\includegraphics[width=0.35\textwidth]{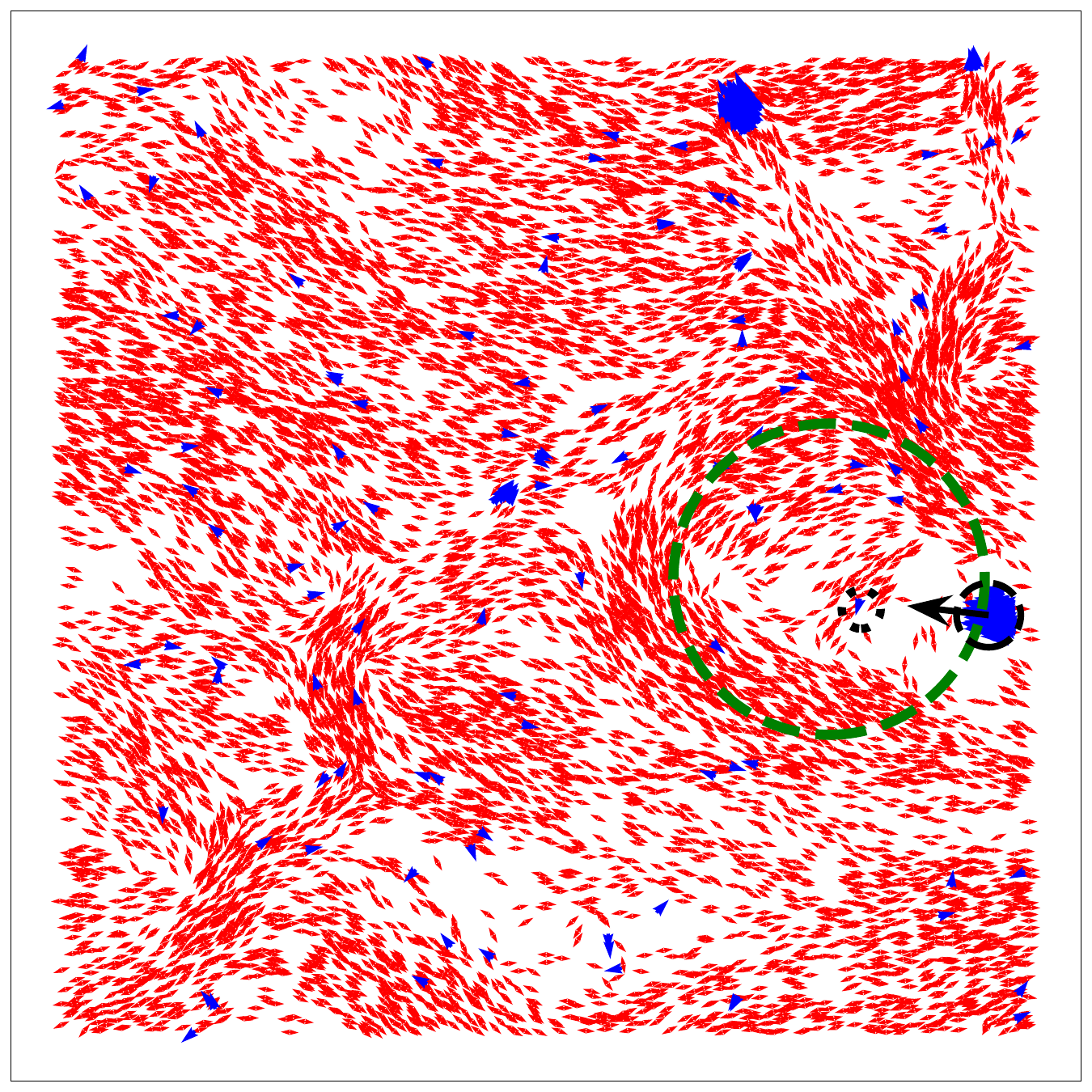}}
~
\subfloat[]{\includegraphics[width=0.35\textwidth]{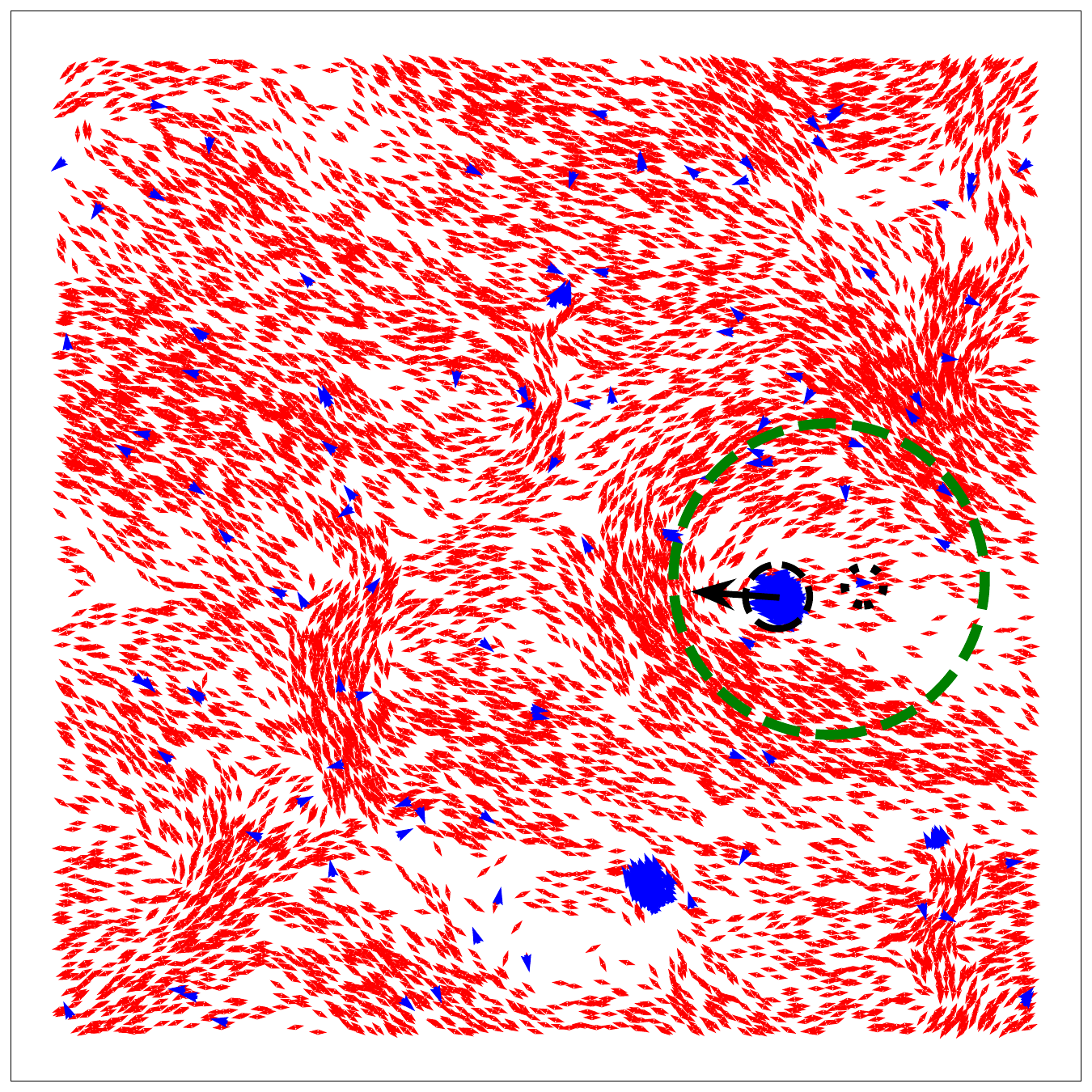}}

\subfloat[]{\includegraphics[width=0.35\textwidth]{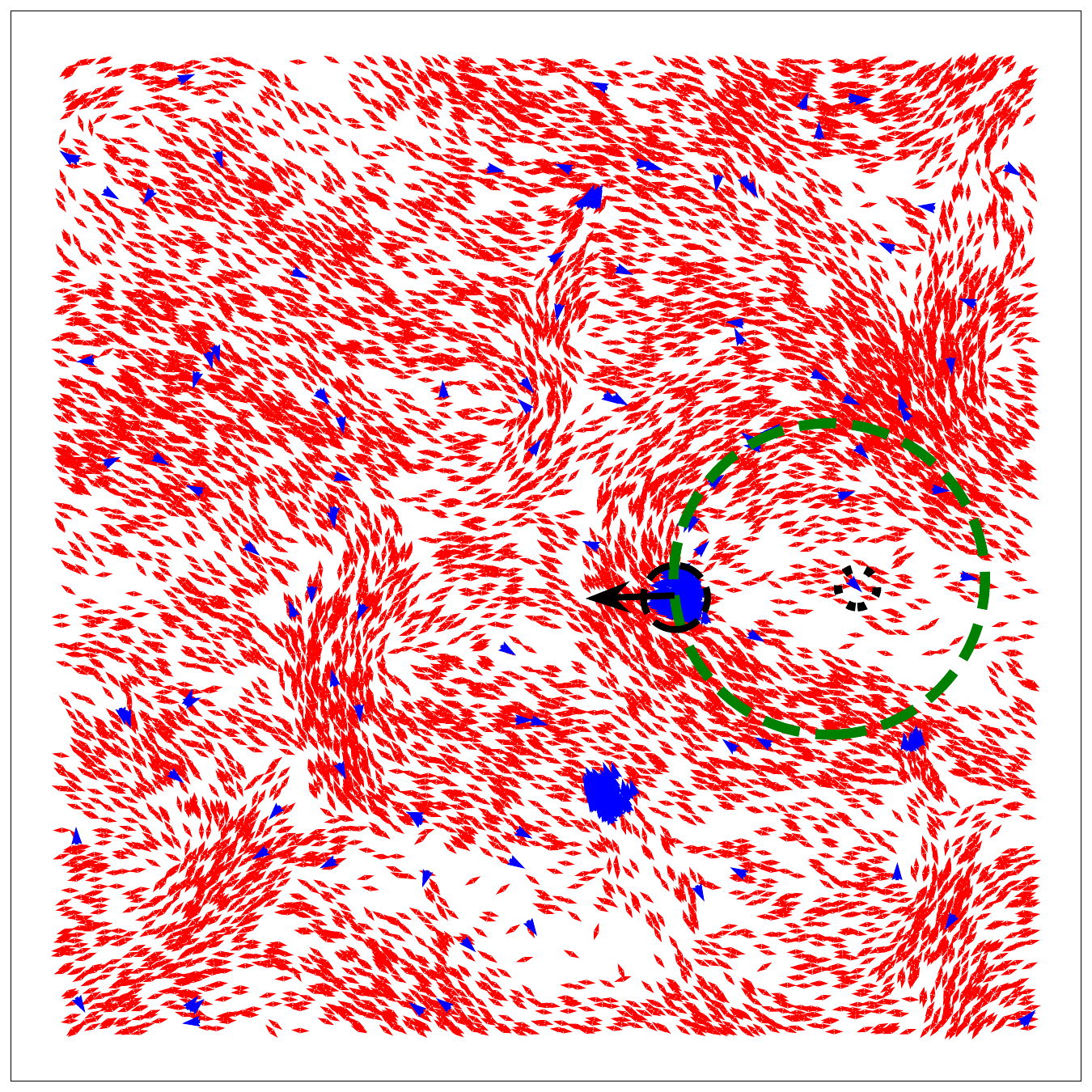}}
~
\subfloat[]{\includegraphics[width=0.35\textwidth]{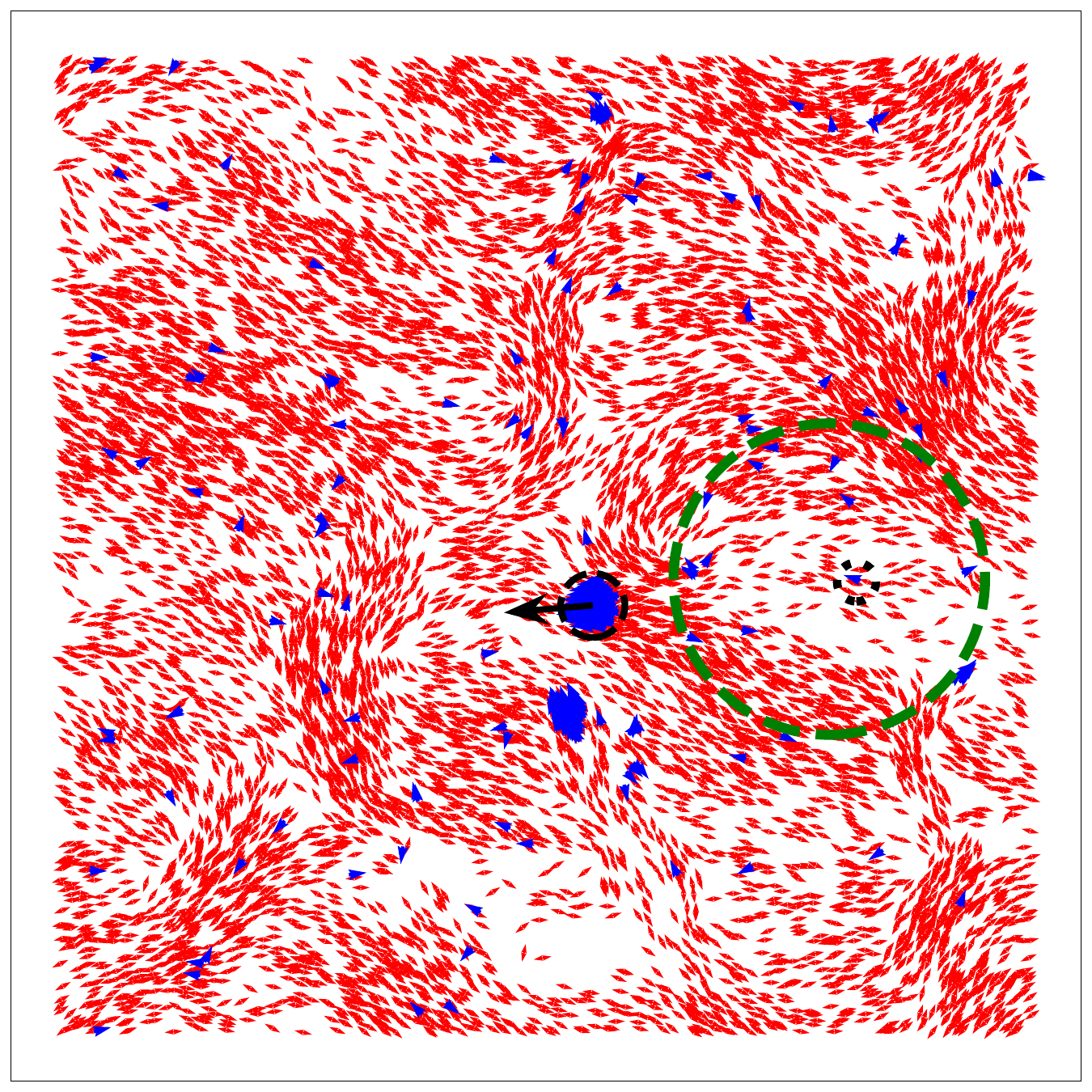}}

\caption{(color online) This figure illustrates an interaction between a Polar cluster and a strong swirl structure in State-III of the mixture. The motion of a Larger Polar particle cluster is tracked and   black arrow shows its direction of motion at a particular instant. The boundary of the circular defect is highlighted by a dashed green circle. Snapshots (a)-(d) exhibit the dynamic evolution of the system over time. Remarkably, in State-III, the larger polar cluster remains largely unaffected by the vortex structure. It successfully breaks through the vortex, restoring nematic ordering along its wake, showcasing the dominance of polar particle clusters in State-III of the mixture.	Throughout all the snapshots, a dotted circle is utilized to mark an isolated polar particle, which becomes entrapped at the center of the vortex. The remaining parameters are consistent with those shown in FIG.3 in the main text.}
\label{fig:G1}
\end{figure}
\FloatBarrier

\subsection{Swirl structures are intrinsic in the system}\label{sec:sm5}
Here, we discuss the intrinsic nature of the swirl structure with the help of the time series of SP, $\lambda$, and its probability density function (PDF), $f(\lambda)$ vs. $\lambda$. The details of the calculation of $f(\lambda)$ is similar to that of OPDF described in the Appendix.B in the main text.\\ 
The plots for the same are shown in FIG.\ref{fig:h1}. In State-I, the nematic is strongly ordered, and the swirly character of the nematic orientation field is minimal except at the edge of the stripes. This is represented by the a small mean value of $\lambda$. Further, the small fluctuations in $\lambda$ about the mean, which leads to a sharp $f(\lambda)$, implies that this state persists in the steady state of the system in State-I. In State-II, the mean value of $\lambda$ is much larger, which implies the presence of strong bend and swirl structures in the orientation field of apolar particles. The large fluctuations of $\lambda$ about the mean, leading to a broader $f(\lambda)$, can be interpreted as the dynamic nature of the swirl structures in the system, i.e., in the steady state, the swirl structures undergo formation, dissociation, and reformation. Moving on to State-III, the mean value of $\lambda$ decreases whereas the range of fluctuations around the mean is much larger. This implies the weak and transient nature of swirl structures of apolar particles in comparison to State-II. FIG.\ref{fig:h1} clearly shows that the mean value of $\lambda$ in State-II is approximately $2-2.5$ times larger than in  State-I. Compared to State-III, the value of $\lambda$ comes out to be $1.2-1.3$ times larger in State-II, and the fluctuations are much smaller. The argument presented here clearly shows that the swirl structures are much stronger and persist much longer in State-II compared to the other two States.

\FloatBarrier
\renewcommand{\thefigure}{S\arabic{figure}}
\begin{figure}[htp]
	\begin{center}
		\includegraphics[width=0.80\linewidth]{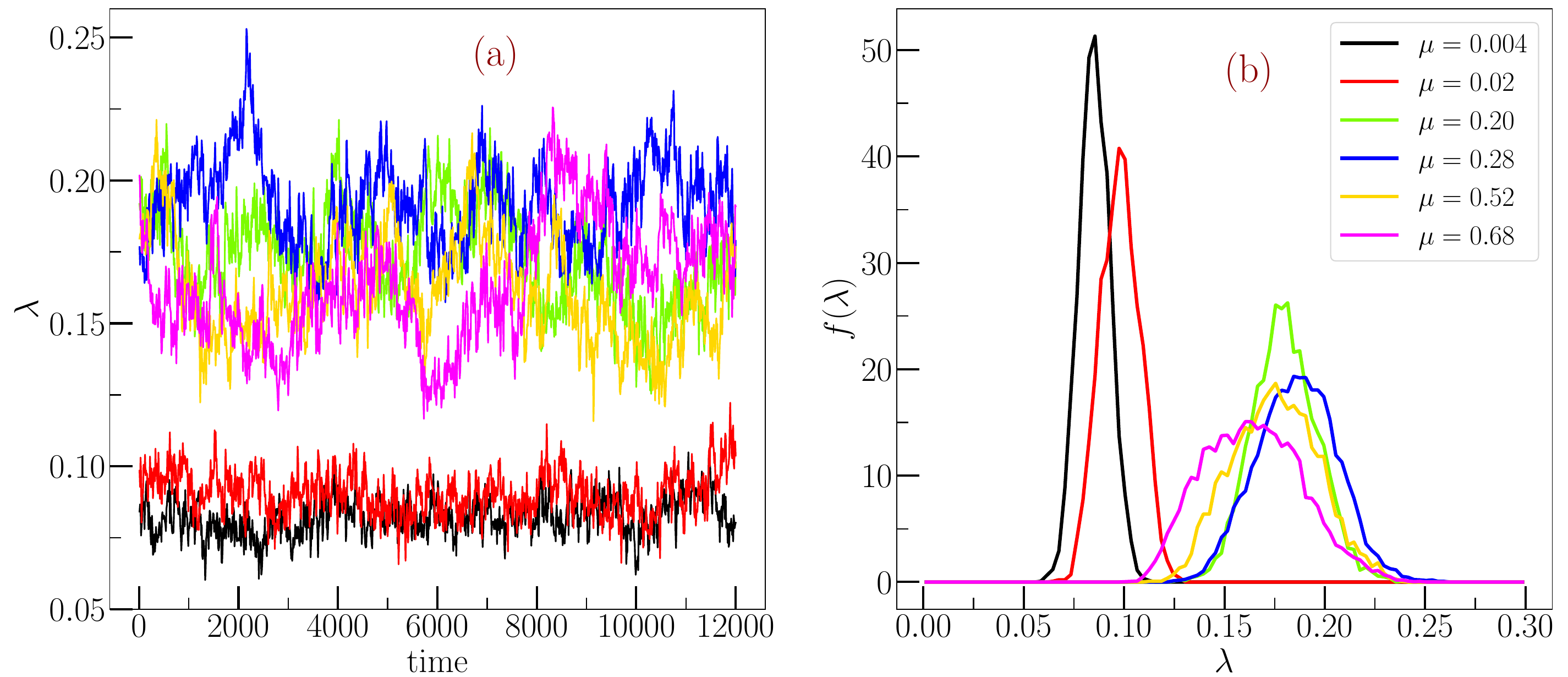}
		\caption{(Color online) The figure illustrates the characteristics of the SP, $\lambda$, in the steady state of the system in the three states. The (a) panel displays the the time series of SP, $\lambda$, in steady state, and the (b) panel shows the PDF of $\lambda$. In both (a) and (b) panels different colors are used for different $\mu$ values in three different states : State-I ($\mu =$ $0.004$ $\&$ $0.02$), State-II ($\mu =$ $0.20$ $\&$ $0.28$), and State-III ($\mu =$ $0.52$ $\&$ $0.68$). The parameters are same as FIG.3 in main text. }
		\label{fig:h1}
	\end{center}
\end{figure}
\FloatBarrier

\subsection{Adding chirality to the polar particles}\label{sec:sm6}
 Our study shows that the motile nature of the polar particle promotes the formation of dynamic swirl structures in active nematics. Till now, we have examined the effect of achiral polar particles on the active nematics. However, in nature, most of the natural microswimmers are chiral \cite{berg1990chemotaxis,diluzio2005escherichia,lauga2006swimming,schmidt2008non}. Hence, to understand the effect of the chirality of microswimmers on the system, we conducted additional simulations with chiral polar particles.
 The modified orientation update equation of the polar particles in the presence of chirality is the following: 
\begin{equation}
	\theta_{i,p}(t+1) = \bigg< \theta(t) \bigg>_{R_{0,p}} + \omega + \zeta_{i,p}(t)
\end{equation}
where the terms on the right-hand side, apart from the newly introduced second term, remain consistent with our original model. This second term represents the chirality effect, with $\omega$ denoting the chirality coefficient that can be both positive and negative, representing right and left handed chirality, respectively.\\
The introduction of chirality adds a rotational bias to the movement of polar particles, introducing a new rotational length scale proportional to $\frac{v_p}{\omega}$.  This creates a complex interplay between the activity of particles and chirality, potentially leading to a range of exciting features within the system \cite{liebchen2017collective,semwal2024macro,ventejou2021susceptibility}. Our preliminary analysis indicates that for certain ranges of activity and chirality, there is a noticeable enhancement in the stability of swirled structures. In FIG.\ref{fig:l1}, we illustrate through a sequence of snapshots how a cluster of polar particles moves in a circular trajectory within the swirl structures, significantly extending the persistence of these structures.\\
The comprehensive examination of the full spectrum of effects on the dynamics of our system due to chirality remains an area for future investigation.

\FloatBarrier
\renewcommand{\thefigure}{S\arabic{figure}}
\begin{figure}
\centering

\subfloat[]{\includegraphics[width=0.25\textwidth]{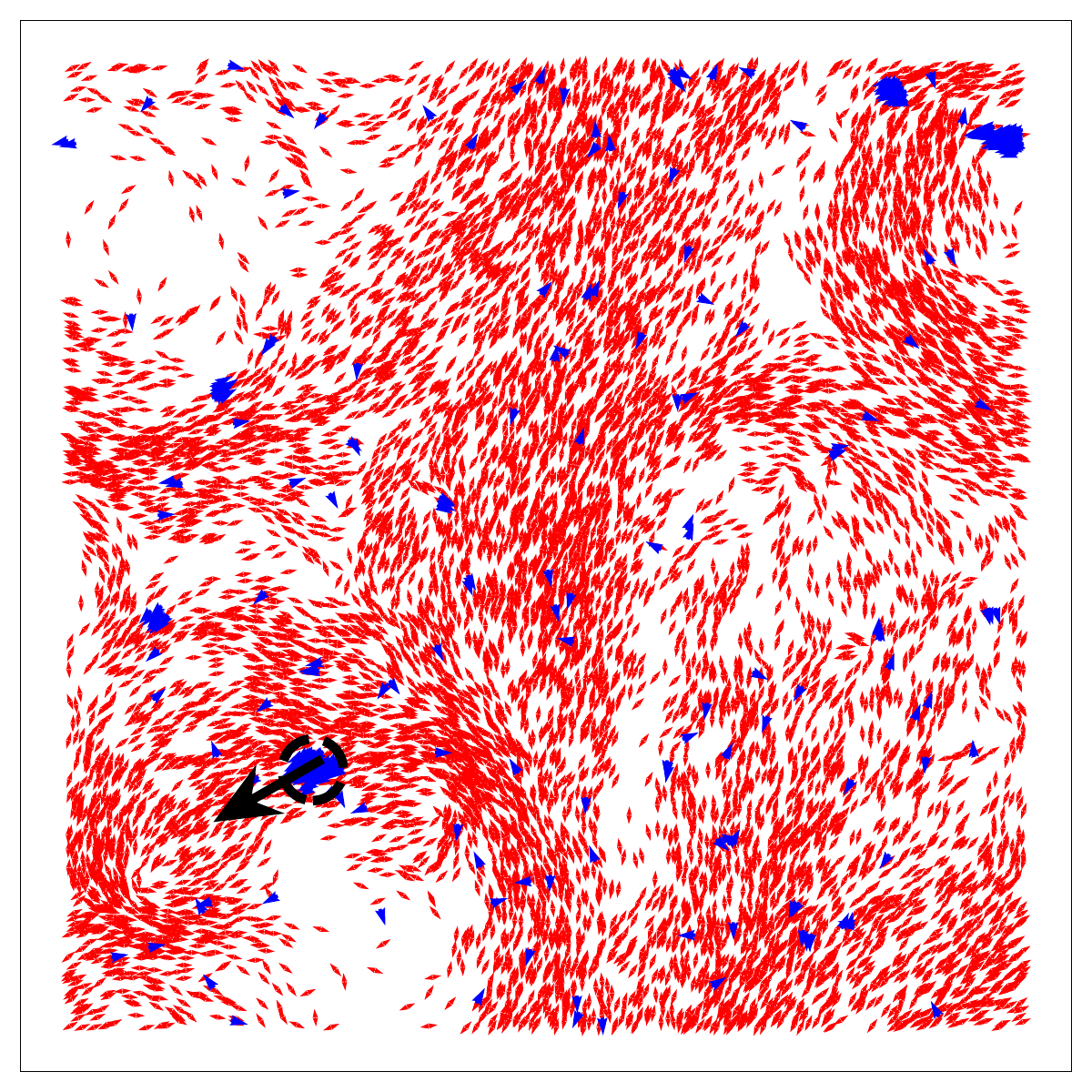}}
~
\subfloat[]{\includegraphics[width=0.25\textwidth]{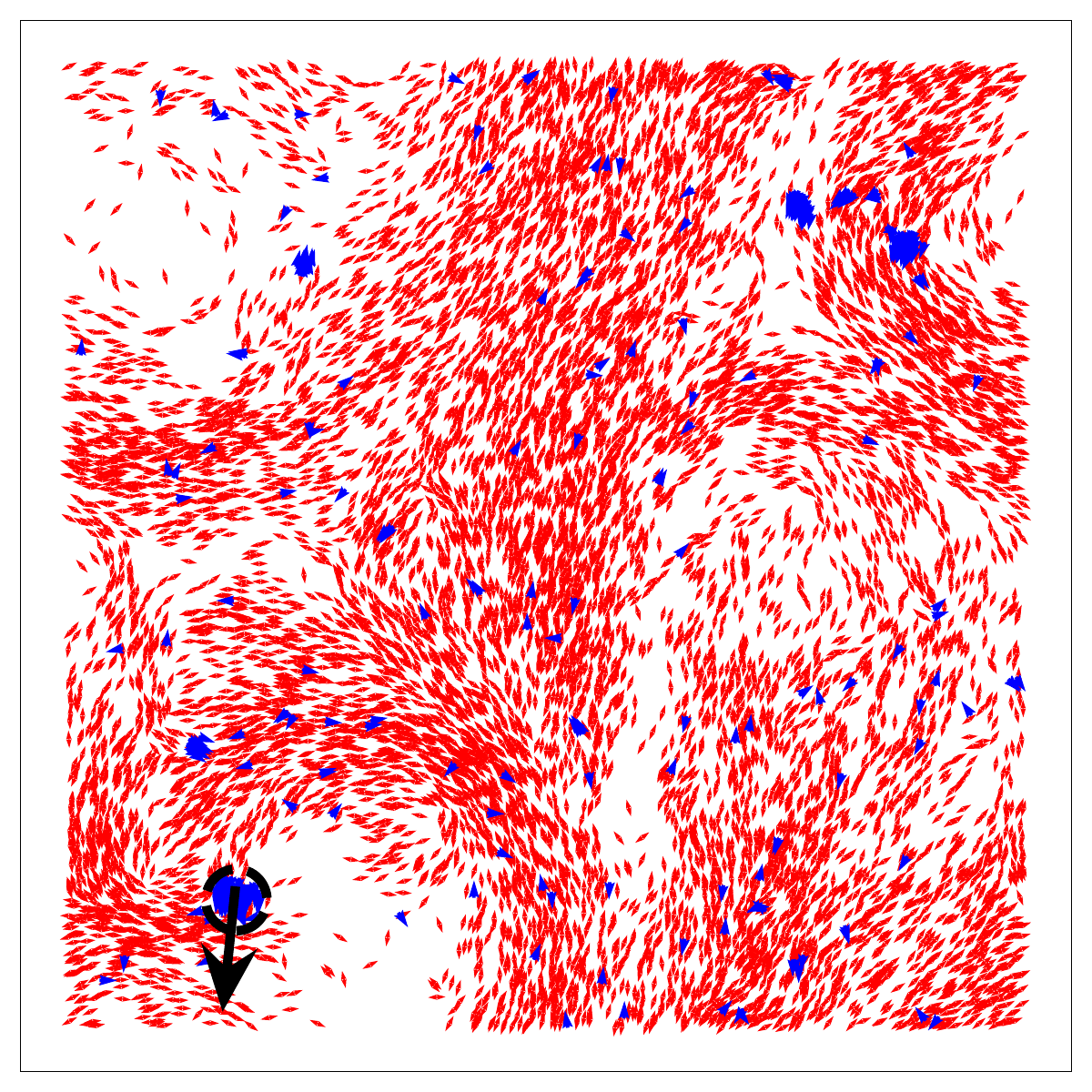}}
~
\subfloat[]{\includegraphics[width=0.25\textwidth]{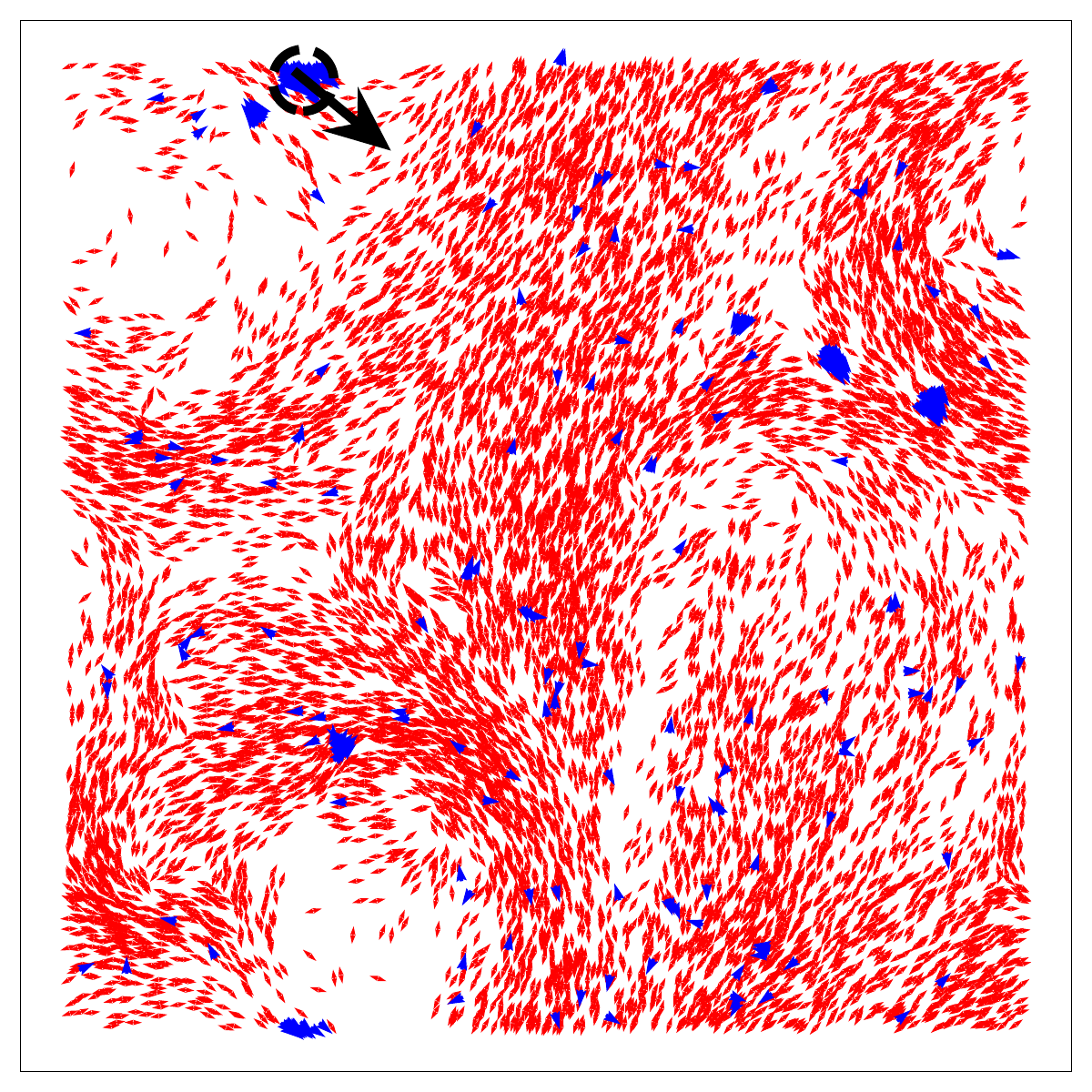}}

\subfloat[]{\includegraphics[width=0.25\textwidth]{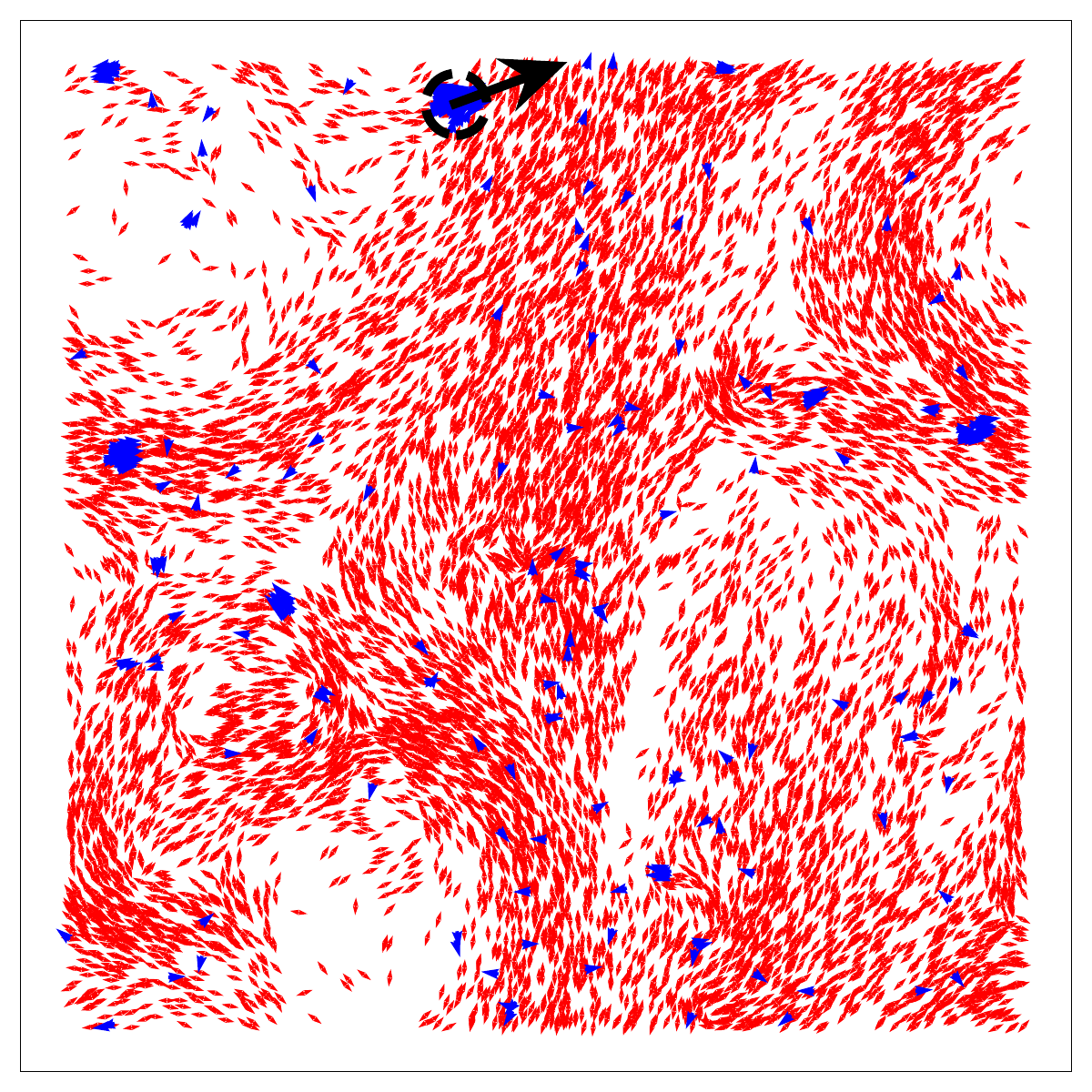}}
~
\subfloat[]{\includegraphics[width=0.25\textwidth]{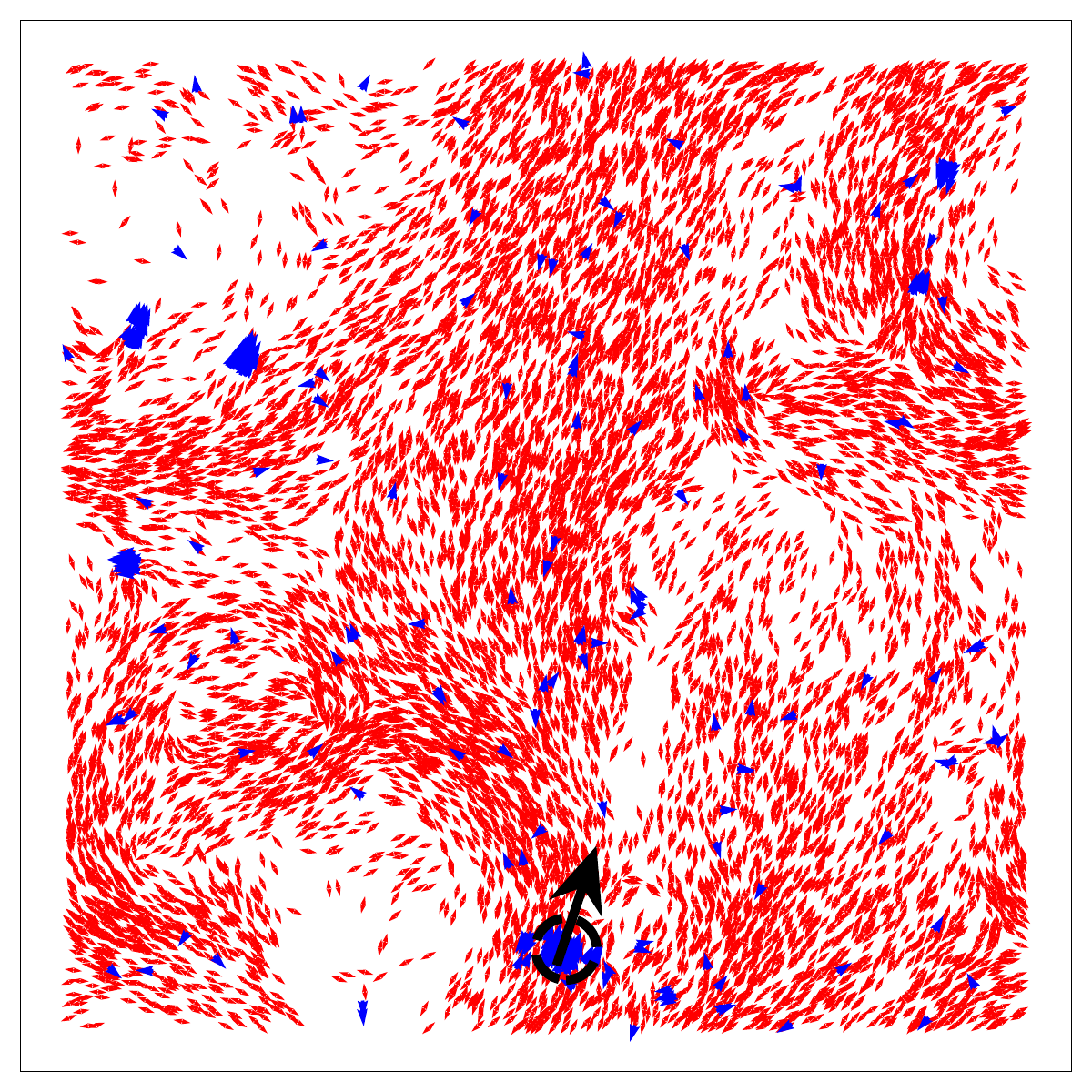}}
~
\subfloat[]{\includegraphics[width=0.25\textwidth]{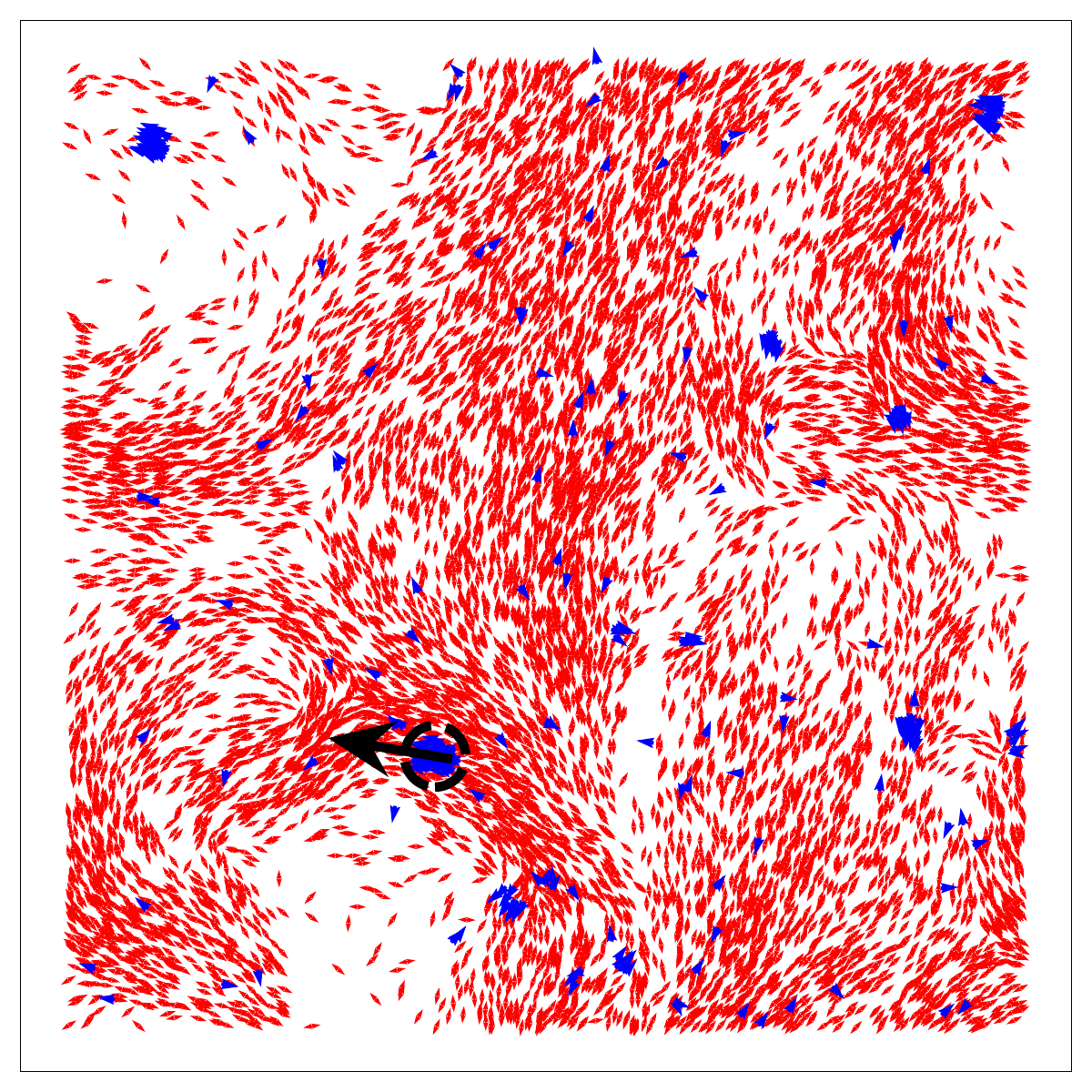}}

\caption{(color online) This figure highlights the impact of chirality on the stabilization of swirl structures, illustrated through snapshots (a)-(f) taken at successive time intervals. These images depict a polar cluster (marked by black dashed circle and the direction of motion of the cluster is shown with black arrow) as it travels in a counter clockwise  circular motion within the swirl structures, significantly enhancing their  persistence for $\mu = 0.68$ and $\omega = 0.01$. Rest of the parameters are same as in FIG.3 in the main text.}
\label{fig:l1}
\end{figure}
\FloatBarrier

 \bibliographystyle{elsarticle-num} 
 \bibliography{cas-refs}





\end{document}